\documentclass[10pt,preprint]{aastex}        
\usepackage{graphicx,amssymb,natbib,longtable,lscape}
\usepackage{times,datetime}

\shorttitle{ALFALFA Dwarf Galaxies}
\shortauthors{S. Huang}

\begin{document}

\title{Gas, Stars and Star Formation in ALFALFA Dwarf Galaxies\footnote
{Based on observations made with the Arecibo Observatory 
and the NASA Galaxy Evolution Explorer (GALEX). 
The Arecibo Observatory is operated by SRI International under a cooperative agreement with the 
National Science Foundation (AST-1100968), and in alliance with Ana G. M\'endez-Universidad Metropolitana, 
and the Universities Space Research Association.
GALEX is operated for NASA by the California Institute of Technology under NASA contract NAS5-98034.}}
\author{Shan Huang, Martha P. Haynes, Riccardo Giovanelli\altaffilmark{1}}
\affil{Center for Radiophysics and Space Research, Space Sciences Building, Cornell University, Ithaca, NY 14853.}
\email{shan@astro.cornell.edu, haynes@astro.cornell.edu, riccardo@astro.cornell.edu}
\author{Jarle Brinchmann\altaffilmark{2}}\affil{Sterrewacht Leiden, Leiden University, NL-2300 RA Leiden, 
       The Netherlands} \email{jarle@strw.leidenuniv.nl}
\author{Sabrina Stierwalt\altaffilmark{3}}\affil{Spitzer Science Center, 
       California Institute of Technology, 1200 E. California Blvd., Pasadena, CA 91125} \email{sabrina@ipac.caltech.edu}
\author{Susan G. Neff\altaffilmark{4}}\affil{NASA GSFC, Code 665, Observational Cosmology Lab, 
        Greenbelt, MD 20771} \email{susan.g.neff@nasa.gov}

\begin{abstract} 
We examine the global properties of the stellar and HI components
of 229 low HI mass dwarf galaxies extracted from
the ALFALFA survey,
including  a complete sample of
176 galaxies with  HI masses $< 10^{7.7} {\rm M_\odot}$ and HI line widths $<$ 80 km s$^{-1}$.
SDSS data are combined with photometric properties derived from GALEX 
to derive stellar masses ($M_*$) and star formation rates (SFRs) by
fitting their UV-optical spectral energy distributions (SEDs).
In optical images, many of the ALFALFA dwarfs are faint and of low surface brightness;
only 56\% of those within the SDSS footprint have a counterpart in the
SDSS spectroscopic survey. A large fraction of the dwarfs have high specific star
formation rates (SSFRs) and estimates of their SFRs and $M_*$ obtained by
SED fitting are systematically smaller than ones derived via standard formulae
assuming a constant SFR.
The increased dispersion of the SSFR distribution
at $M_* \lesssim 10^8 {\rm M_\odot}$ is driven by a set of dwarf galaxies that
have low gas fractions and SSFRs; some of these are dE/dSphs in the Virgo cluster.
The imposition of an upper HI mass limit yields the selection of a sample
with lower gas fractions for their $M_*$ than found for the overall ALFALFA population.
Many of the ALFALFA dwarfs, particularly the Virgo members, have HI depletion
timescales shorter than a Hubble time. An examination of the dwarf galaxies within the full
ALFALFA population in the context of global star formation laws
is consistent with the general assumptions that gas-rich galaxies
have lower star formation efficiencies than do optically selected populations
and that HI disks are more extended than stellar ones.
\end{abstract} 

\keywords{galaxies: dwarf -- galaxies: evolution -- galaxies: fundamental parameters -- 
radio lines: galaxies -- galaxies: star formation -- surveys
}

\section{Introduction} 

Principal aims of current studies of galaxy formation and evolution include the exploration of the interplay 
between the gaseous and stellar components of galaxies and the mechanisms which trigger the 
conversion of gas into stars. During the last decade, wide area surveys such as the Sloan
Digital Sky Survey (SDSS) and the Galaxy Evolution Explorer (GALEX) have enabled statistical studies of 
star formation (SF) in the local universe.  For example,
galaxies which are currently forming stars, the so-called ``blue-cloud galaxies'' in the color-magnitude diagram 
\citep{Baldry2004}, occupy a relatively narrow 
``star-forming sequence'' in a plot of specific star formation rate ($SSFR=SFR/M_*$) versus 
stellar mass $M_*$ \citep{Brinchmann2004, Salim2007}, 
with the SSFR declining as the stellar mass increases. 
Such a trend suggests that the galaxy's stellar mass regulates the overall 
star formation history (SFH), at least at intermediate masses. 
This star-forming sequence breaks down above $\sim 10^{10}~\rm{M_\odot}$, 
where the ``red sequence'', occupied by massive galaxies having lower values of the SSFR, becomes more prominent. 
The importance of the gaseous component is 
reflected in the Kennicutt-Schmidt law which relates the gas column density ($\Sigma_{gas}$) to the SFR surface 
density ($\Sigma_{SFR}$). A super-linear slope is sometimes 
reached, e.g. $\sim 1.4$ as in \citet{Kenn1998b}, indicating that the star formation efficiency is 
higher in regions of higher gas surface density. However, the empirical relations among galaxy properties 
which are derived from such surveys apply to the galaxy populations which dominate them, typically the
more massive and luminous galaxies. The same relations may not apply to dwarf or low surface brightness (LSB)
galaxies. Of particular relevance to this work, in such objects the environment where star formation occurs
may be quite different.

Compared to the optically bright and massive systems which dominate the SDSS, gas-rich dwarf galaxies 
are often underrepresented in samples selected by stellar mass. Often the optical emission arising
in such systems is very blue, patchy and of very low surface brightness or small in extent.
However, gas-rich, low mass, low metallicity, low optical surface brightness galaxies are important 
to the study of star formation because the processes by which gas is converted into stars within 
such systems may mimic those which occurred in the early universe. As the most chemically unevolved 
systems within the present-day galaxy population, the faintest dwarfs represent unique laboratories for 
understanding star formation and galaxy evolution in extreme environments, that is, in regimes of low metallicity, 
low dust content, low pressure, low shear, and low escape velocity \citep{Begum2008}. 

Several recent works suggest that the star formation in dwarf galaxies may proceed quite differently from 
that in large spirals.  Based on a sample of very local dwarf irregular galaxies, 
the Faint Irregular Galaxies GMRT Survey 
\citep[FIGGS,][]{Begum2008, Roy2009}, found a lower average $\Sigma_{SFR}$ than would be expected 
from the Kennicutt-Schmidt law \citep{Kenn1998b}. Moreover, 
no threshold density is observed below which star formation is completely turned off. 
Recently, \citet{Lee2007} explored the distribution of the SSFR against absolute magnitude 
for a complete sample of $\sim$300 star-forming galaxies within 11 Mpc of the Milky Way, 
from the 11Mpc H$\alpha$ UV Galaxy Survey (11HUGS). 
In addition to confirming the transition in star formation activity at the high mass end, those authors
found a second transition, with low luminosity dwarf galaxies ($M_B \gtrsim -15$) having a very large spread 
in their SSFRs. This second transition suggests that the star-forming behavior may be distinct at the
lowest mass range. 
After showing that other potential drivers are not able to explain the magnitude of observed 
systematics, \citet{Lee2009} suggest that the over-prediction of the SFR by the UV flux
compared to that estimated from H$\alpha$ in dwarf systems is consistent with an 
IMF deficient in the most massive stars. 
However, those authors also point out that it is possible that some combination of effects may 
conspire to produce the observed trend, and thus the requirement of systematic variations in the IMF can be avoided. 

The 11HUGS sample is complete in HI mass only above $2 \times 10^8~
\rm{M_\odot}$, becoming rapidly incomplete at smaller HI masses.
To develop further the current understanding of how the gas supply regulates star 
formation in the lowest mass systems, a larger sample of extreme dwarf galaxies is needed. 
Making use of the Arecibo L-band Feed Array (ALFA), the on-going Arecibo 
Legacy Fast ALFA (ALFALFA) extragalactic HI line survey is specially designed to 
identify low mass, gas rich objects in the local universe \citep{Giovanelli2005}. Because
of its combination of wide areal coverage, sensitivity, and velocity resolution, 
ALFALFA has already detected more than 400 galaxies with HI masses $M_{HI} < 10^8 
\rm{M_\odot}$ \citep{Haynes2011}. While star formation is more directly linked to
the molecular interstellar component, the detection of CO
in low-metallicity dwarfs is difficult \citep[][and references therein]{Leroy2005}, 
suggesting further that CO no longer traces H$_2$ well. Furthermore, 
in many gas-rich dwarf galaxies, the HI component
dominates both the gas as well as the baryonic mass \citep[e.g.][]{Leroy2007}. 
The combination of HI parameters from ALFALFA with
complementary multi-wavelength data contributed by SDSS and GALEX provides an ideal 
dataset to investigate the abundance and distribution of gas-rich dwarfs and to explore the relations
among their gas content, stellar populations and star formation properties. 

This paper is organized as follows. In \S\ref{sample} we define our HI-selected sample and present
its basic gas properties. In \S\ref{data} we present the supplementary SDSS and GALEX data, especially 
our re-processed UV photometry. The directly measured colors and selected spectroscopic 
behavior are also briefly examined. In \S\ref{prop} we describe how we utilize SED fitting techniques 
to obtain physical parameters for the dwarfs, e.g., $M_*$ and the dust extinction-corrected SFR. 
We discuss in \S\ref{SFgas} the relations between gas, star and star formation in dwarfs, and 
how they compare to the overall ALFALFA HI-selected
population. A summary is presented in \S\ref{summary}. 

All the distance-dependent quantities in this work are computed assuming $\Omega=0.3$, $\Lambda=0.7$ and
$H_0=70$ km s$^{-1}$ Mpc$^{-1}$, and a \citet{Chabrier2003} IMF is adopted. 

\section{Sample selection}
\label{sample}

In this section, we use the 40\% ALFALFA catalog \citep[$\alpha$.40:][]{Haynes2011} to define 
two HI-selected dwarf galaxy samples, one of which is complete in HI mass
and velocity width ({\it s-com}). The second sample ({\it s-sup}) is less restrictive in those parameters but
supplements the first through the availability of deeper GALEX NUV/FUV observations. We discuss here the 
selection of these two samples and their HI properties. In \S\ref{sed}, we will define a third 
sample {\it s-sed} as a subset of dwarfs among the {\it s-com} and {\it s-sup} samples. 

\subsection{The ALFALFA-SDSS parent sample} 

Begun in 2005, the ALFALFA survey has been using the 7-beam ALFA receiver to 
conduct a blind search for HI sources with $cz < 18000$ km s$^{-1}$ over 
7000 deg$^2$ of high galactic latitude sky \citep{Giovanelli2005}. 
The targeted regions cover the sky visible to Arecibo $0 <$ Dec. $< +36^\circ$ in both
the spring ($07^h30^m <$ R.A. $< 16^h30^m$) and fall ($22^h00^m <$ R.A. $< 03^h00^m$) night sky.
With a median $cz$ of $\sim$8200 km s$^{-1}$, ALFALFA for the first time samples the HI population
over a cosmologically fair volume, and is expected to detect $\sim$30,000 extragalactic HI-line sources 
out to redshifts of $z\sim0.06$. As a second generation wide area HI survey, ALFALFA is designed
to greatly improve on the HI census derived from previous results.
For example, ALFALFA is 8 times more sensitive than the HI Parkes All-Sky Survey 
\citep[HIPASS][]{Barnes2001}, with 4 times the angular resolution and 
3 times the velocity resolution,
all of which are essential to the discovery of the lowest HI mass objects. In particular,
HIPASS detected only 11 objects with 
$M_{HI} < 10^{7.5} {\rm M_\odot}$ \citep{Zwaan2005} whereas ALFALFA is detecting
hundreds of such low mass systems. For example, at the 
distance of the Virgo Cluster, ALFALFA is sensitive
down to $\sim 3\times 10^7 {\rm M_\odot}$ for sources with S/N $\sim$6.5 \citep{Giovanelli2007,
Haynes2011}.
In addition, HI source positions derived from ALFALFA can be determined with a median 
accuracy of about 20$^{\prime\prime}$ \citep[see Eqn 1 of][]{Haynes2011}, 
allowing the identification of optical/UV counterparts for the vast majority of HI sources 
without the need for follow-up synthesis mapping. While source confusion
within the 3.5$^\prime$ arcmin beam can affect sources at large distance, it has 
little effect on the identification of nearby dwarfs, except when they are located in close
proximity to giant neighbors. 

The current ALFALFA survey, ``$\alpha.40$'',  
covers $\sim40\%$ of the final survey area, and includes
 a catalog of 15855 HI sources, 15041 of which are extragalactic \citep{Haynes2011}. 
The remainder have no optical counterparts (OCs) and lie at low velocities which are consistent
with Galactic phenomena, e.g., as high velocity clouds (HVCs). ALFALFA 
HI detections are further categorized by source reliability:
Code 1 sources are reliable 
extragalactic detections with high signal-to-noise ratio ($S/N \gtrsim 6.5$)
while Code 2 sources, also known as ``priors'', have lower S/N ($4.5 \lesssim S/N \lesssim 6.5$) but coincide with an 
OC of known optical redshift matching the HI measurement. The HVCs are identified
as Code 9 objects. Further details are given in \citet{Haynes2011}.

HI masses in units of solar mass are obtained from the relation
$M_{HI} = 2.356 \times 10^5D^2\ S_{int}$, where $D$ is the distance in Mpc 
and $S_{int}$ is the integrated HI line flux density in units of Jy km s$^{-1}$. 
In the local universe, distance determinations suffer significantly from the 
uncertainty introduced by a galaxy's peculiar velocity. In order to
minimize the HI mass error introduced by the uncertainty in distances, we
adopt a peculiar velocity flow model which incorporates both primary
distances available from the literature and secondary distances derived
from the SFI++ survey \citep{Springob2007}. The flow model derived by
\citet{Masters2005} is adopted for galaxies with $cz_{CMB} <$ 6000 km s$^{-1}$,
while distances for more distant objects are derived from redshifts
in the CMB rest frame.
Primary distances are assigned to individual galaxies 
wherever available from the literature, and following the method discussed in
\citet{Springob2007}, galaxies identified as members of groups and clusters
are placed at the distance to their assigned hierarchical unit. 
We have been conservative in ambiguous cases, assigning larger distances where
a choice is given to avoid the inclusion of higher mass galaxies in the
present analysis. 

All of the spring sky coverage of ALFALFA and part of the fall sky survey region overlap
the footprint of the SDSS Legacy Survey, thereby allowing a direct cross-match of the two. As
part of the ALFALFA catalog production process, the HI detections have been crossed matched to 
SDSS DR7 photometric objects for 12470 of the $\alpha.40$ HI detections \citep{Haynes2011}. 
Through SED fitting to the five SDSS 
photometric bands (see \S \ref{sed}), we are able to derive additional basic
properties of the full $\alpha.40$-SDSS HI selected parent sample. 
A more detailed discussion of the $\alpha.40$-SDSS-GALEX sample in general, as well as the selection 
effects characteristic of the $\alpha.40$ survey will be presented in \citet{Huang2012}.
As discussed in \citet{Haynes2011}, the identification of OCs
to the $\alpha.40$ HI sources and the
cross match to the SDSS DR7 is not a perfect process; in individual cases, the wrong
counterpart may have been selected, the SDSS photometry may be bad etc. However, 
the cross match with the SDSS DR7 allows us to make a first statistical study of 
the relationships between gas, stars and star formation, and provides us with a 
parent sample of gas-rich galaxies within which we can
explore the distinctiveness of the lowest HI mass systems. 

\subsection{A complete HI-selected dwarf sample} 

The studies which infer global properties derived from the SDSS main galaxy catalog are 
highly biased against the inclusion of dwarf irregular galaxies because of the magnitude 
and surface brightness limits on the SDSS spectroscopic targets ($r \lesssim 17.77$ and 
$\mu_{r, 50} \lesssim 23.0$ mag arcsec$^{-2}$). Since $M_{HI}/L_{opt}$ increases 
with decreasing $L_{opt}$, HI-selected samples are more inclusive of star-forming galaxies than optical 
samples of similar depth. Because of their relatively young stellar populations 
and low dust contents, gas-rich dwarfs are typically blue and often patchy in optical appearance; at the
same time, they are often extended and diffuse in HI. 
Since the cold gas is the fuel needed to sustain star-formation, a blind HI
survey of sufficient depth, like ALFALFA, is especially effective in identifying star forming
systems at the low mass end, and hence should offer a full census of star forming galaxies in
the local universe. 

In order to identify a sample of low mass, gas-rich dwarfs, we have
applied selection criteria to the $\alpha.40$ catalog as follows:
(i) ALFALFA detection code = 1 or 2 (reliable sources and priors, but no HVCs); 
(ii) $\log M_{HI}<7.7$; 
(iii) velocity width of the HI line, $W_{50}<80$ km s$^{-1}$; 
(iv) the optical images were visually inspected to eliminate the ones without optical counterparts, 
those which appear to be more massive but HI-deficient galaxies.
Following the detailed analysis of the HI mass error 
in \citet{Martin2010} and \citet{Haynes2011}, the $\log M_{HI}$ error in the $7.5$ bin 
is $\sim 0.2$ dex. Hence, requirement (ii) ensures that we are unlikely to miss 
dwarfs with $\log M_{HI}<7.5$ due to their HI mass error. 
Criterion (iii) helps to insure that we 
include only truly low mass systems. 
Based on criterion (iv), 2 gas-poor face-on giant 
galaxies (UGC~7622 = NGC~4469 and UGC~7718 = NGC~4526) have been removed. Both have SDSS $r$-band 
absolute Petrosian magnitudes brighter than $-18$ and are early type spirals situated in the
Virgo cluster; their HI masses and velocity widths are unusually low, probably due to interaction
within the cluster environment. 
Additionally, extragalactic HI sources without OCs 
(38 of them) are dropped. As discussed by \citet{Haynes2011}, the majority of those are 
part of the extended HI structures in the 
Leo region: the Leo Ring and Leo Triplet \citep{Stierwalt2009} or are similar
fragments associated with nearby groups of galaxies. With these objects removed, the final complete 
ALFALFA dwarf galaxy sample, referred to hereafter as {\it s-com}, contains 
176 galaxies. 

The designation of `complete' for this HI-selected sample emphasizes that
it is a complete subset of the $\alpha.40$ catalog.
Because the ALFALFA sensitivity depends not only on the integrated
flux but also on the profile width, there is no simple translation of a limiting flux to the 
lower limit on the HI mass but the completeness can be well characterized
\citep{Haynes2011}. Of particular relevance here, we note that,
at the mean $W_{50}=36.8$ km s$^{-1}$ of the {\it s-com} galaxies, 
$\alpha.40$ is 90\% complete to $\log M_{HI}=7.1$ and 25\% complete to $\log M_{HI}=6.9$, 
within a distance of 11Mpc. In comparison, the 11HUGS
sample is complete in $\log M_{HI}$ only to $8.3$ \citep{Lee2009b} within the same distance.
While we include all the low HI mass detections in the {\it s-com} sample out to a  
distance of $\sim$ 30~Mpc, the $\alpha.40$ completeness limit at that distance is well above 
the HI mass upper limit of the {\it s-com} sample, $\log M_{HI}=7.7$. Thus, {\it s-com} sample 
is not complete in a volume-limited sense, but it does probe the extreme low 
HI mass tail of the $\alpha.40$ catalog. 

\subsection{Supplementary dwarf galaxies with GALEX data}

Since ALFALFA is an on-going survey, its catalog of HI detections continues to grow with time.
Similarly, the simultaneous undertaking of the GALEX satellite mission has provided
some opportunity to obtain images in the NUV and FUV bands for early
ALFALFA detections, at least until the GALEX FUV channel failure in 2009.
Because exploration of the population of galaxies which define the low mass end of the HI mass function (HIMF)
has always been one of the main goals of ALFALFA, we proposed to obtain GALEX MIS 
(Medium Imaging Survey) level FUV and NUV observations of low HI mass targets,
based on early releases of the ALFALFA catalog, in
GALEX cycles 3, 4 and 5 (GI3-84, GI4-42 and GI5-2). As ALFALFA has progressed, the identification
of the lowest HI mass population has likewise been an ongoing process, extending to lower HI masses
as its catalog of HI sources has grown.
Hence, the complete $\alpha.40$ low mass sample {\it s-com} 
as defined above is more restrictive in HI mass than our GALEX target 
dwarf galaxy lists which were based on early ALFALFA catalogs. 
Fortunately, although the criteria for the GALEX target 
selection derived from the early ALFALFA catalogs were less restrictive 
in terms of HI mass and velocity width, the ALFALFA-based GALEX targets
are nonetheless of relatively low mass. Of the 77 galaxies for which we acquired GALEX
FUV/NUV observations, 24 overlap with the strictly complete sample {\it s-com}. The remaining
53 galaxies have somewhat higher HI masses (see discussion in \S \ref{HIprop}); we refer to this
supplementary sample as {\it s-sup}. While the {\it s-sup} sample is not complete in any sense, 
the availability of GALEX MIS-depth imaging in both FUV and NUV bands for its galaxies 
allows us to explore with better statistics the low HI mass systems
so that we can test for trends (or thresholds) with HI mass at the low mass end of the HIMF. 

\subsection{HI properties of the ALFALFA dwarf sample} 
\label{HIprop}

The combination of the two samples {\it s-com} and {\it s-sup} yields a final ALFALFA-selected 
set of 229 low HI mass and low velocity width dwarf galaxies upon which we base the analysis
presented here. Table \ref{table:dwarf} presents the relevant UV, SDSS  and HI properties for them.
Columns are as follows:
\begin{itemize}

\item Column(1): ALFALFA catalog identifier (also known as the AGC number).

\item Columns(2) and (3): J2000 position of the OC assigned to the HI source.
 
\item Columns(4) and (5): The adopted FUV and NUV magnitudes, with their associated error, respectively,
as derived via our reprocessing of the GALEX images (see \S\ref{Gdata}).

\item Column(6): The $r$-band {\it modelmag} with its associated error, from the SDSS pipeline.

\item Column(7): The $u-r$ color with its associated error, from the SDSS pipeline.
 
\item Column(8): The SDSS code sFlag indicating the quality of SDSS photometry as defined in \S\ref{SDSSdata}. 

\item Column(9): The adopted distance with error, in Mpc.

\item Column(10): The logarithm of the HI mass and its error, taken from the $\alpha.40$ catalog \citep{Haynes2011}.

\item Column(11): The logarithm of the stellar mass and its error, derived from SED fitting (see \S\ref{smass})

\item Column(12): The logarithm of the SFR and its error, in solar masses per year (see \S\ref{SFR})

\end{itemize}

\begin{figure*}
\center{
\includegraphics[scale=0.8]{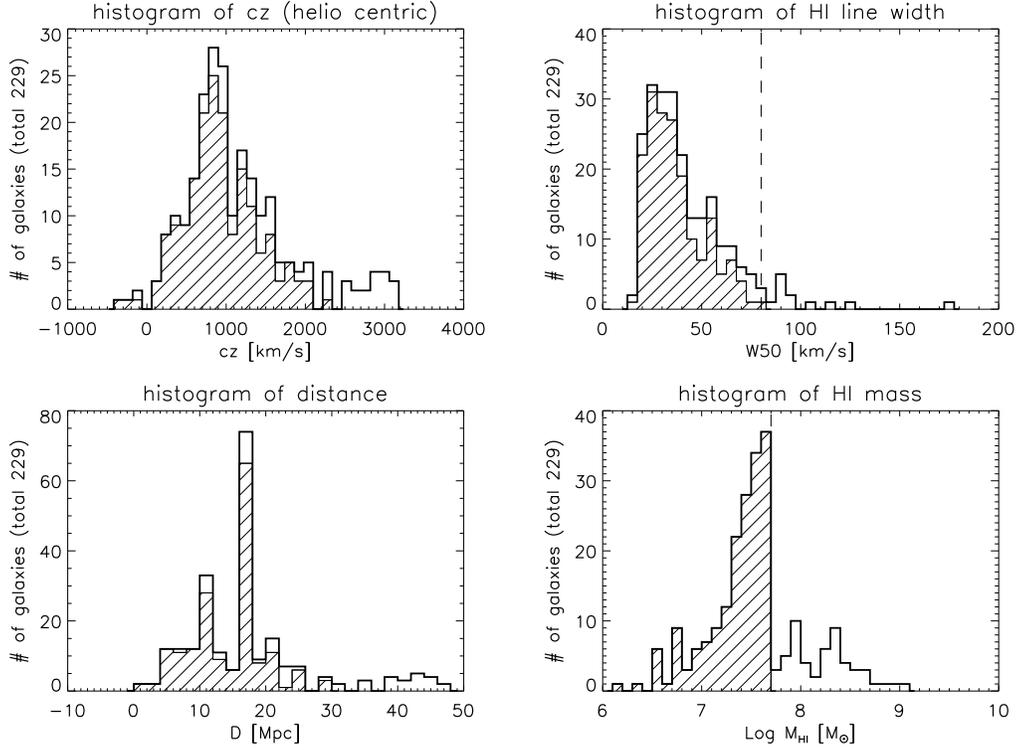}
}
\caption[]{HI properties of the full ALFALFA dwarf sample. In all panels, the
shaded area identifies the galaxies in the complete 
HI mass/velocity width-limited sample {\it s-com} (total = 
176) while the unfilled area represents the additional galaxies with GALEX 
photometry in the {\it s-sup} sample (total = 53). 
{\it Top\ left:} the distribution of HI line recessional velocity $cz_{HI}$.
{\it Top\ right:} the distribution of HI line velocity width $W_{50}$. 
The dashed vertical line corresponds to the $W_{50}$
cutoff ($W_{50}<$ 80 km s$^{-1}$) imposed on the {\it s-com} sample. 
{\it Bottom\ left:} the distribution of adopted distance to the ALFALFA
dwarfs. The peak in the distribution at 16.7 Mpc reflects the assignment
of 37 galaxies to membership in the Virgo cluster.
{\it Bottom\ right:} the distribution of HI mass. 
The dashed vertical line corresponds to the $M_{HI}$ cutoff 
($M_{HI} < 10^{7.7} {\rm M_\odot}$) imposed on the  {\it s-com} sample. 
}
\label{fig:sample5}
\end{figure*}

Figure \ref{fig:sample5} shows histograms of the recessional velocity $cz$, adopted distance,
observed HI line width and logarithm of the HI mass for the combined ALFALFA dwarf sample, with shaded areas
identifying the complete sample, {\it s-com} and the open areas indicating the additional
{\it s-sup} galaxies. 
As expected since the lowest HI masses are detected only nearby, most galaxies lie within 
the Local Supercluster at $cz<$ 3000 km s$^{-1}$. The peak in the distance distribution 
at 16.7 Mpc arises from the assignment of Virgo membership to a significant number of 
the dwarf galaxies. A cross match with the VCC catalog \citep{Binggeli1985} shows that 
there are 37 Virgo members belonging to the dwarf sample defined here 
(35 in {\it s-com} and 2 in {\it s-sup}). 
The dashed vertical line in the HI line width histogram corresponds to the 
adopted line width cutoff, $W_{50}<$ 80 km s$^{-1}$. 
The mean uncertainty on the line width measurement is 6.7 km s$^{-1}$. 
Note that only one galaxy in the {\it s-com} sample is included in the last bin below this cutoff, 
suggesting that the low HI mass and narrow line width criteria are consistent: we are not missing 
a population of high line width but still low mass dwarfs because our line width cutoff is 
set to be too narrow, and the low HI mass criterion is more important
than the narrow line width one in the definition of the {\it s-com} sample. 
In agreement with this point, a quick check of the {\it s-sup} galaxies reveals the fact that 
they all have $\log M_{HI} > 7.7$, although by the definition of the {\it s-com} subset, it is possible 
that galaxies in the {\it s-sup} sample could have $\log M_{HI} < 7.7$ 
but $W_{50}>$ 80 km s$^{-1}$. In another words, the supplemental sample {\it s-sup} does
not include low mass dwarf galaxies 
which are excluded from the {\it s-com} one through the restriction to the velocity width,
$W_{50}<$ 80 km s$^{-1}$. As anticipated,
galaxies in the {\it s-sup} sample span a wider range of $W_{50}$ and
distance than the stricter low HI mass sample. 
In the histogram of $\log M_{HI}$, the low mass tail extends to $\log M_{HI} \sim$6;
the sharp edge at $\log M_{HI} = 7.7$ 
(our upper limit for {\it s-com}) reflects the fact that the {\it s-sup} galaxies represent only
a small subset of the $\alpha.40$ galaxies in the higher HI mass range. Figure 2
of \citet{Haynes2011} shows the similar distributions for the full $\alpha.40$  catalog.

\begin{figure*}
\center{
\includegraphics[scale=0.9]{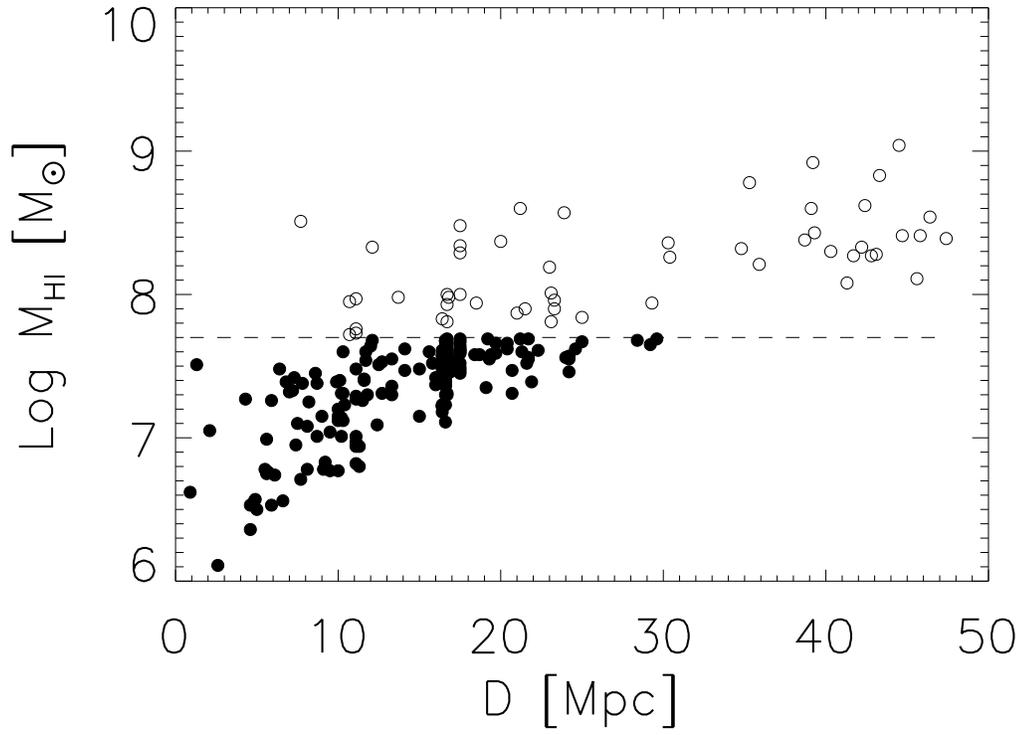}
}
\caption[]{Spaenhauer diagram showing HI mass versus distance for the
ALFALFA dwarf galaxy sample. Filled circles denotes the lowest HI mass
{\it s-com} members; 
whereas open ones identify the supplementary {\it s-sup} galaxies. 
The horizontal dashed line corresponds to the $M_{HI}$ cutoff 
at $M_{HI} = 10^{7.7} {\rm M_\odot}$ 
imposed on the {\it s-com} sample.}
\label{fig:plotHI}
\end{figure*}

Previous studies of dwarf galaxies have focused mainly on optically selected samples and
contain relatively few objects with $\log M_{HI} < 7.7$. For example,
the FIGGS sample of \citet{Begum2008} contains only 41 galaxies with 
$\log M_{HI} < 7.7$, compared to 176 in the ALFALFA {\it s-com} sample.
The majority of the FIGGS targets lie within 8 Mpc and have been selected
from existing optical surveys. Similarly, the volume-limited 11HUGS
sample is complete in $M_{HI}$ to $2\times10^8 {\rm M_\odot}$ \citep{Lee2009b}. 
In fact, only seven of the {\it s-com} plus {\it s-sup} galaxies are included in the 11HUGS catalog; 
at the same time, 54 of the ALFALFA dwarfs lie at distances of less than 11 Mpc.
Figure \ref{fig:plotHI} shows a Spaenhauer plot of HI mass versus distance for the ALFALFA dwarfs, 
with filled circles denoting the {\it s-com} members. The lower edge of the distribution represents the 
ALFALFA sensitivity limit \citep{Haynes2011}. 
Compared to the similar plot in Figure 1 from \citet{Lee2009b}, the distribution of ALFALFA dwarfs 
is shifted towards the lower HI mass range, just where the deviation of the 
UV-based SFR from that inferred from H$\alpha$ is more likely to show up \citep{Lee2009}. 
Benefiting from its improved sensitivity and angular and spectral resolution, the
ALFALFA survey catalog
allows us to draw a statistically significant sample 
of the lowest HI mass galaxies in the local universe.

\section{Data}
\label{data}
In this section, we describe
the GALEX and SDSS datasets used, including reprocessed UV photometry from GALEX. 
In addition, the UV-to-optical color and emission-line diagnostics critical to the
appraisal of star formation in the dwarf galaxies are discussed. 

\subsection{Targeted GALEX observations}

In dusty starbursting galaxies, a dominant fraction of the UV emission may be obscured 
by dust and reprocessed at FIR wavelengths \citep{Treyer2007}. In red sequence galaxies,
older evolved stars make significant contributions to the UV luminosity \citep{Wyder2007}. 
However, the ALFALFA-selected dwarf galaxies are 
likely to suffer less from extinction and their young stellar populations contribute the bulk of the UV light. 
The FUV luminosity $L_{FUV}$ is generally thought
to give the most robust measure of the SFR in individual galaxies with 
low total SFRs and low dust attenuation. Dwarfs are known to be low in metallicity and dust content,
so that IR indicators of the SFR, which are calibrated via massive spirals, may be less reliable. 
In contrast to H$\alpha$ emission, FUV photons primarily originate in the more abundant 
and relatively longer-lived 
population of B-stars so that the FUV flux is not as vulnerable to stochastic effects.
Furthermore, FUV photons are emitted directly from the stellar photospheres, 
and thus do not suffer from possible uncertainties in the photoionization of 
the gas in low density media \citep{Lee2011}. 

In order to explore the UV properties of the faint and low surface brightness
galaxies in the ALFALFA dwarfs, we examined all available moderate exposure (MIS-depth) 
GALEX images coincident with them, including both ones from our own GI programs as well as 
others available in the GALEX archive.
GALEX simultaneously imaged the sky in the FUV (effective wavelength of 1516~$\rm \AA$) and NUV 
(effective wavelength of 2267~$\rm \AA$), with a circular field of view of $\sim1.2^\circ$ in diameter 
\citep{Morrissey2007}. The images were processed through the GALEX pipeline, and the intensity 
maps with a 1.5$^{\prime\prime}$ pixel scale were retrieved. With typical exposure times
of $\sim$1500 sec, the images reach limiting magnitudes of $\sim$22.7 mag in both the FUV and NUV,
corresponding to surface brightness limits of $\sim$27.5 mag arcsec$^{-1}$ 
or a SFR of $\sim10^{-3}\ \rm{ M_\odot\ yr^{-1}\ kpc^{-2}}$ \citep{Lee2011}.
As discussed in \S\ref{sample}, all 53 {\it s-sup} galaxies were included in our  
GALEX GI programs, though five of them  were only observed in the NUV due to the failure of the FUV detector. 
Since the final {\it s-com} sample was extracted from the $\alpha.40$ catalog
after the FUV channel was completely turned off, we also searched the GALEX archive for any additional
MIS-depth images with adequate coverage of ALFALFA dwarf galaxies. Among the final ALFALFA-GALEX/MIS 
sample of 77 {\it s-com} galaxies, seven were observed in the NUV only.

Using a standard ellipse fitting extraction of the magnitudes (GALPHOT; see below),
67 of the 70 extreme dwarfs in the {\it s-com} sample are clearly detected in the FUV band.
Two of the remaining three are extremely faint and LSB in the FUV, but magnitudes are still 
measurable in concentric apertures (ELPHOT; see below). Furthermore, $\it all$ of the sources 
in the higher HI mass supplementary {\it s-sup} sample which have FUV images are detected in FUV.
Thus, only $\it one$ out of the 118 dwarfs in the combined {\it s-com} plus {\it s-sup} sample 
with FUV MIS level images is a non-detection. A similarly high detection rate was found by 
\citet{Lee2011}. Only 22 of the 390 galaxies in their 11HUGS sample 
observed by GALEX were not detected in FUV. About half of these are galaxies classified as faint dwarf 
ellipticals/spheroidals (dE/dSph) and lack any evidence of recent star formation; nearly
all the others 
were found in images of exposure times less than 200 sec. Those authors concluded that,
despite the variable, episodic or bursty star formation histories of dwarfs, 
the fluctuations in the SFR do not go to zero on timescales comparable to the lifetimes of 
UV emitting stars ($\sim$100 Myr). Furthermore, \citet{Lee2011} propose the need to examine the possible
complete cessation of SF in low luminosity systems via an HI selected sample probing 
masses down to $10^7~{\rm M_\odot}$, exactly what the ALFALFA {\it s-com} sample is.
The presence of HI selects against the very gas-poor dE/dSph population, so our even 
higher FUV detection rate is not unexpected. This result clearly suggests that
virtually all HI-bearing dwarf galaxies exhibit some level of recent star formation.
 
\subsection{GALEX photometry} 
\label{Gdata}

Because we rely mainly on the FUV luminosity to infer SFRs, it is essential to obtain accurate
FUV photometry. Given the typical faint, LSB and patchy nature of the UV emission of dwarf
galaxies, extra attention must be paid to the extraction of magnitudes. The standard 
GALEX pipeline, which is based on the SExtractor
code \citep{Bertin1996}, suffers from shredding if multiple star forming sites are resolved. 
It also suffers from blending if foreground stars or UV-bright
background galaxies are viewed in projection with the target galaxy. Additionally, the background determination 
matters more than that in high surface brightness regions, since the background subtraction uncertainty,
rather than the photon noise, dominates in LSB regions \citep{Gil2007}. For these reasons, we developed our 
own tools to perform the photometric extraction on the GALEX images of the ALFALFA dwarf galaxies. 

For each of the target galaxies, we extracted a portion of the intensity map retrieved from the MAST website.
Further reduction of the image was performed within the IRAF
\footnote{IRAF is distributed by the National Optical Astronomy Observatory, 
which is operated by the Association of Universities for Research in Astronomy (AURA) under cooperative agreement 
with the National Science Foundation.}/STSDAS environment using a set of 
scripts developed for previous I-band imaging surveys undertaken by our group, 
referred to as the GALPHOT package \citep{Haynes1999},
and appropriately modified to accommodate the GALEX images. 
Because the morphology of galaxies in the FUV is not 
necessarily the same as in the NUV, and because foreground stars are often much brighter in the NUV images, 
we elected to work on the two channels separately rather than adopt a single identical 
set of apertures. 

A constant value associated with the sky background was subtracted using a procedure that 
allows the user to mark boxes that are free of bright stars and galaxies on each frame. The boxes 
are placed so that they surround the target galaxy but are far enough from it to avoid the 
influence of faint extended emission. In contrast to the circumstances applied to
optical images, the UV sky is so dark that the pixel values in sky boxes follow a Poisson 
distribution rather than a Gaussian one. The subtracted sky value in the UV case is the mean 
intensity obtained within the sky boxes after iteratively clipping 
out the pixels whose values are more than 3$\sigma$ above the mean value, a process which removes the 
faint stars and galaxies within the sky boxes. Usually this process converges quickly after 1 or 2 clipping
cycles. 

\begin{figure*}
\center{
\includegraphics[scale=0.7]{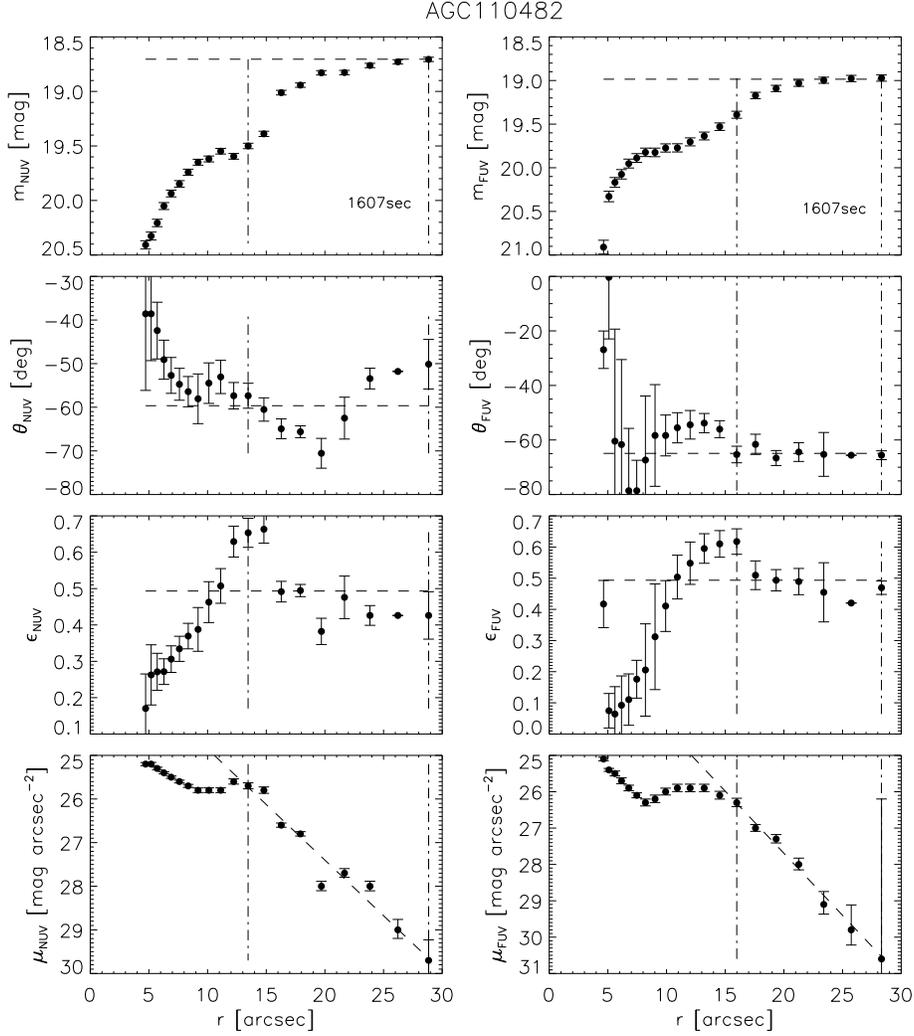}
}
\caption[]{Example of the UV isophotal fitting result for AGC~110482 = KK~13, a galaxy which 
shows star formation in two UV bright knots. The enclosed magnitude ($m$), 
position angle ($\theta$), ellipticity ($\epsilon=1-b/a$), and surface brightness 
($\mu$) of the elliptical isophote as a function of the semi-major axis of 
the isophote ($r$), are plotted in rows from top to bottom.
Results for the NUV image are shown in the left panels, for the FUV image on the right. 
Numbers at the corners of each plot in the top row give the exposure times in each channel. 
The marked disk region is denoted by the vertical dash-dotted lines.
Horizontal dashed lines represent the adopted magnitude $m_{8r_D}$ (flux within 8$r_D$) 
in the $m$ plot, and final values determined from the mean of data points in 
the disk region in the $\theta$ and $\epsilon$ plots. 
Dashed lines in the $\mu$ plots show the linear fit to the light profile in 
the adopted disk region, assuming and exponential drop off in surface brightness. 
}
\label{fig:SB110482}
\end{figure*}

\begin{figure*}
\center{
\includegraphics[scale=0.7]{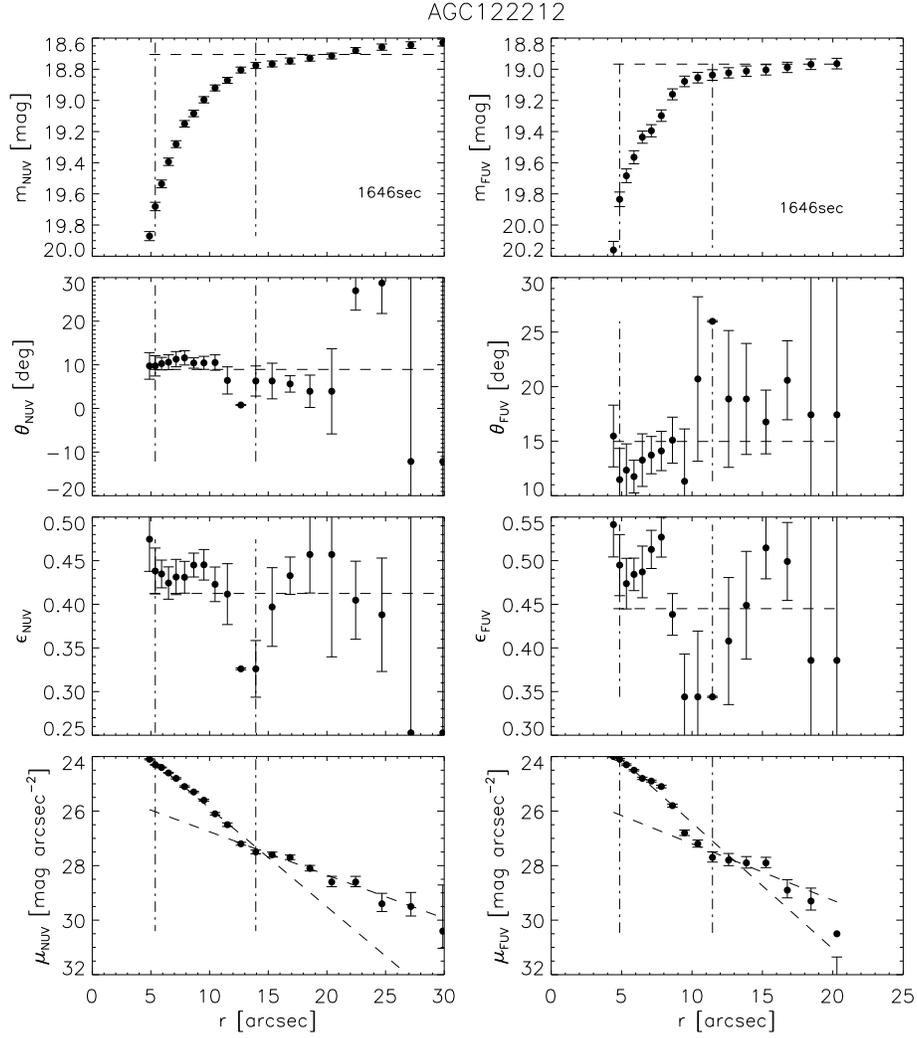}
}
\caption[]{Example of the isophotal fitting result for AGC~122212 fit best by a combination
of a two disks. The symbol definition is the same as that in Figure \ref{fig:SB110482}. 
This galaxy has a shallower outer disk in both the NUV and FUV images. Separate linear 
functions are fit to the inner and outer regions; the inner fit determines the disk 
scale length and the outer fit is used for extrapolating the profile. 
}
\label{fig:SB122212}
\end{figure*}

\begin{figure*}
\center{
\includegraphics[scale=0.7]{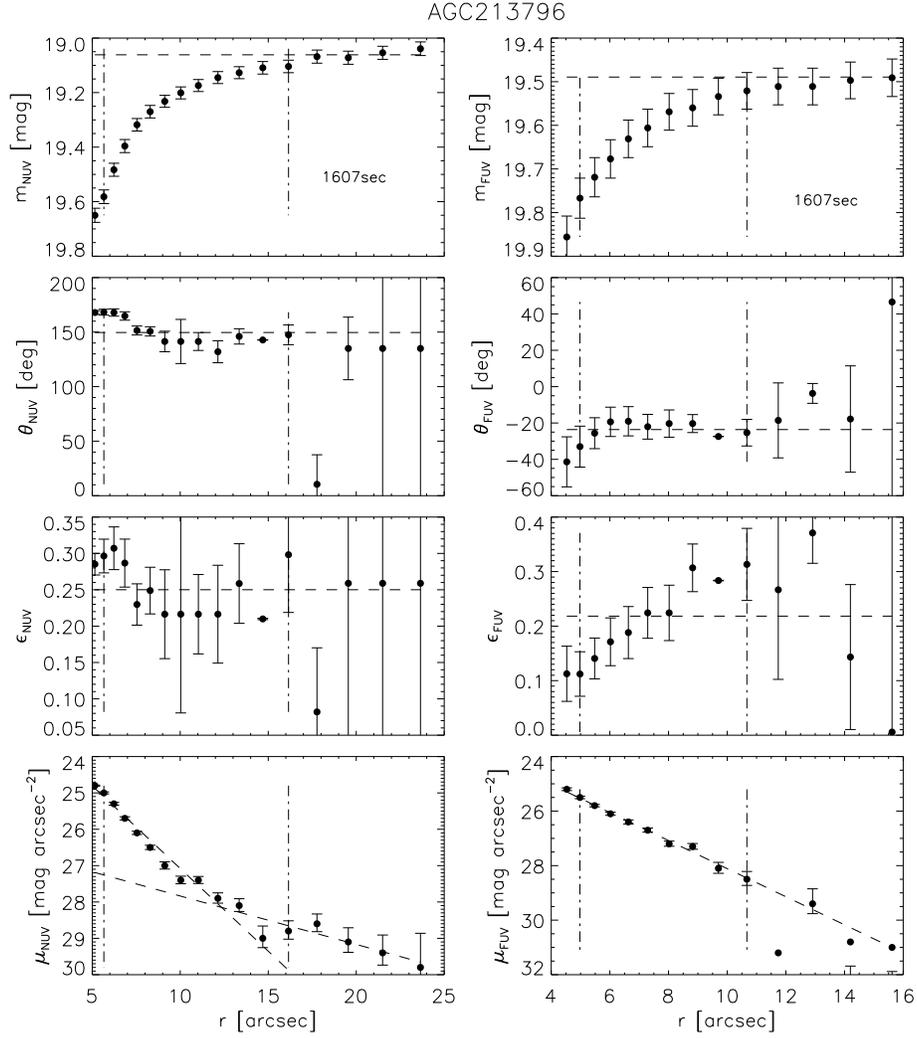}
}
\caption[]{Example of the isophotal fitting result for AGC~213796.
The symbol definition is the same as that in Figure \ref{fig:SB110482}. This galaxy exhibits a 
two-disk structure only in the NUV, either because the emission in the FUV image 
is too weak to trace the outer disk, or the extended outer disk is not actively 
forming stars. The traceable FUV disk matches the inner NUV disk. Similar results 
are observed in a total of 15 galaxies which are assigned to this category. 
}
\label{fig:SB213796}
\end{figure*}

\begin{figure*}
\center{
\includegraphics[scale=0.7]{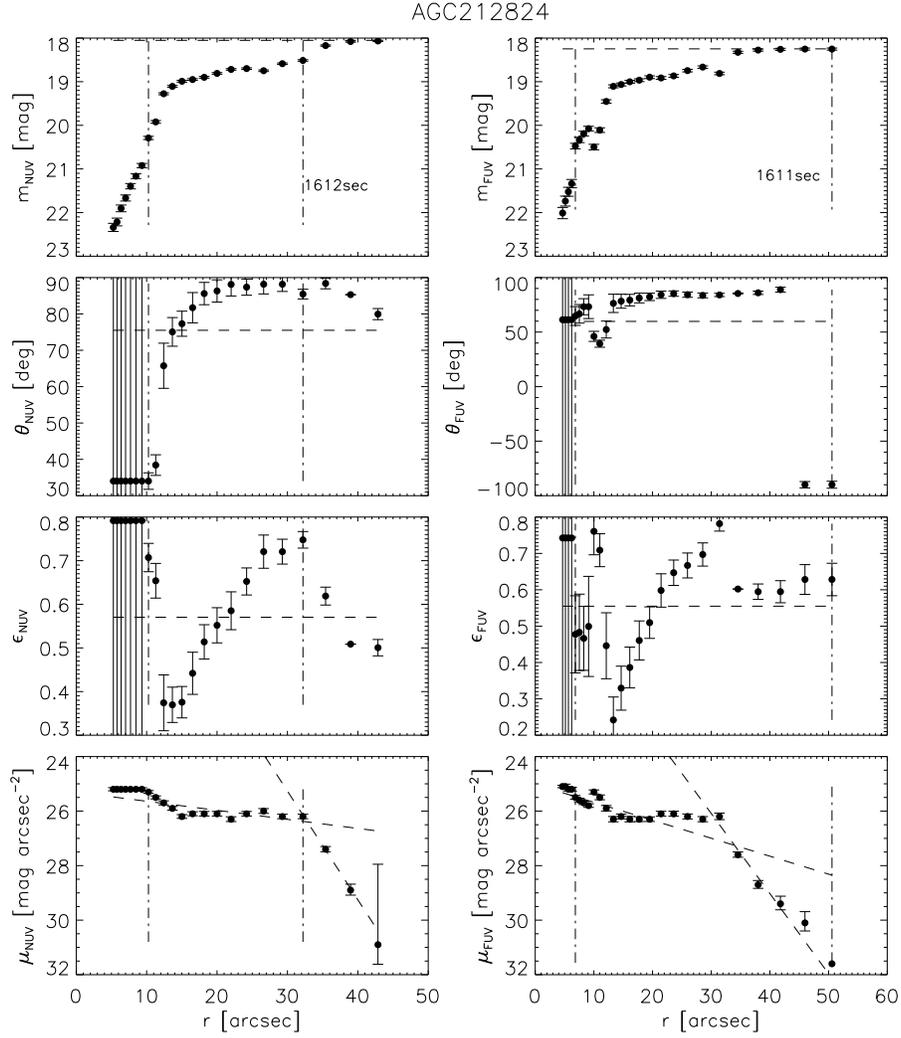}
}
\caption[]{Example of the isophotal fitting result for AGC~212824 = KK~100, which has an outer disk
that is steeper than the inner one. The symbol definition is the same as that in Figure \ref{fig:SB110482}. 
This object has the highest HI mass of any galaxy
in the supplementary {\it s-sup} sample, log $M_{HI}$ = 9.04. The flattening of the UV profile may be due 
to higher extinction in the center regions or, alternatively to a higher 
efficiency of star formation in the inner disk.
}
\label{fig:SB212824}
\end{figure*}

\begin{figure*}
\center{
\includegraphics[scale=0.5]{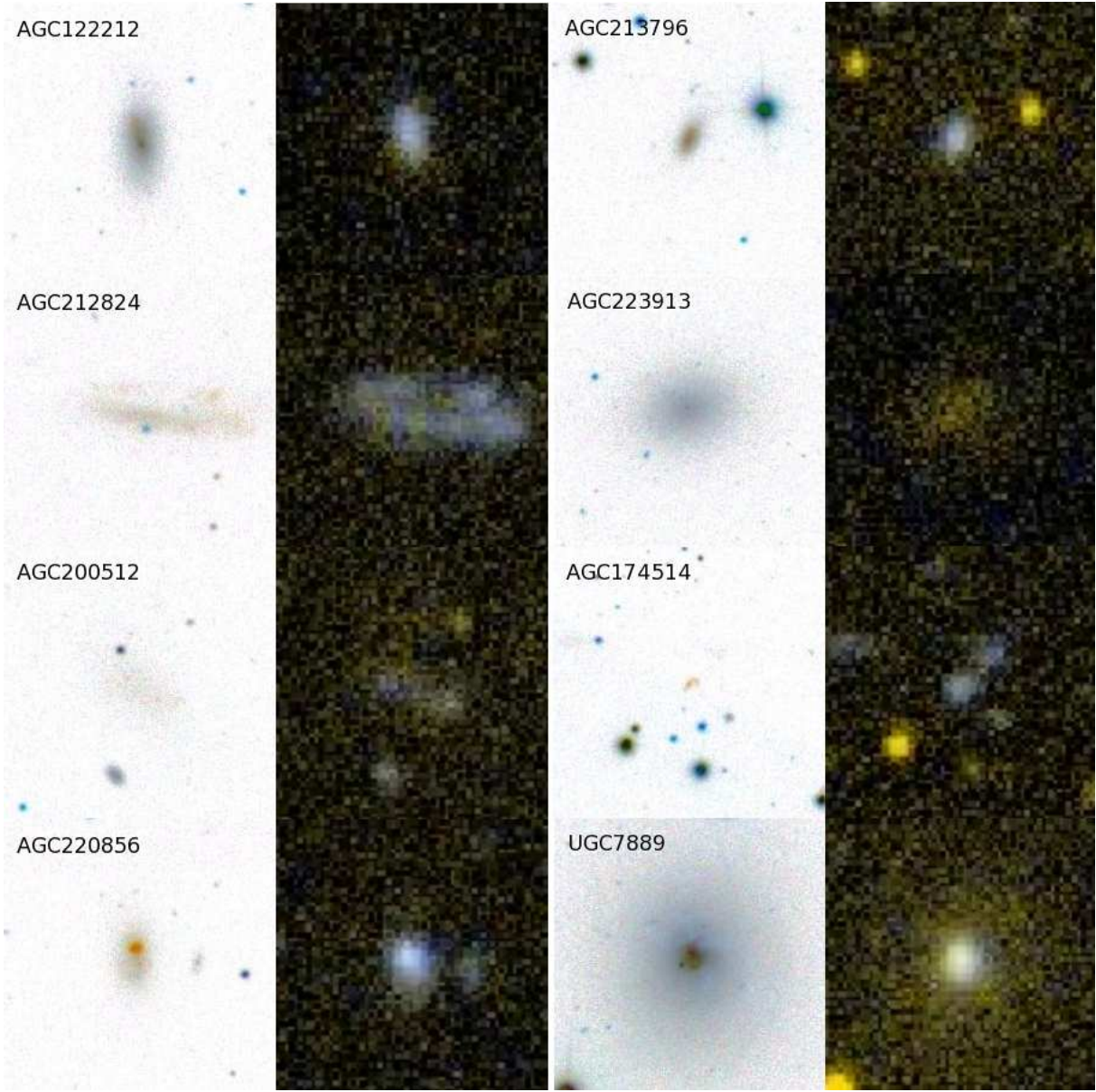}
}
\caption[]{Gallery of ALFALFA dwarf galaxies discussed in text. Inverted SDSS images, 1.5$^{\prime}$ on a side, 
	are shown to the left of the GALEX images (not inverted) of the same galaxies. 
{\it First\ row:} AGC~122212 (left) and AGC~213796 (right) both belong to the {\it s-com} sample and 
	have shallower outer disks in NUV. 
	The latter has no FUV outer disk. 
{\it Second\ row: } AGC~212824 - KK~100(left) is patchy and has a steeper UV outer disk. 
	It is also the most HI massive galaxy in the ALFALFA dwarf sample, belonging to the {\it s-sup} set. 
	AGC~223913 = VCC~1649 (right) is a dE/dSph in Virgo with 
	the lowest SSFRs among the {\it s-sed} set (see its definition in \S\ref{sed}), 
	i.e. it SF is suppressed. 
{\it Third\ row:} AGC~200512 = LeG~6 (left) is LSB and thus is 
	shredded by the GALEX pipeline measurement in FUV. 
	AGC~174514 (right) has the highest $f_{gas}$ value and bluest $u-r$ color among the {\it s-sed} galaxies. 
	Faint patches nearby may be associated with the same object.
{\it Bottom\ row:} AGC~220856 = VCC~1744 (left) and UGC~7889 = NGC~4641 (right) are both BCDs in Virgo. 
	The former suffers from serious emission line contamination and has the highest $b$-parameter 
	value among the {\it s-sed} objects. }
\label{fig:img}
\end{figure*}

To clean the regions over which the galaxy photometry is to be derived, we first used an 
automatic procedure to mask UV sources at least 2 galaxy radii away from its center, 
and then masked by hand sources within 2 radii deemed to be unrelated to the galaxy itself.
Starting from an initial guess marked by hand, elliptical surface brightness 
contours were then fitted to the cleaned images, using the STSDAS package ISOPHOTE, 
outwards to the radius at which the fitting fails to converge, and inwards to the seeing limit.
This process yields the azimuthally-averaged surface brightness profile as well as 
the variation with semi-major axis of the ellipse centroid, its position angle 
and ellipticity; these ellipses are then used later as the apertures for the photometric extraction
and allows for the interpolation of masked regions. The disk portion of the surface brightness 
profile is then fit by a linear function, 
using an interactive procedure that allows specification of the inner and outer disk radii, 
as discussed by \citet{Giovanelli1994} and \citet{Haynes1999}. 
Figures \ref{fig:SB110482} to \ref{fig:SB212824}
illustrate examples of the isophotal fitting result for four representative cases.
Selected SDSS and GALEX images of them are shown in Figure \ref{fig:img}. 

Figure \ref{fig:SB110482} shows the result for AGC~110482 = KK~13, a 
patchy dwarf galaxy whose FUV flux is dominated by two bright knots. 
The GALPHOT isophotes are centered on the brighter knot in the inner
region but the ellipse fitting treats the merged light from the two knots 
as a single disk at large radius. 
This process results in a second peak in the surface brightness profile, when the 
ellipses reach the center of the fainter knot, but at large radii,
the surface brightness profile falls exponentially as expected, yielding a valid total magnitude.

This process of fitting the disk assumes that the surface brightness profile 
of dwarf irregulars follows an exponential falloff \citep{Hunter2010}
so that disk scale lengths can be determined. However, dwarfs sometimes show multiple disk components, 
with clearly different slopes in inner and outer regions. In such cases, we fit two linear functions 
to each portion individually, and use the inner fit to determine the general disk properties, 
e.g. scale length, position angle, etc, whereas the outer fit is used to extrapolate the surface
brightness profile beyond the outermost measured isophote. Among the 130 ALFALFA dwarfs observed 
in GALEX (12 with no FUV exposure), 28 have such double disks. The majority of those (20/28) 
have shallower outer disks in the NUV (e.g. AGC~122212 in Figure \ref{fig:SB122212} and 
its images in Figure \ref{fig:img}), 
but 15/20 of them show no outer disk in the FUV 
(e.g. AGC~213796 in Figure \ref{fig:SB213796} and images in Figure \ref{fig:img}). 
Of the 8 blue compact dwarfs (BCDs) studied by \citet{Hunter2010}, 
5 also have NUV double exponentials, all of which drop more shallowly in the outer disk. 
Those authors argue that the shallower outer profile represents the underlying stellar 
population, while the steeper inner profile is dominated by the centrally concentrated and
intense recent star formation. Consistent with this interpretation, the shallow outer disks 
are frequently not in evidence in the FUV. Further evidence for this scenario arises from
the fact that the disk scale length is smaller in the FUV band compared to that seen in the NUV,
again suggesting that recent star formation is more centrally concentrated than is the overall
stellar disk, i.e., the active star-forming region is shrinking. 

On the other hand, of the 29 dwarf irregulars studied by \citet{Hunter2010}, eight have clear double exponential 
profiles in the NUV; in only one of those is the outer exponential shallower than the inner one. 
Similarly, most of the ALFALFA dwarfs with steeper outer disks (7/8) 
belong to the higher $M_{HI}$ {\it s-sup} sample; it is possible that some may be
more massive galaxies with more extinction in the inner regions. An example of this case is
shown in Figure \ref{fig:SB212824} for AGC~212824 = KK~100 (see its images in Figure \ref{fig:img}) 
which has the highest $M_{HI}$ in our sample. 
Another possible explanation for the apparent flattening of the UV profile was 
proposed in \citet{Boissier2000}, namely
that the SFR in the inner disk has been higher than the infall of the gas, leading to a progressive 
consumption of gas towards the center. In the outer parts however, star formation is 
less efficient, and the infall of gas proceeds on longer timescales. As a result, the gas reservoir of 
the outer disk is not exhausted, and the shape of the exponential profile is preserved. 
The steeper outer disk may thus be more evident in the gas rich dwarf irregulars, while
the lower $M_{HI}$  {\it s-com} dwarfs which are gas-poor relative to the overall ALFALFA  
population (see \S\ref{fgas}) frequently show flatter outer UV profiles.

Once ellipses are fit, magnitudes are calculated using the IRAF 
routine POLYPHOT to measure the total flux within the ellipses and the disk fits are
used to extrapolate beyond the measured isophotes. Following \citet{Haynes1999},
several sets of magnitudes are recorded, including ones at fixed isophotal 
levels, partial magnitudes integrated to a certain number of disk scale lengths, and 
asymptotic magnitudes extrapolated to infinity. The extrapolation 
helps to recover the LSB outer emission below the sky level. While
magnitudes extrapolated to infinity are adopted by the 11HUGS studies \citep{Gil2007}, we
follow the discussion in \citet{Haynes1999} and adopt
in this work a total magnitude computed at a radius of eight disk scale lengths 8$r_d$. 
Note that for the majority of the galaxies, a radial extent of 8$r_d$ lies beyond the 
outermost radii marked as defining the disk region, except in a few cases that the 
UV emission is compact and the image is particularly
deep (e.g. AGC~225852 in the NUV and AGC~220609 in the FUV). 
To determine the error in the total magnitude, following \citet{Salim2007}, we add the zero-point 
calibration errors of 0.052 (FUV) and 0.026 (NUV) mags to the poisson errors, 
and the uncertainty in the determination of the sky level, which is formally the rms noise 
of the pixel values in the sky boxes after masking the sources contained within them. 
We do not account for other types of errors, e.g. the flat-fielding errors, which are improved 
in the products of the latest version of the GALEX pipeline (GR6). 

Because of their extreme LSB in the UV, two galaxies, AGC~201970 = LeG~18
and AGC~223913 = VCC~1649 (see its images in Figure \ref{fig:img}), fail to produce convergence of
fitted isophotal ellipses. 
The former one is also LSB in the SDSS images, but AGC~223913,  a dE/dSph in Virgo, is only faint in 
the UV (see \S\ref{cmd} for more discussion). 
For these two objects, the
GALEX pipeline extracts a handful of unmatched faint sources in the NUV and FUV at the location
of the target galaxy. We assign concentric elliptical apertures with increasing 
semi-major axis by hand to extract a curve of growth and adopt their magnitudes 
from those measured in the outermost aperture. 

Only one galaxy, AGC~220483 = VCC~628 is too faint in the FUV to yield a reliable magnitude;
it is also undetected by the GALEX pipeline to a limiting magnitude $\sim$22.7 mag.
It is a dwarf irregular (Im) associated with the Virgo B Cluster but moving at very high velocity;
it exhibits LSB also at optical wavelengths.  
It remains the lone object in our sample of ALFALFA dwarfs with MIS-depth GALEX FUV imaging to
exhibit no traceable FUV emission.

\begin{figure*}
\center{
\includegraphics[scale=0.8]{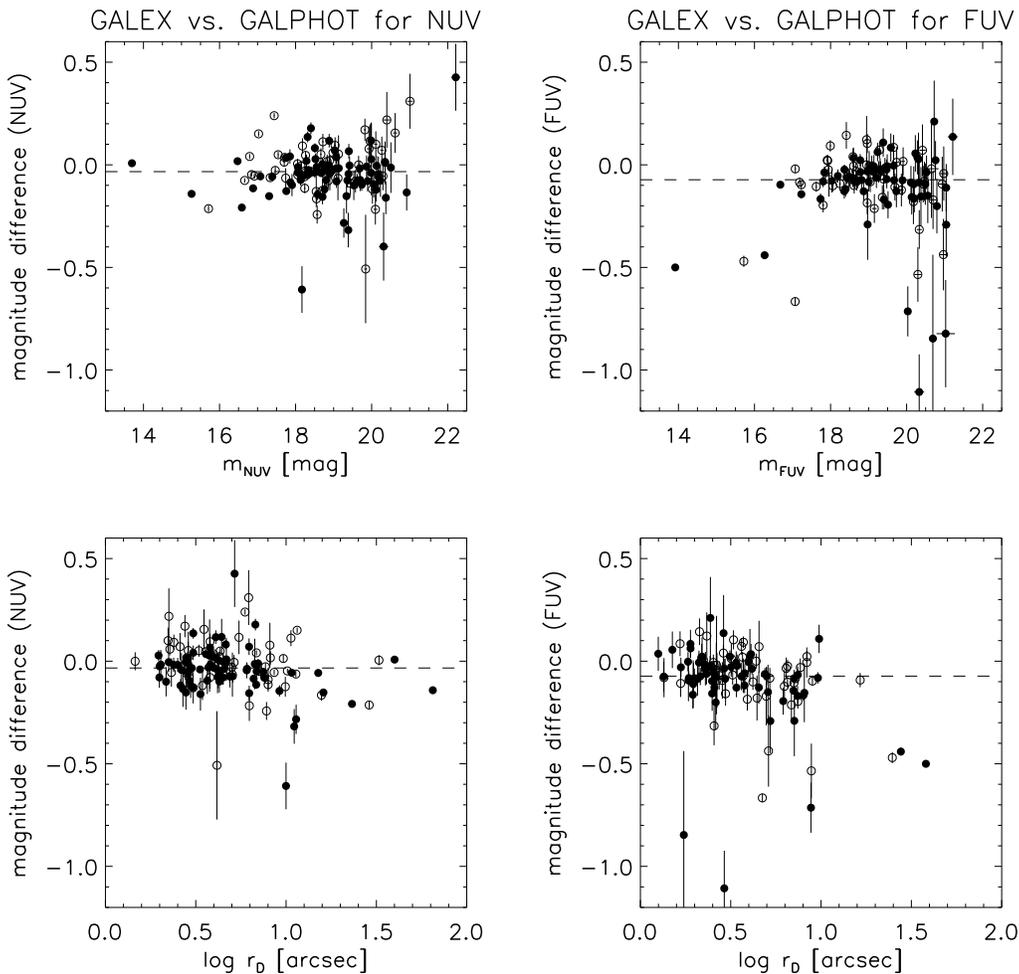}
}
\caption[]{Comparison of the total UV magnitudes of the ALFALFA dwarf sample 
obtained by using the GALPHOT package with those derived from the GALEX pipeline. 
The magnitude difference is defined as (mag$_{GALPHOT}$-mag$_{GR6}$). 
Results for the NUV channel are shown in the left panel, for the FUV on the right. 
Filled circles denote the galaxies within the complete {\it s-com}
sample and open ones, the additional {\it s-sup} objects. 
The galaxies which are completely missed by the pipeline are excluded from these plots. 
${\it Upper:}$ The magnitude difference as a function of our measurements. 
	In general, the GALPHOT photometry measures a brighter magnitude for a significant 
	number of galaxies, especially in the FUV.
${\it Lower:}$ The magnitude difference as a function of disk scale length r$_D$ 
	on a logarithmic scale. A weak trend is seen that the pipeline becomes ineffective at detecting 
	all emission in the more extended galaxies. Disk scale lengths are smaller on average in the FUV, 
	i.e., the SFR is more centrally concentrated. 
}
\label{fig:plotUVmag}
\end{figure*}

A comparison between our GALPHOT/ELPHOT-derived magnitudes and the GALEX pipeline magnitudes (GR6) 
is shown in the upper row of Figure \ref{fig:plotUVmag} for the NUV (left) and FUV (right) 
channels respectively. The magnitude difference is defined as (mag$_{GALPHOT}$-mag$_{GR6}$).
Filled and open circles denote the galaxies in the {\it s-com} and {\it s-sup} samples
respectively.
In addition to AGC~220483 = VCC~628, which was not detected in the FUV as discussed above, 
several galaxies are excluded from these plots because they were not detected by the 
GALEX pipeline in either one band or the other: 
AGC~205097 in the NUV (2 NUV-bright stars are nearby), 
AGC~112505 in the NUV (blend with 1 nearby NUV-bright star and also very close to UGC~1176 = DDO~13), 
AGC~223913 in the FUV (= VCC~1649, dE in Virgo, LSB in the NUV, invisible in the FUV by eye;
SDSS and GALEX images shown in Figure \ref{fig:img}).
In addition to these non-detections, several clear outliers are seen in Figure \ref{fig:plotUVmag}. 
The outliers in the FUV plot have a brighter magnitude obtained by the
`GALPHOT' ellipse fitting than that measured by the GALEX pipeline. 
Close inspection shows that they are mostly extended irregulars 
with large disk scale lengths and patchy UV emission; they suffer from significant shredding by the 
GALEX pipeline, e.g. UGC~4415 and AGC~122206 in the NUV; AGC~200512 in the FUV 
(SDSS and GALEX images are shown in Figure \ref{fig:img}); 
AGC~201970 in both the NUV and FUV. Several galaxies lie at such nearby distances that
they have resolved HII regions visible in the FUV 
(e.g. UGC~12613 with a distance of only 0.9 Mpc; UGC~5373 = Sextans B at 1.3 Mpc). 
After excluding the non-detections and outliers which arise from blending or shredding by the
pipeline process, the magnitudes we derive are in reasonable agreement with the 
pipeline magnitudes. The improved GR6 pipeline magnitudes agree better with our results, 
especially in the NUV. 

The bottom row of Figure \ref{fig:plotUVmag} shows the same magnitude difference versus 
the disk scale length r$_D$ on a logarithmic scale. As evident in Figure \ref{fig:plotUVmag},
our photometric extraction technique, which is specifically designed to capture all of 
the low surface brightness flux, yields brighter magnitudes for a significant number of galaxies. 
The median magnitude difference is $-0.035$ and $-0.073$ mag in the NUV and FUV, respectively. 
A weak trend is seen such that, as the galaxies become more extended (larger r$_D$), 
the GALEX pipeline misses the outer LSB UV emission and hence underestimates the
true UV magnitude. For the remainder of this work, we use our own measurements of GALEX UV 
magnitudes, denoted as $m_{NUV}$ and $m_{FUV}$. 

\subsection{SDSS data} 
\label{SDSSdata}

In addition to the UV photometry from GALEX, we use optical data from the SDSS Data Release 7 
\citep{Abazajian2009}.
The SDSS database provides photometry in five bands ($u,~g,~r,~i,~z$) and spectroscopic follow-up 
for most galaxies with $r<17.77$. For the present work, we use the SDSS archival measurements 
to derive the properties of the underlying stellar population, but because
of issues of shredding and blending similar to those found with the GALEX magnitudes, significant
caution is applied. 
Given the higher resolution of the SDSS imaging and the higher density of bright 
stars, the problems associated with magnitude estimation are worse in the optical than in the UV.
To check for such issues, we inspected visually each galaxy and the magnitudes derived
from the SDSS pipeline, and adopted the photometric properties associated with the
best photometric crossmatch along with its reported quality code, sFlag (see below). 
As suggested by the SDSS team, {\it modelmag} magnitudes are adopted to derive 
stable colors while still capturing most of the total light.

Of the 
229 galaxies in the {\it s-com} and {\it s-sup} samples, 
24 lie outside of the SDSS Legacy
Survey footprint; most of these are found in the fall sky region of ALFALFA 
($22^h <$ R.A.$< 3^h$, $0 <$ Dec. $< +36^\circ$). 
It is well known that the SDSS photometric pipeline is optimized for small, high surface brightness 
objects \citep{West2010}, not the clumpy and LSB dwarf galaxies typical of our sample.
In fact, of the 205 ALFALFA dwarfs in the SDSS sample, only
44 ALFALFA dwarfs are extracted as single photometric 
objects by the SDSS pipeline (sFlag=``oly'' as given in Column 8 of Table \ref{table:dwarf}). 
Since multiple peaks are usually evident in the light profiles of patchy dwarfs (the ``parent'' object), 
these are often de-blended by the standard pipeline into pieces (the ``child'' objects). 
Accurate magnitudes can be recovered in those cases where the ``parent'' object contains 
all (or nearly all) of the emission associated with the galaxy and is not blended with any other nearby objects. 
53 of the ALFALFA dwarfs belong to this category (sFlag=``par''); in these cases, the magnitudes of 
the associated {\it parentID} are adopted. For the remaining objects for which the ``parent''
objects still suffer from blending, we examined the magnitudes of each ``child'' object
within the region of the target galaxy. According to \citet{West2010}, roughly 
75\% of galaxies have more than 90\% of their flux contained in the brightest child. 
We used the magnitudes of the brightest child where it is brighter than the second brightest child 
by at least 3 magnitudes, implying that it contains the vast majority of the galaxy's flux. 
55 of the 205 ALFALFA-SDSS dwarfs fit this category (sFlag=``domi'') so that the SDSS magnitudes 
contribute a satisfactory lower limit of the optical flux. 
The remaining 53 galaxies have photometry too uncertain to be used further, usually because
the parent object is blended with 
other sources or has no dominant child (sFlag=``pbphot''). 
It is relevant to note that most are dropped for one of two specific reasons: either (1) because
they are very patchy or (2) because they are contaminated by the presence of nearby or superposed stars.
The latter reason has nothing to do with the galaxy itself, but the first is particularly
common for this sample of dwarf galaxies. 
We checked the distribution of HI properties for the excluded objects, and find that
they span the full range in distance, line width, and HI mass evident in Figure \ref{fig:sample5}, 
suggesting that the inclusion of a criterion of acceptable SDSS photometry introduces no extra 
bias in terms of HI parameters. Although we have not reprocessed the SDSS photometry as did \citet{West2010}, 
the careful visual inspection technique applied here effectively avoids the shredding or 
blending of the SDSS magnitudes which are the main causes of the failure of the SDSS
pipeline to deliver reliable photometry for diffuse, patchy and/or LSB galaxies.  

However, because of the typical faintness and 
LSB optical appearance of the ALFALFA  dwarfs, the percentage of the photometric objects 
that are targeted for SDSS spectroscopy is quite low: only 
114/205 have a counterpart in the
SDSS spectroscopic survey. This circumstance further validates the importance of
HI-selection for deriving a complete sample of dwarfs in the local universe. 
Furthermore, only 101 of the 114 spectroscopic targets are included in the MPA-JHU DR7 
release of SDSS spectral measurements 
\citep[http://www.mpa-garching.mpg.de/SDSS/DR7/, ][]{Brinchmann2004}, 
which contains reprocessed line flux, etc. We confirm that all these 101 dwarfs are star-forming or 
low S/N star forming galaxies, following the classification in \citet{Brinchmann2004}. 
In comparison with the SDSS-selected population, the HI-selected dwarf sample 
includes a much smaller representation of galaxies containing AGNs or are classified as `non-star-forming'; 
this finding also implies that the SDSS magnitudes are more likely to be contaminated by line emission. 
Similarly, a quick inspection shows that all the ALFALFA dwarf galaxies 
with SDSS $D_n(4000)$ measurements have a value below 1.6, the demarcation in the bimodal distribution 
derived in \citet{Kauffmann2003}. This result suggests that these dwarfs have been actively forming stars 
throughout their history. 

\subsection{UV-to-optical colors of the ALFALFA dwarfs}
\label{cmd}

The global color-magnitude diagram (CMD) is a powerful tool for the assessment of 
the basic properties of a sample of galaxies. The bimodal nature of the field galaxy 
CMD is well demonstrated with the large datasets available from recent large scale
galaxy surveys, in particular SDSS \citep{Baldry2004}. It is clear that galaxies separate
into the `red sequence' of early type galaxies which show little or no evidence of ongoing 
star formation (corresponding to the low SSFR portion in the SSFR versus $M_*$ diagram), 
and the `blue cloud' of star-forming spirals (corresponding to the star forming sequence 
in that diagram). Compared to the traditional optically-based CMDs, 
a CMD constructed from a UV-to-optical color provides a more powerful diagnostic 
\citep{Wyder2007, Schimin2007, Salim2007}. 
By contrasting the recent star formation, as indicated by the UV light, to the total past 
star formation, as indicated by the optical light, the UV to optical color provides
a more concrete diagnostic of a galaxy's SFH \citep{Salim2005}. 

\begin{figure*}
\center{
\includegraphics[scale=1]{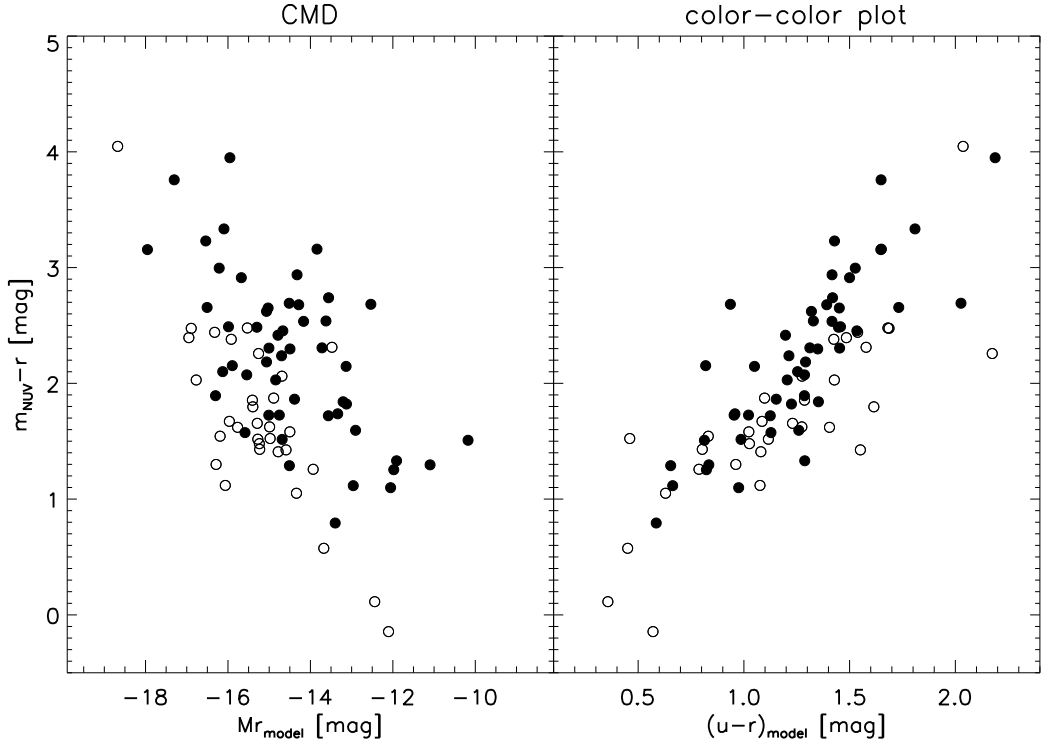}
}
\caption[]{The UV-to-optical colors of the ALFALFA dwarfs. Filled circles identify dwarfs 
   in the lowest HI mass complete {\it s-com} sample, while open ones represent the additonal {\it s-sup} galaxies. 
{\it Left:} Color-magnitude diagram. Unsurprisingly, nearly all the ALFALFA dwarf 
        galaxies are less luminous than $M_r \simeq -18$, 
	and lie below the blue cloud -- red sequence divider quoted by \citet{Salim2007}, $m_{NUV}-r \lesssim 4$. 
{\it Right:} Color-color plot. Within the blue color range occupied by the sample overall, 
        the $m_{NUV}-r$ color is well correlated with the $u-r$ color, 
	with the former probing a larger dynamical range, $\delta(m_{NUV}-r)/\delta(u-r)\sim2$. 
}
\label{fig:CMD}
\end{figure*}

To produce the CMD shown in Figure \ref{fig:CMD}, the magnitudes given in 
Table \ref{table:dwarf} have been corrected for foreground reddening. 
For the SDSS bands, we adopted the pipeline extinction correction. 
For the GALEX bands, we used $E(B-V)$ values based on the FIR DIRBE maps of \citet{Schlegel1998}, 
the \citet{Cardelli1989} extinction law with $R_V=A_V/E(B-V)=3.1$, and 
$A(\lambda)/E(B-V)=8.24$ for the FUV and 8.2 for the NUV bands respectively, following \citet{Wyder2007}. 
Of the 229 galaxies listed in Table \ref{table:dwarf},
51 belonging to the complete {\it s-com} sample and 32 to the higher $M_{HI}$ supplementary {\it s-sup} set
have both acceptable SDSS photometry (sFlag not ``pbphot'') and NUV magnitudes,
making it possible to estimate a color ($m_{NUV} - r$). Their distribution 
in the combined UV-optical CMD is
shown in the left panel of Figure \ref{fig:CMD}, with filled circles denoting objects
in the {\it s-com} sample and open ones representing the additional {\it s-sup} galaxies. 

The first obvious difference of this work from most previous studies is the much fainter absolute 
magnitude range probed by the ALFALFA dwarfs. For example, as clear 
from Figure 1 of \citet{Salim2007}, the majority of galaxies in that study have 
$-24 \lesssim M_r \lesssim -18$, whereas nearly all of the ALFALFA dwarfs are less 
luminous than $M_r\simeq-18$, with the mean $M_r \sim -15$. 
Secondly, HI selection results in no obvious color bimodality in dwarfs, because nearly all HI-bearing
galaxies are blue, actively star-forming systems; HI selection is highly biased against the red sequence 
\citep{West2009, Haynes2011}. Based on the distribution of their full GALEX-SDSS matched catalog, \citet{Salim2007} 
identified blue cloud galaxies as those with $m_{NUV}-r \lesssim 4$. 
Given the color-magnitude relation, fainter galaxies are bluer on average 
and the dividing criterion shifts to $m_{NUV}-r \sim 3.5$ at $M_r \sim -17$, according to Figure 1 in \citet{Kim2010}. 
As evident in Figure \ref{fig:CMD},
virtually all the ALFALFA dwarf galaxies are below the division of $m_{NUV}-r = 4$. 
Only three objects, UGC~6245, AGC~223913 = VCC~1649 and AGC~222297 = VCC~180 (the latter two early type
galaxies associated with the Virgo Cluster) have $m_{NUV}-r >3.5$, lying in the green valley 
or blue edge of the red sequence. 
At fixed $M_r$,
galaxies with lower $M_{HI}$ (the filled circles) have, on average,
redder colors, indicating that they also exhibit a lower gas fraction ($f_{gas}$, 
defined as $M_{HI}/M_*$ throughout this work); this trend confirms the general association of
lower $f_{gas}$ and redder color (see also \S\ref{fgas}). 
As a result, in the range of $-18 \lesssim M_r \lesssim -10$, a population of extremely 
low gas fraction dE/dSph would have $M_{HI}$ lower than 
the detection limit of ALFALFA; they would sit on the red sequence in this low luminosity 
range, but are not included in this sample. This point is further discussed in
the case of the Virgo cluster by
\citet{Hallenbeck2012}. In addition, 
there is a trend for lower luminosity objects to have bluer colors, 
as is also found in studies of large samples of massive galaxies \citep{Wyder2007}. 
As is the case with the full ALFALFA sample \citep{Haynes2011, Huang2012},
the CMD shown in Figure \ref{fig:CMD} confirms that, in comparison with large optical-UV samples,
an HI-selected dwarf sample is highly biased against red sequence galaxies.

The right panel of Figure \ref{fig:CMD} presents a color-color plot, with the horizontal axis
showing the optical color $(u-r)$. Not surprisingly, there is a strong
correlation such that as a galaxy becomes redder in $m_{NUV}-r$, it also becomes redder in $u-r$. 
As found by \citet{Wyder2007} for blue galaxies, the two colors are very well correlated with 
a slope of $\delta(u-r)/\delta(m_{NUV}-r)\sim0.5$; this slope is roughly consistent for the ALFALFA dwarf
sample. Those authors also found that, 
for galaxies with colors redder than $m_{NUV}-r\approx3.5$, the $u-r$ color begins 
to increase less quickly with $m_{NUV}-r$ than for bluer $u-r$ colors. 
The ALFALFA sample shows no obvious change in slope, as expected since most lie on the 
blue side of the division.

We also examine the $m_{FUV}-m_{NUV}$ color, a sensitive probe of
the rate of current star formation. There are 117 galaxies (69 belonging to the
complete {\it s-com} sample and 48 to the supplementary {\it s-sup} one) which have detectable 
magnitudes in both the FUV and the NUV. 
As described in \citet{Gil2007}, late-type spiral and irregular `blue-cloud' galaxies can be roughly 
separated from early-type `red-sequence' galaxies using a division at $m_{FUV}-m_{NUV}=0.9$. 
According to that demarcation, 92\% of our galaxies fall on the blue side. 
72\% of the combined ({\it s-com} plus {\it s-sup}) sample have $m_{FUV}-m_{NUV}<0.5$; among the
lower HI mass {\it s-com} galaxies, this percentage drops to 64\%. 
Furthermore, the median UV color is 0.36 mag for the {\it s-com} plus {\it s-sup} galaxies 
(0.42 mag for the {\it s-com} and 0.31 mag for the {\it s-sup} galaxies). Evaluation of the
K-S statistic shows that the probability that the UV color distributions of 
the two samples ({\it s-com} vs. {\it s-sup}) 
are drawn from the same underlying distribution is only 2.6\%. 
The galaxies with the lowest HI mass ({\it s-com}) are 
{\it redder} than comparable galaxies with higher HI masses, that is, 
the HI mass cutoff imposed on the {\it s-com} results in a sample with redder UV colors on 
average at {\it fixed optical luminosity} (see also \S\ref{fgas}). 

\begin{figure*}
\center{
\includegraphics[scale=0.8]{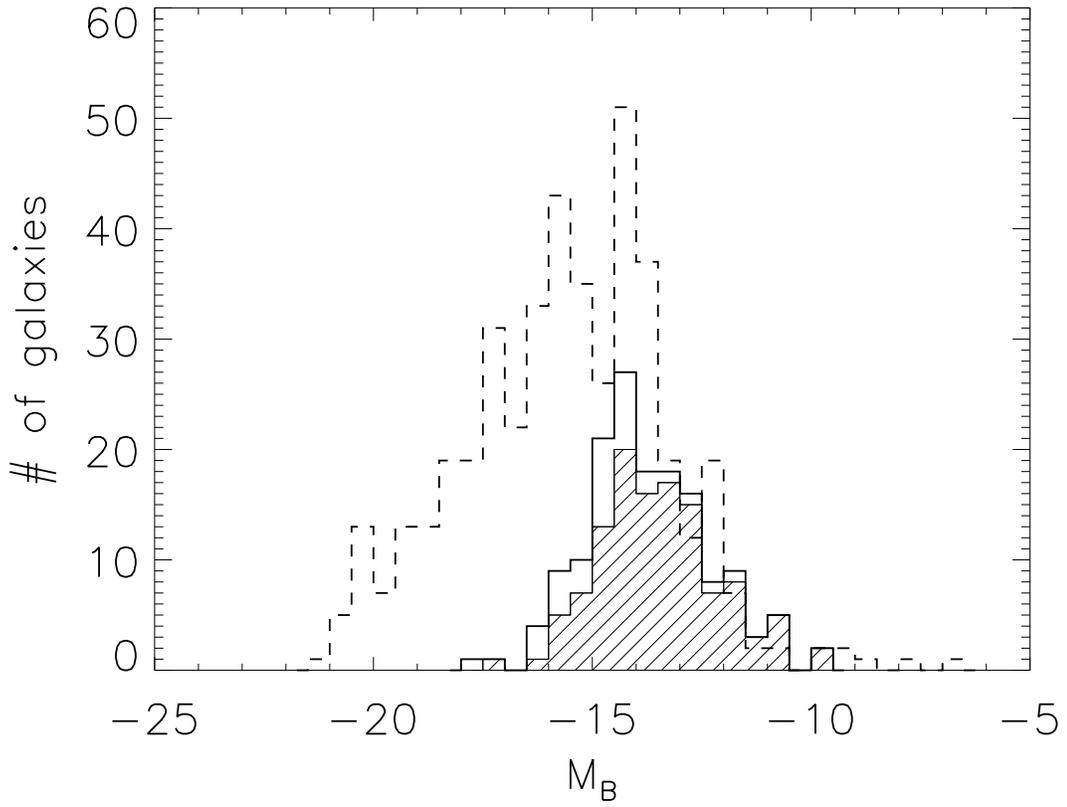}
}
\caption[]{$B$-band absolute magnitude distribution of the 152 galaxies with acceptable SDSS magnitudes. 
Solid histograms indicate the ALFALFA dwarf galaxies, with the shaded area tracing the {\it s-com} sample and 
the open area, the {\it s-sup} galaxies. 
In comparison, the dashed histogram shows the distribution for the 11HUGS sample. 
Although the  {\it s-com} sample extends to lower HI mass and its galaxies have, on average, 
lower gas fractions, both ALFALFA and 11HUGS probe to comparably faint optical absolute magnitudes.
}
\label{fig:absMag}
\end{figure*}

In comparison, the sample studied by \citet{Lee2011}, mainly drawn from the 11HUGS, has 
79\% of the galaxies with $m_{FUV}-m_{NUV}<0.5$, and a median UV color of 0.29 mag, 
bluer than the ALFALFA dwarf galaxies, especially the {\it s-com} ones. Figure \ref{fig:absMag}
examines more closely the distribution of absolute magnitude $M_B$ for the ALFALFA dwarf
and 11HUGS samples. The solid histograms trace the distribution of the 152 ALFALFA dwarfs 
with acceptable SDSS magnitudes (see \S\ref{SDSSdata}), with the shaded area identifying 
the complete {\it s-com} sample and the open area, the {\it s-sup} galaxies. 
To convert the SDSS bands to $B$-band, the transformation equation derived by 
Lupton (2005) is used, $B = g + 0.3130(g - r) + 0.2271$. For comparison,
the dashed histogram represents the parent 11HUGS sample with their $B$-band absolute magnitudes 
drawn from Table 3 of \citet{Kennicutt2008}. 
Although the ALFALFA {\it s-com} sample extends to considerably lower HI mass, 
the faintest optical magnitudes probed by both ALFALFA and 11HUGS are similar.
Relative to the 11HUGS galaxies, the
lower gas fractions typical of the {\it s-com} dwarfs are consistent with their redder colors. 

One of the ALFALFA dwarfs, AGC~223913 = VCC~1649 (see its images in Figure \ref{fig:img}) 
even has $m_{FUV}-m_{NUV}=3$, implying that it is in 
a quiescent stage of star-formation. It is a dE/dSph in Virgo, whose red UV color could 
result from a recent quenching of star formation activity, possibly due to gas loss by 
ram pressure stripping \citep[e.g.][]{Boselli2008}. This object and the other early type dwarfs
detected by ALFALFA in the
Virgo cluster are discussed in more detail elsewhere \citep{Hallenbeck2012}.

\section{The derivation of the physical properties of ALFALFA dwarf galaxies}
\label{prop}

In this section, we use the datasets discussed above to
derive the physical properties of the ALFALFA dwarf galaxies ({\it s-sed}): their stellar
masses, internal extinction, metallicity and SFR.

\subsection{SED fitting}
\label{sed}

While studies of resolved stellar populations provide the best means to determine
the star formation histories of galaxies, it is not yet feasible to
conduct such studies on the majority of the ALFALFA dwarfs. Hence we
analyze their integrated light and global properties by SED fitting, 
following the method of \citet{Salim2007}, with a slight change in the prior distribution of 
the effective optical depth in V band, $\tau_V$ (see \S\ref{Afuv}). 
Of the full ALFALFA dwarf
sample, reprocessed FUV and NUV magnitudes and acceptable quality 
SDSS pipeline photometry as described above
are available for a total of 74 galaxies, 45 belonging to the more restrictive {\it s-com} 
sample and 29 to the additional {\it s-sup} set. 
The full likelihood distributions of parameters are derived for this combined sample,
referred to hereafter as {\it s-sed}, by fitting the seven observed SDSS ($u~g~r~i~z$) 
and GALEX ($FUV~NUV$) bands to an extensive library of model SEDs \citep{Gallazzi2005}, 
generated using the B\&C 03 
stellar population synthesis code \citep{Bruzual2003}. Dust is accounted for with the 
\citet{Charlot2000} two-component model to include attenuation from both the diffuse 
interstellar medium (ISM)
and short-lived (10Myr) giant molecular clouds. 
It is proposed to produces consistent treatment 
for both the H$\alpha$ and UV continuum attenuation. 
However, we do note that recent studies show that the extinction curve may be a function of 
stellar mass, SSFR, axis ratio and stellar surface mass density, etc 
\citep[e.g.][]{Johnson2007, Wild2011}. A \citet{Chabrier2003} IMF is assumed, 
and random bursts are allowed to be superimposed on a continuous SFH. 

A concern for the study of HI-selected galaxies is whether the parameter space used
for building the library is set wide enough to cover the intrinsic values of the metal-
and dust-poor ALFALFA dwarfs. For example, following \citet{Salim2007}, the effective 
$V$-band optical depth lies in the range $0 \leqslant \tau_V < 6$, and the $\mu$ factor, 
i.e., the fraction of the optical depth that affects stellar populations older than 10 Myr, 
varies from 0.1 to 1. A commonly-used prescription by
\citet{Calzetti1997} adopts $E(B-V)_{star} = 0.44 E(B-V)_{line}$ (see also discussion in \S\ref{Afuv}).
Furthermore, the metallicity of the stellar population is uniformly 
distributed between 0.1 and 2 $Z_\odot$. 
In particular, we can anticipate that in some cases, the emission lines may be so 
strong that the observed colors may deviate from the adopted line-free continuum model 
\citep{Salim2007, West2009}, e.g. the H$\alpha$ emission may dominate the $r$-band luminosity 
to offset the observed $r-i$ color bluewards of the models. However, outliers
affected by line emission can be identified in a color-color plot, e.g. $r-i$ versus $g-r$. 
As a result, the $\chi^2$-value of the best-fitting 
model, $\chi^2_{best}$, would have a mean larger than that predicted by the degree 
of freedom of the fitting. We checked the distribution of $\chi^2_{best}$ and found 
it is generally good, implying that the library does, in fact, reproduce most of the observed SEDs. 
However, there is a tail of large $\chi^2_{best}$ objects, and we confirm by visual inspection
that their deviation is likely caused by strong line emission 
(e.g. AGC~220856 = VCC~1744 and AGC~223390 = VCC~274, two Virgo BCDs; 
the SDSS and GALEX images of the former are included in Figure \ref{fig:img}). 
Additionally, although we already excluded the ``shredded'' sources, there remain a few objects with 
suspicious SDSS magnitudes, e.g. AGC~205165 (problematic de-blending of a superposed star), 
AGC~191791 = LSBC D634-03 (unusual color in the SDSS image), which also lead to large $\chi^2_{best}$. 
For these reasons, we emphasize that the SED fitting results should be interpreted 
only in a statistical sense, with these large $\chi^2_{best}$ objects being less reliable. 
The galaxies with $\chi^2_{best}<10$ are noted by a `*' over the AGC numbers in Table \ref{table:dwarf}. 

Compared to SED fitting to the optical SDSS bands only, fitting with the addition of 
the GALEX UV bands yields better constraints on the parameters, especially on the dust 
optical depth and the SFR over a timescale of 100 Myr, comparable to the lifetime 
of UV bright stars \citep{Salim2005}. 
We use the likelihood-weighted average as our nominal estimate 
of the logarithm of the parameter value, rather than the mode, to avoid sensitivity to binning, 
and $1/4$ of the 2.5-97.5 percentile range as a proxy for what would have been the uncertainty estimate 
in the Gaussian distribution \citep{Kauffmann2003b,Salim2007}. An extra term which accounts for
distance uncertainties is added for distance-dependent quantities, e.g. the stellar mass and SFR. 
This term always dominates the SED fitting error (characterized by 
the probability distribution function, PDF) for the nearby dwarfs in the stellar mass estimates, 
but not for the errors in the SFRs. Note that
systematic uncertainty is not included in the error estimate. 
When the UV bands are excluded, the median uncertainty in the $\log M_*$ estimate is 
0.217; in $\log \tau_V$, it is 0.455; in $\log SFR$, it is 0.323. 
When the UV data are incorporated, those values decrease to 
0.212, 0.367, 0.277, 
corresponding to median improvement of
1\%, 21\% and 7\% respectively. While the constraint on the stellar mass is least improved,
the incorporation of the UV data is critical to the other parameters, both of which
are more sensitive to changes in the UV luminosity.   

\subsection{Stellar mass} 
\label{smass}

Based on the SED fitting error term characterized by the PDF, the
$M_*$ is the best constrained parameter in the SED fitting. 
The logarithm of $M_*$, in units of solar mass, is listed in Column (11) of Table \ref{table:dwarf}.
$<\log M_*>$= 7.40 for the {\it s-sed} dwarfs (
7.28 for the lower HI mass {\it s-com} galaxies and 7.58 for 
the additional {\it s-sup} objects), significantly lower than those of typical SDSS or GALEX samples 
\citep{Brinchmann2004, Salim2007}. 
This result further confirms that the imposition of an HI line width cutoff
effectively eliminates most of the HI deficient but high luminosity objects. 
The only two galaxies with $\log M_*>8.5$ are UGC~6245, 
a low luminosity SB0 galaxy viewed almost face-on so that its true 
velocity width may be much larger, and UGC~7889 (= NGC~4641), a BCD in Virgo. 
The former has the highest stellar mass
$\log M_*=9.52$ and a low gas fraction, but is left in the {\it s-sup} sample 
because of the availability of its GALEX image from our GI program (see \S\ref{sample}). 

Another widely used method to estimate the stellar mass from optical magnitudes employs
the relation of mass to light ratio versus color as calibrated by \citet{Bell2003}. 
However, the SFH adopted by \citet{Bell2003} does not fully
account for the impact of the bursty behavior seen in nearby dwarfs \citep{Lee2009b}. 
Gas-rich dwarfs have $b$-parameters ($b \equiv SFR/<SFR>$, the current SFRs normalized to the average 
past SFRs) which are high on average (see \S\ref{SFS}). As a result, application of
the standard \citet{Bell2003} calibration would produce stellar masses that
are more massive for the same optical color. Additionally,
at low redshift, H$\alpha$ emission may contribute to the $r$ band flux and [OIII] 
and H$\beta$ to that at $g$ band. On the other hand, the $g-r$ color is 
largely unchanged \citep{West2009} and the $i$ band contains almost no emission lines. 
Therefore, following \citep{West2009}, we also calculate the stellar mass
using the $i$ band luminosity and the $g-r$ color. We apply K-corrections using the IDL code ``kcorrect'' 
($v4\_1\_4$), described in \citet{Blanton2007} and estimate internal extinction
corrections based on equation (12) 
in \citet{Giovanelli1997}. The latter has little effect on the stellar mass estimate, 
since when the luminosity is increased, the color also becomes bluer by an associated amount 
such that the two effects cancel each other \citep{Bell2003}. 

The two stellar mass estimates, i.e., that derived from \citet{Bell2003} calibration versus 
that derived from SED fitting, agree 
for the most massive galaxies with $M_* \gtrsim 10^{10} {\rm M_\odot}$ and low $b$-parameters, 
but systematic deviations between the two become non-negligible at the low mass end. 
While the \citet{Bell2003} method gives a median $\log M_*$=7.73 for the {\it s-sed} ALFALFA dwarfs 
(converted to Chabrier IMF), SED fitting yields a considerably lower median of 7.45 corresponding 
to a factor of two difference. 
Furthermore, because the most gas rich galaxies (i.e. higher $M_{HI}$ for their $M_*$) 
with higher $b$-parameters in this stellar mass range are excluded from the dwarf sample 
in this work (see \S\ref{SFgas}), the deviation is even larger among the full low
stellar mass ALFALFA population \citep{Huang2012}. 
We find a discrepancy of similar magnitude to that found by \citet{Wyder2007}, $\sim$0.4 dex on average at 
$10^8 {\rm M_\odot}$, and we confirm that the galaxies which are estimated to be more massive 
by the \citet{Bell2003} method have generally higher $b$-parameters. 
To incorporate reasonably the effects of stochastic star formation, bursty behavior in dwarf galaxies 
should be a key feature of modeling the SEDs. 

\subsection{Internal extinction}
\label{Afuv}

Although the FUV luminosity provides a more robust estimate of the recent SFR in dwarf galaxies
than optical measures, 
it is subject to significant uncertainties because of the required dust attenuation corrections
which themselves are subject to large scatter produced by the large range of possible dust content, 
dust distribution, and geometry relative to the stars and ISM in galaxies. 
Even at $0.02 Z_\odot$, I Zw 18 contains a non-negligible amount of reddening, determined from 
optical observations to be equivalent to $A_V = 0.5$ \citep{Cannon2002}. 
In the dwarf galaxy SBS 0335-052 ($0.025Z_\odot$), $A_V$ values as high as 20 to 30 have been 
suggested from MIR observations \citep{Thuan1999}. 
On the other hand, arguing that dwarf galaxies are expected to be extremely dust poor, \citet{Roy2009} chose to neglect 
internal extinction entirely in their study of the FIGGS dwarf galaxy sample. Based on that
assumption, they found a lower average $\Sigma_{SFR}$ for the FIGGS dwarfs than would be expected 
from the Kennicutt-Schmidt law \citep{Kenn1998b}. 
Similarly, \citet{Hunter2010} took $E(B-V) \sim 0.05$ and \citet{Lee2009} found $A_{H\alpha} < 0.1$ 
for the faintest galaxies. 
Here, rather than ignoring extinction, we 
have attempted to quantify its contribution by SED fitting using the optical/UV photometry. 

\citet{Salim2007} already have pointed out the difference between the 
effective optical depth in V band derived from emission-line fitting, 
$\tau_{V, H\alpha}$, and that derived from SED fitting, $\tau_{V, UV}$, 
as a function of stellar mass. 
$\tau_{V, H\alpha}$ arises to first order from the Balmer decrement, whereas the latter 
is mainly constrained by the UV spectral slope. At lower masses, $\tau_{V, UV}$ is higher than $\tau_{V, H\alpha}$, 
but the situation is reversed at the high mass end. According to those authors, 
$\tau_{V, UV}-\tau_{V, H\alpha} \sim 0.25$~dex (equivalent to 0.27 mag in $A_V$) at $\log M_* = 8.5$. 
This finding can be linked to that found by \citet{Wild2011} that $\tau_{V, cont}/\tau_{V, line}$ strongly 
increases with increasing SSFR, i.e., the galaxies with high SSFRs have a higher fraction of 
diffuse dust and their dust is more centrally concentrated. Adopting the same prior distribution 
of $\tau_V$ and $\mu$ as in \citet{Salim2007}, we find, for the {\it s-sed} sample,
mean values of $A_{FUV}$ = 1.47 mag and $A_{NUV}$ = 1.08 mag; the
the Balmer decrement yields systematically smaller values of $A_{FUV}$, assuming the Calzetti law (see below).  
In one object, UGC~7889 (= NGC~4641; see its images in Figure \ref{fig:img}), 
the $A_{FUV}$ value reaches 6.9 mag. This object is one of the few BCDs in Virgo detected
by ALFALFA and, among the dwarf {\it s-sed} sample, it
is one of the reddest in $m_{FUV}-m_{NUV}$ color, has one of the highest
stellar masses, has a low gas fraction and is relatively metal-rich. 
Given its UV color, the high $A_{FUV}$ value of UGC~7889 derived by the SED fitting 
is consistent with that predicted by the $IRX-\beta$ relation (see below) as
calibrated for starburst galaxies \citep{Meurer1999}, but other methods
such as the Balmer decrement and the $IRX-SFH-color$ relation (see below), give quite lower extinction estimates. Given the possible degeneracy of its
red color arising from either high extinction or strong SF quenching in the Virgo environment,
it is probable that the large extinction implied by the SED fitting for this galaxy is unreliable. 

Another approach which we suggest is relevant here anticipates that extinction should be lower in less luminous 
and face-on systems with low metallicities \citep[e.g.][]{Giovanelli1995, Xiao2011}.
\citet{Salim2007} showed that the derived $\tau_{V, UV}$ is sensitive to the assumed prior distribution 
of $\tau_V$ in the model library. Therefore, to constrain better $\tau_{V, UV}$, we first try to estimate 
$\tau_V$, following \citet{Giovanelli1995}, and trim out the models with unrealistically high extinctions. 
With this improvement in the SED fitting, the mean values for the ALFALFA dwarf {\it s-sed} sample become 
$A_{FUV}$ = 1.33 mag and $A_{NUV}$ = 0.96 mag, with formal uncertainties of 
0.42 and 0.32 mag, respectively. In particular, the $A_{FUV}$ value is reduced to 2.36 mag for UGC~7889. 
Since the $A_{FUV}$ estimates are reduced by 0.14 mag on average, the $\log SFR$ is reduced correspondingly 
by 0.06~dex (see values in \S\ref{SFR}). As noted previously (\S\ref{smass}),
such improvement has little effect on estimates of stellar mass.

To convert the extinction to an estimate of the dust mass, we adopted the eqn (44) from \citet{Popescu2011}, 
$M_{dust} = 0.9912 [{\rm M_\odot pc^{-2}}] h_S^2\tau_B^f$, where $h_s$ is the stellar disk scale length 
in parsec and $\tau_B^f$ is the face-on optical depth in B-band. SDSS g-band values are used 
to approximate values in the B-band. Following \citet{Giovanelli1994} and \citet{Shao2007}, 
the observed disk scale lengths and optical depths are converted to face-on values. 
The resulting dust to HI gas mass ratio is 0.003 on average for the {\it s-sed} galaxies 
(0.002 for the {\it s-sup} sample and 0.004 for the lower HI mass {\it s-com} population);
this ratio is below the mean value of 0.007 quoted in \citet{Draine2007}. 
In comparison, a dust-to-gas ratio of 0.002 was determined in the nearby starbursting 
dwarf galaxy NGC~4214 \citep{Lisenfeld2011}. We also note that
\citet{Boissier2004} found that the dust-to-gas ratio is proportional to $(Z/Z_\odot)^{0.88}$. 

In order to assess fairly the accuracy of our results, we have examined
correlations between the FUV extinction $A_{FUV}$ and various relevant quantities, in particular
those for which relations have been established for samples of more massive galaxies.
We find no significant trend with axial ratio, mainly because our sample includes
no high axial ratio systems; nearly all show $\log a/b < 0.5$. 
As discussed in \S\ref{HIprop}, it is unlikely that we are missing 
edge-on dwarfs because of the line-width cutoff. Rather, the observed distribution 
implies that dwarf galaxies are intrinsically thicker or that  
the distribution of SF sites within them are quite irregular, making the
tracing of their disks more complicated. 
Similarly, \citet{Xiao2011} suggested that dust reddening is not so sensitive to 
axial ratio at low metallicity as at high metallicity. 
We have also looked for trends in $A_{FUV}$  with $M_*$ and $f_{gas}$. 
The trend seen by \citet{Salim2005} that attenuation increases with stellar mass,
is not evident among the ALFALFA dwarf galaxies; the latter of course include
no really massive galaxies.
However, in comparison with Figure 2 of \citet{Salim2005}, we find that the highest value of
$A_{FUV}$ reached by gas-rich galaxies (as identified by high $f_{gas}$) is lower than that 
estimated for relatively gas-poor galaxies. Such a result
is reasonable assuming that the galaxies which are gas-rich for their stellar masses are less evolved,
have lower dust content and metallicity, and thus exhibit less attenuation. 

Other common approaches to estimating extinction exploit the infrared wavelengths. In this case however, the
Spitzer archive contains too few of the ALFALFA dwarfs to permit a direct measurement of the 
infrared excess (IRX) incorporating MIPS wavelengths.
Instead we examine how $A_{FUV}$ varies with UV color ($\beta$), equivalent to the $IRX-\beta$ relation
following \citet{Salim2007}.
In comparison with the relation given for normal star-forming galaxies by those authors,
the ALFALFA dwarfs only marginally follow the trend, 
with the more massive {\it s-sup} members showing less scatter about the relation.
A number of the lowest HI mass {\it s-com} galaxies fall
below the relation for normal galaxy samples; their $A_{FUV}$ would thus be 
over-predicted by the standard $IRX-\beta$ relation 
given their UV color, possibly because of their different SFH and/or dust geometry.
We also find that $A_{FUV}$ is better correlated with the 
$m_{FUV}-m_{NUV}$ color than with the $m_{NUV}-r$ color. 

In addition, we use the the spectroscopic Balmer lines to compare the observed 
flux ratios in the H$\alpha$ to H$\beta$ lines with the theoretical value (2.86 for Case B recombination) 
to estimate $A_{H\alpha}$. We assume that $A_{FUV}=1.8A_{H\alpha}$, following the Calzetti law \citep{Calzetti2000}, 
with the caveats about large variations in UV extinction curves and aperture effects. 
We approximate the ratios using the line flux measurements from the MPA-JHU DR7 catalog (see \S\ref{SDSSdata}). 
This method is applied to 42 galaxies which are common to both the {\it s-sed} sample and the MPA-JHU catalog. 
Surprisingly, this crude application of the Balmer decrement method gives systematically smaller extinction estimates, 
with a mean $A_{FUV}$ = 0.64 mag when we adopt the Calzetti law. Given that the $A_{FUV}$ values are presumably 
dominated by attenuation by the diffuse ISM whereas the emission lines probe the local attenuation within 
the clouds in which they originate, one might expect the opposite result as is found in 
massive galaxies \citep{Calzetti2000}, and incorporated into the factor of 1.8 between the FUV and H$\alpha$
extinctions.
However, we note that \citet{Wild2011} also found a strong decrease in emission line-to-continuum 
attenuation ratio with increasing SSFR, in agreement with the overall high 
SSFRs of the ALFALFA dwarf sample relative to the massive galaxies (see \S\ref{SFS}) and likely indicative of
the variation of the dust geometry with the SSFR. 
Furthermore, there is large uncertainty in the UV extinction curve so that the constant in 
the $A_{FUV}=1.8A_{H\alpha}$ relation suffers from large uncertainty and is most likely underestimated. 
For the low stellar surface mass density galaxies ($\mu_*<3\times10^8~{\rm M_\odot kpc^{-2}}$) 
with $A_V=1$, \citet{Wild2011} found 0.3 mag 
more attenuation in the NUV compared to the Calzetti law which is derived for starburst galaxies. 
Similarly, \citet{Buat2011} found a steeper dust extinction curve in ordinary star-forming galaxies 
than found in starburst galaxies by \citet{Calzetti2000}. 
In addition, the bump strength of the 2175~{\rm\AA} dust feature is known to vary. 
Incorrectly assuming a dust curve with no 2175~{\rm\AA} feature (e.g. the Calzetti law for starburst galaxies) 
would lead to an underestimate in the amount of dust, in turn leading to the incorrect conclusion that the 
stellar populations are extremely metal poor \citep{Wild2011}. 
In particular, the steep FUV rise and small 2175~{\rm\AA} feature in dwarf galaxies, resembling the 
observed extinction curve of some regions in the LMC, may due to different dust grain 
properties in low mass galaxies \citep[e.g.][]{Galliano2003}. 

As a final approach, we have explored the extended $IRX-SFH-color$ relation based on 
the $m_{NUV}-r$ color and the $D_n(4000)$ measurement derived in \citet{Johnson2006}, 
which is commonly used by local GALEX studies \citep{Wyder2007, Schimin2007}; application
of this relation to the ALFALFA dwarfs 
predicts a much larger extinction $A_{FUV}$ in many of the lowest mass {\it s-com} galaxies. 

At the end of this analysis of the various methods of estimating the internal
extinction, we conclude that (a) SED-fitting using the UV continuum is the most 
reliable currently available tool to obtain $A_{FUV}$ values for the ALFALFA dwarfs, 
but that (b) there is large uncertainty in the extinction estimates. 
It is inappropriate to ignore completely a correction for 
extinction at UV wavelengths (the mean of $A_{FUV}$ is
1.33 mag), although attenuation in the dwarfs 
is smaller than that found in more massive SDSS galaxies, e.g. 2.03 mag in 
\citet{Salim2005}. We note that taking attenuation into account
can bring the dwarf galaxies in the FIGGS study of \citet{Roy2009} 
closer to the Kennicutt-Schmidt star formation law obtained for more massive galaxies. 
However, it will not lead to a change in the slope of the relation. 
As was found by \citet{Roy2009}, the relationship between the HI density and the SFR 
in gas-rich dwarfs is steeper than the Kennicutt-Schmidt star formation law.

\subsection{Metallicity}
\label{Z}

Standard recipes used to convert FUV luminosities to SFRs \citep[e.g.][]{Kenn1998a} 
are derived on the assumption of solar metallicity stellar populations, $Z_\odot$. 
However, in the general population of star-forming galaxies, the gas-phase metallicity $Z$ of the ISM 
is known to increase with stellar mass, e.g. \citep{Tremonti2004}. 
This dependence is consistent with simple closed-box models \citep[e.g.][]{Garnett2002}, whereby high gas fraction 
galaxies are relatively unevolved and thus have experienced low metal enrichment. Especially in low mass systems,
ISM blow-out can follow episodic star formation events. 
The adoption of an incorrect $Z$ to estimate the SFR through the observed FUV luminosity 
can significantly bias the result, as a relative deficiency of metals would result in higher temperatures in the 
photospheres of forming stars and thus in a greater number of UV photons produced per unit stellar mass. 

It is thus legitimate to question the validity of SFRs derived for low $M_*$ systems such as the ALFALFA dwarfs, 
if an account for their presumed low $Z$ is not considered. Short of specific measurements through optical 
spectroscopy, the stellar population metallicity may be estimated via SED fitting. 
Unlike other well constrained parameters such as $M_*$ however, $Z$ is poorly constrained by SED fitting; in fact
deriving $Z$ from SED fitting often yields a likelihood distribution with multiple peaks. While
we would normally expect that the mean estimates of metallicity $<Z>$ derived for larger samples are more reliable, 
we note that the \citet{Salim2005} sample yields $Z\simeq Z_\odot$, while for the ALFALFA dwarf {\it s-sed} sample 
we obtain $0.48 Z_\odot$ ($0.44 Z_\odot$ for the low HI mass {\it s-com} galaxies and $0.54 Z_\odot$ for the 
additional {\it s-sup} ones). Therefore, for our estimates of SFR, we adopt Kennicutt's relation, with a correction of 
1.1 as given by \citet{Hunter2010}, appropriate for a Salpeter IMF and $Z\sim0.4 Z_\odot$.

The stellar population metallicity is correlated with gas-phase metallicity, though with large scatter 
\citep{Gallazzi2005}. Given the shallow potential well of dwarf galaxies, the gas-phase metal enrichment 
can be easily lost by outflow. Oxygen abundance measures from the 
MPA-JHU DR7 release of SDSS spectral \citep{Tremonti2004} is available for 58 ALFALFA dwarf galaxies, 
with a mean of $12+\log(O/H) = 8.26$ (8.27 for 44 low HI mass {\it s-com} members and 8.22 for 14 {\it s-sup} galaxies). 
The oxygen abundances can be converted to gas-phase metallicities in units of solar metallicity 
($Z_\odot = 0.02$), adopting a value of $12+\log(O/H)_\odot = 8.66$ \citep{Asplund2004}. 
The mean of the gas-phase metallicity for the ALFALFA dwarf galaxies corresponds to $0.4 Z_\odot$. 
However, according to the mass-metallicity relation derived in \citet{Tremonti2004}, 
based mainly on more massive galaxies $8.5 < \log M_* < 11.5$, the stellar mass of $\log M_*=7.4$ corresponds to 
$12+\log(O/H) = 7.78$. Almost all of the ALFALFA dwarfs lie above this relation. 

\subsection{Star formation rates}
\label{SFR}

As a next step, we derive the global SFR for the ALFALFA dwarf galaxies 
both from the FUV luminosity alone,  $SFR_{FUV}$, and by SED fitting, $SFR_{SED}$.
The latter is averaged over the last 100 Myr, comparable to the lifetimes of UV bright stars.  
Compared to SFRs derived from the H$\alpha$ luminosity, both $SFR_{FUV}$ and $SFR_{SED}$ are 
sensitive to very low levels of star formation and hence are particularly relevant
for application to the ALFALFA dwarf sample.

$SFR_{SED}$ values are given, in logarithmic units of ${\rm M_\odot\ yr^{-1}}$,
in Column (12) of Table \ref{table:dwarf} for the ALFALFA {\it s-sed} galaxies. 
The mean is 
$-2.18$ ($-2.48$ for the lower HI mass {\it s-com} sample and $-1.71$ for the {\it s-sup} galaxies), 
within a minimum of $\sim -4$. In comparison
with values found for typical SDSS samples \citep{Salim2005}, the ALFALFA dwarfs overall
show very low values of SFR. 
In such a regime where only a handful of O-stars are formed 
over timescales comparable to their lifetimes (i.e., a few million years), H$\alpha$ 
emission will appear weak or even absent. The SFRs inferred from H$\alpha$ deviate from 
those of FUV at $\log SFR \simeq -2$, but such stochastic effects may begin to 
play a role below $\log SFR \simeq -3$, according to \citet{Lee2009}. 
Low mass stars may form continuously
but at a very low rate so that no change in the IMF is required. However,
\citet{Lee2009} have shown that other factors, including internal dust 
attenuation, stellar model uncertainties, metallicity, ionizing photon loss and starbursts, 
if considered alone, are insufficient to explain the amplitude of the underestimate of the SFR
derived from H$\alpha$ relative to that implied by the UV luminosity. 
They also suggested that an IMF deficient in high mass stars is consistent with their results. 
While the current dataset is insufficient to provide conclusive insight, the ALFALFA dwarfs with 
their very low SFRs, will provide an 
ideal opportunity to test the hypothesis of a non-universal IMF. 

To obtain $SFR_{FUV}$, a commonly used standard conversion factor from the dust-corrected 
FUV luminosity into the SFR is given by \citet{Kenn1998a}, assuming 
solar metallicity, a Salpeter IMF,  and a constant SFH over
at least the past $\sim 100$ Myr in the stellar mass interval of $0.1 M_\odot - 100 M_\odot$: 
\begin{equation}
\label{K98Salpeter}
	SFR [{\rm M_\odot\ yr^{-1}}] = 1.4\times10^{-28}L_{\nu} [\rm{ergs\ s^{-1}\ Hz^{-1}}],
\end{equation}
where the FUV flux is derived from the AB magnitude \citep{Lee2009}: 
	$f[{\rm mJy}]=10^{(23.9-m_{FUV})/2.5}$. 
Taking into account the sub-solar metallicity 
(see $\S$\ref{Z}), and adopting a Chabrier IMF which predicts a smaller SFR for 
the same luminosity than a Salpeter IMF by a factor of 1.58 \citep{Salim2007}, 
equation (\ref{K98Salpeter}) becomes 
\begin{equation}
\label{K98Chabrier}
	SFR [{\rm M_\odot\ yr^{-1}}] = 0.81\times10^{-28}L_{\nu} [\rm{ergs\ s^{-1}\ Hz^{-1}}]. 
\end{equation}

\begin{figure*}
\center{
\includegraphics[scale=0.8]{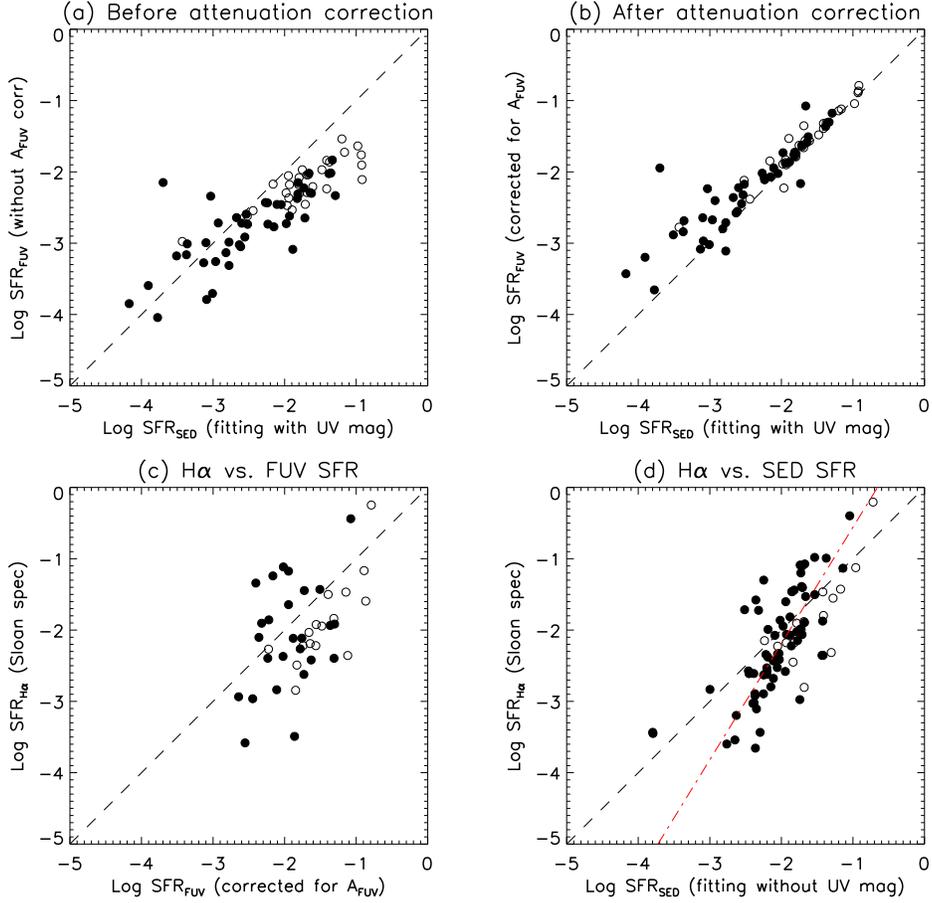}
}
\caption[]{Comparison of SFR measures for the ALFALFA dwarf sample with open circles identifying
        the lower HI mass {\it s-com} galaxies and filled ones, the additional {\it s-sup} galaxies in each panel. 
{\it Panel~(a):} $SFR_{FUV}$ obtained by the standard conversion from FUV luminosity, 
	(Eqn \ref{K98Chabrier})
	without attenuation correction, 
	versus $SFR_{SED}$ derived from SED-fitting. 
{\it Panel~(b):} Same as panel~(a) except that $A_{FUV}$, derived by SED fitting, have been
        applied to $SFR_{FUV}$.  
	The FUV magnitudes tend to over-predict SFRs below $\sim 10^{-2}$ M$_\odot$ yr$^{-1}$, 
	which may due to the bursty nature of star formation in dwarfs. 
{\it Panel~(c):} $SFR_{FUV}$ compared with $SFR_{H\alpha}$, derived from the 
        SDSS spectroscopic data (MPA-JHU DR7 measurements). 
        Though our aperture correction tends to overestimate $SFR_{H\alpha}$, 
        the deficiency of $SFR_{H\alpha}$ is still visible. 
        We are limited by the small number of the {\it s-sed} galaxies with usable SDSS data. 
{\it Panel~(d):} SFRs derived from SED fitting only to the SDSS bands compared with $SFR_{H\alpha}$. 
	A weak trend exists that H$\alpha$ under-predicts the SFR below $\sim 10^{-2}$ M$_\odot$ yr$^{-1}$. 
	The dash-dotted line is the linear fit to the data points, 
	with a slope steeper than the one-to-one dashed line. 
}
\label{fig:SFRcom}
\end{figure*}

The upper panels in Figure \ref{fig:SFRcom} shows a comparison of the two measures, $SFR_{FUV}$ 
(obtained via equation \ref{K98Chabrier}) and $SFR_{SED}$, 
before (panel a) and after (panel b) extinction corrections are applied, for all the ALFALFA {\it s-sed} galaxies. 
Filled circles denote the low HI mass {\it s-com} members, 
whereas open ones identify the additional {\it s-sup} galaxies. 
As evident in panel (a), in the absence of an extinction correction, the
FUV luminosity underestimates the SFR, because, although small, 
the extinction at UV wavelengths is not negligible. 
Moreover, the deviation from the one-to-one relation (dashed line)
is greater at the higher SFR end, suggesting that attenuation is larger for higher SFRs. 
Applying the extinction corrections $A_{FUV}$ derived from the SED fitting (as discussed in $\S$\ref{Afuv})
gives the result seen in panel (b): the scatter about and deviation from the
one-to-one relation is significantly reduced. The relatively tight correlation evident here 
confirms that the stellar population synthesis model we used is consistent 
with what was used in \citet{Kenn1998a}. 

However, as is still evident in panel (b), the 
FUV luminosity alone tends to overestimate the SFR compared to estimates derived from SED fitting, 
especially for the lowest SFRs, i.e. SFR below
$\sim10^{-2}$ M$_\odot$ yr$^{-1}$. Although we do not compare the $SFR_{FUV}$ with $SFR_{H\alpha}$ here,
we note that this threshold is close to the critical value identified by \citet{Lee2009}. 
Similarly, \citet{Salim2007} found that, since their sample is characterized by
a wide variation in SFH, 
the conversion factor from FUV luminosity to SFR is smaller than that which would apply for a constant SFH. 
Compared to their sample, the ALFALFA dwarfs are currently forming more stars relative to the 
accumulated stellar masses (see \S\ref{SFS}). It is possible that the $b$-parameters become 
systematically larger in dwarf systems, i.e. a higher fraction of their stars have been formed more recently. 
For the same amount of observed FUV luminosity, the SFRs averaged over 100 Myr 
would be smaller than those derived if the SFRs had remained constant. If so, the standard 
\citet{Kenn1998a} conversion factor from $L_{FUV}$  to $SFR_{FUV}$ 
would be too large in the case of dwarf galaxies. 
On the other hand, a bursty SF behavior would have less impact on the $SFR_{H\alpha}$, 
because the SFR is more likely to remain constant during the much shorter life times of O-stars. 
These results suggest not only that the $SFR_{H\alpha}$ values may be suspicious \citep{Lee2009}, 
but so also may be the estimates of $SFR_{FUV}$ derived using standard conversions
obtained for samples of more typical and more massive galaxies. 

At this point, we do not have available H$\alpha$ images, but we can make a crude attempt 
to quantify the SFRs with the H${\alpha}$ line flux derived from the SDSS spectra, $SFR_{H\alpha}$, 
for those galaxies with such measures.
We again use the standard conversion factor calibrated in 
\citet{Kenn1998a}, taking into account the metallicity and IMF difference. 
The MPA-JHU reprocessed H$\alpha$ line flux is corrected for attenuation, 
given by the Balmer decrement as in \S\ref{Afuv}. We correct for the aperture effect by using the ratio of 
the fluxes corresponding to the entire galaxy $r$-band magnitude and the magnitude through the fiber in the same band, 
following \citet{Hopkins2003}. The aperture correction implicitly assumes that the emission measured 
through the fiber is characteristic of the galaxy as a whole. However, in almost all of the ALFALFA dwarfs 
the $u-r$ colors are bluer within the fiber than across the galaxy, a fact which can be explained if
the bluest star-forming peaks in surface brightness are usually the targets of the spectroscopic observations. 
As a result, our crude scaling by the $r$-band light should overestimate $SFR_{H\alpha}$. 
However, the bottom panels in Figure \ref{fig:SFRcom} still show a systematic deficiency in $SFR_{H\alpha}$, 
relative to both $SFR_{FUV}$ (panel c) and $SFR_{SED}$ (panel d), albeit with large scatter. As evident
in panel (c), the number of galaxies in the {\it s-sed} sample and with SDSS spectra is quite limited.
In panel (d), we attempt to improve the statistics by dropping the requirement of UV data and 
thus calculating a $SFR_{SED}$ by fitting only to the SDSS bands. Albeit crude,
the best linear fit line to the data points (the red dashed-dotted line) has a slope larger than 1, 
suggesting that while the $SFR_{FUV}$ could be an overestimate due to stochastic effects, 
we cannot rule out the possible systematic deficiency of $SFR_{H\alpha}$, 
compared to $SFR_{SED}$, and thus a non-universal IMF. 
We are in the process of obtaining
H$\alpha$ images for this sample and plan to present a more comprehensive 
comparison of SFR measures in a future work. 
	
\section{Stars, star formation and gas in the ALFALFA dwarfs}
\label{SFgas}

Numerous previous works have explored the relationships between stars, star formation
and gas in galaxies, and in this section, we 
explore how the ALFALFA dwarf galaxies, specifically the {\it s-sed} sample
galaxies, compare to others. In comparison
with results obtained for large samples derived from the SDSS \citep[e.g.,][]{Brinchmann2004},
the imposition of criteria associated with HI-selection and narrow HI line width renders
the ALFALFA dwarf sample a still diverse population of low luminosity blue cloud dwellers.
In the context of examining the {\it s-sed} sample, it is therefore useful to compare it to
a sample of higher mass objects selected by similar HI criteria. To do that, we make use
of a subset of 12154 galaxies contained in the $\alpha.40$ catalog presented in 
\citet{Haynes2011} and which are also contained in the SDSS DR7. 
\citet{Haynes2011} present a cross reference between the $\alpha.40$ and SDSS DR7 databases
which is used as the basis for the parent HI-selected population.
To derive stellar masses and SFRs for the larger sample, we make use here of the 
methodologies described in \S\ref{prop}, but restricted only to the optical SDSS DR7 photometry. 
A more complete discussion of the $\alpha.40$-SDSS-GALEX sample 
will be presented elsewhere \citep{Huang2012}. 

\subsection{The ALFALFA dwarfs on the star-forming sequence}
\label{SFS}

Besides the well-known correlation that the SFR increases with stellar mass \citep[e.g.,][]{Brinchmann2004, 
Bothwell2009}, the `star-forming sequence' in a specific SFR ($SSFR=SFR/M_*$) versus stellar mass 
diagram is identified by the fact that the 
SSFRs of lower mass galaxies appear to be confined to a relatively 
narrow range of SSFR that declines as the stellar mass increases \citep{Salim2007, Schimin2007}. 
At masses above $\sim 10^{10} {\rm M_\odot}$, some galaxies exhibit much lower values of the SSFR; 
i.e., the sequence turns over. The tightness of the `star-forming' sequence strongly suggests
that a galaxy's mass regulates its overall SFH. 
However, a second transition, near $M_B\sim -15$, has been reported \citep{Lee2009b, Bothwell2009}, 
below which the dispersion in SSFRs broadens again: low luminosity, low mass galaxies also show
a wide range of SSFR. Based on a sample of $\sim$1000 galaxies with both H$\alpha$ and HI measures
available in the literature,
\citet{Bothwell2009} have argued that possible causes of this behavior can be categorized under
two main headings; first, the various physical mechanisms that underlie the star formation 
properties of dwarf galaxies not only exhibit a large spread, but also are decoupled from their 
gas contents. A second possibility proposed by those authors is that at low 
luminosities, the SFR is not faithfully traced by the H$\alpha$ luminosity, leading to a 
mischaracterization of the SFR. Since the SSFRs used here are based on SED fitting, the second explanation 
proposed in \citet{Bothwell2009}, i.e., that H$\alpha$ does not adequately trace the 
SFR in dwarfs, is not relevant to this work.
\citet{Lee2009b} have also found that, whereas the dispersion in SSFR(H$\alpha$) increases towards 
the lowest luminosities by $\sim60\%$, the increase is only $\sim25\%$ when SSFR(FUV) is adopted.
Those authors interpreted this difference as evidence that UV measures are less affected 
by purely stochastic variations in the formation of high mass stars. 

\begin{figure*}
\center{
\includegraphics[scale=0.8]{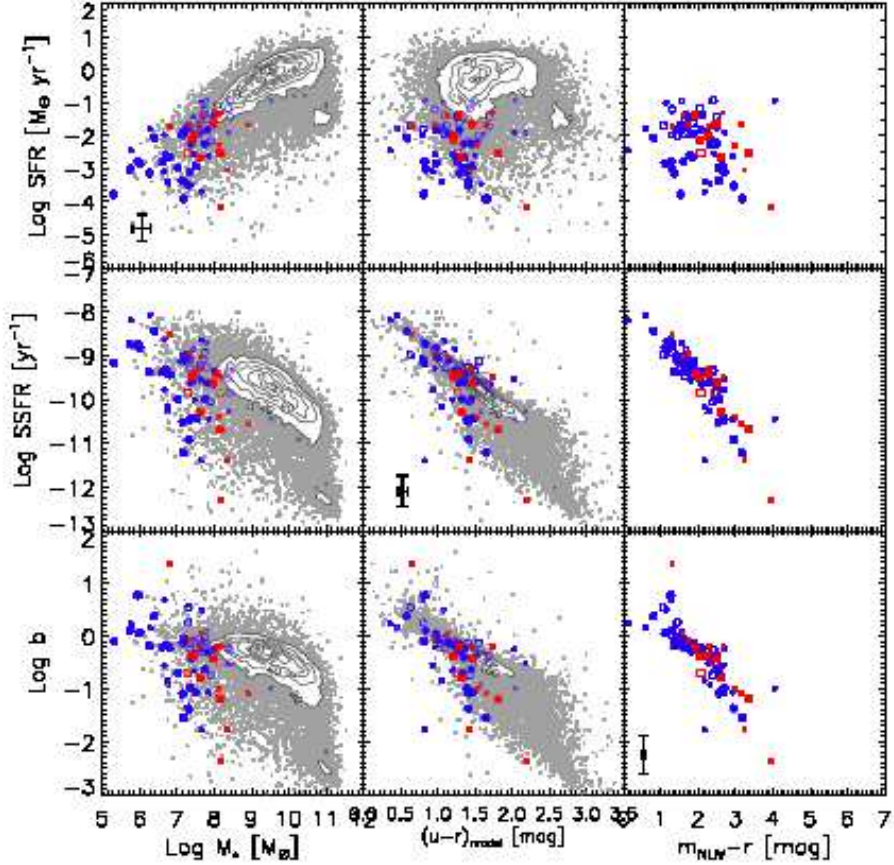}
}
\caption[]{
	The ALFALFA dwarfs on the star forming sequence. 
	Black contours and grey small points depict the distribution for the parent $\alpha.40$-SDSS sample, 
	in high and low number density regions respectively. 
	All the other colored symbols represent the {\it s-sed} galaxies (74 of them) with reliable UV/optical magnitudes 
	so that the SED fitting can be applied. Filled symbols belong to the {\it s-com} (45) 
	while open ones to the {\it s-sup} sample (29); 
	red squares denotes Virgo Cluster members (17) while blue circles are outside of Virgo (57); 
	bigger symbols represent galaxies with SED fitting  $\chi^2_{best}<10$ (46) while smaller ones have $\chi^2_{best}>10$ (28). 
	The typical error bars for the {\it s-sed} galaxies are plotted in the bottom left corners. 
	The horizontal layout shows the run of stellar mass, $u-r$, and the $m_{NUV}-r$ color. 
	In the vertical direction we examine the SFR, SSFR and the birthrate parameter $b$. 
	Note that the $m_{NUV}-r$ range is scaled to be two times that of the $u-r$ color 
	(refer to Figure \ref{fig:CMD}).
}
\label{fig:SFS}
\end{figure*}

Specifically presented for comparison with similar diagrams presented
by other authors \citep[e.g.,][]{Bothwell2009,Lee2011,Salim2007},
Figure \ref{fig:SFS} presents various ways of examining the star formation properties
of the {\it s-sed} galaxies in the context of the parent $\alpha.40$-SDSS sample.
The horizontal layout shows the run of stellar mass,  $u-r$ color, and the $m_{NUV}-r$ color,
from left to right respectively. In the vertical direction, from top to bottom, panels show
the SFR, SSFR and the birthrate parameter $b$. In all panels, the 74 colored symbols represent 
the galaxies in the {\it s-sed} ALFALFA dwarf sample; for those objects, 
physical properties are determined 
by SED-fitting (see \S\ref{sed}) to both the UV and optical magnitudes.
In the left two columns,
black contours and grey small points depict the distribution for the 
parent $\alpha.40$-SDSS sample in high and low number density regions respectively,
derived from SED fitting to the SDSS bands only. Because GALEX photometry is not examined here
for the $\alpha.40$-SDSS sample, the right panels show only the ALFALFA dwarfs.
Filled symbols show the less HI massive {\it s-com} galaxies (45) while open symbols denote
the additional {\it s-sup} ones (29). Red squares identify galaxies which are members of the
the Virgo cluster (17) while blue circles show all the others. 
Bigger symbols represent galaxies with SED fitting  $\chi^2_{best}<10$ (46) 
while smaller ones have $\chi^2_{best}>10$ (28).
Typical error bar estimates for the {\it s-sed} sample are plotted in the bottom left corners of 
Figures \ref{fig:SFS}, \ref{fig:SFE} and \ref{fig:fgas}, with the same symbol definition in these plots. 
Note that the $m_{NUV}-r$ range is scaled to be two times that of the
$u-r$ color (refer to Figure \ref{fig:CMD}). 

In general, the parent $\alpha.40$ sample exhibits the known trends that the SFR increases with 
stellar mass and that a bimodal distribution in SFRs is evident at the highest stellar
masses. However, as is clear in  Figure \ref{fig:SFS}, at low stellar masses, 
the SFR distributions fall below the extrapolation of the ridge defined by the 
massive systems. In particular, the ALFALFA {\it s-sed} dwarfs (the circles and squares) correspond
to the systems of lowest stellar mass and define a low, broad tail of the SFR distribution.
In the SFR versus $u-r$ color plot, while most of the ALFALFA dwarfs lie on the
blue branch, the colors remain almost unchanged despite their varying, but relatively small, SFRs. 

As shown also by \citet{Haynes2011}, the ALFALFA catalog is clearly biased against 
red sequence galaxies. Still, the bimodal nature of the distributions of SSFR is evident
in the left panel of the second row of Figure \ref{fig:SFS}. 
At the high mass end, a relatively small population of red sequence $\alpha.40$ galaxies 
occupies the low SSFR regime. The contours of the blue star-forming sequence
reflect the general trend that the SSFR generally decreases with increasing stellar mass.
At the same time, compared to that in the intermediate mass range ($10^8 - 10^{10} {\rm M_\odot}$),
the dispersion of the SSFR distribution in a given stellar mass bin increases 
below $M_*\sim 10^8 {\rm M_\odot}$. 
The standard deviation of the $\log SSFR$ distribution for galaxies in a given $\log M_*$ bin is 
0.55~dex in $9.25<\log M_*<9.75$, 0.71~dex in $7.75<\log M_*<8.25$ and increases to 0.83~dex in $6.75<\log M_*<7.25$. 
Most of the ALFALFA dwarfs lie in this 
low stellar mass regime, and not surprisingly, they show a large dispersion in SSFR. 
Moreover, because of the HI mass cutoff used to define the ALFALFA dwarf sample, 
it actually includes galaxies with low HI mass for their stellar mass, e.g.,
they are relatively gas-poor in comparison with the parent $\alpha.40$-SDSS sample 
(see \S\ref{fgas}). This selection tends to exaggerate the dispersion of the SSFRs
to low values, and a concentration of small grey points from the parent sample 
with higher SSFRs is visible above the {\it s-sed} galaxies within the same $M_*$ range. 
For the same reason, the additional {\it s-sup} galaxies with their higher gas fractions have higher SSFRs on 
average than the lower HI mass {\it s-com} ones. The few extreme outliers from the star-forming sequence 
with low SSFRs are mostly galaxies drawn from the lowest HI mass {\it s-com} sample (filled circles and squares). 
Several of them are dEs/dSphs in Virgo (e.g. AGC~220819 = VCC~1617 and AGC~223913 = VCC~1649; 
see images of the latter shown in Figure \ref{fig:img}),
among the very few early type Virgo dwarfs detected by ALFALFA \citep{Hallenbeck2012}. 

For galaxies with low SSFRs, a blind HI survey is more efficient 
at the detection of galaxies at both the low and high stellar mass ends \citep{Huang2012}. 
Moreover, a blind H$\alpha$ survey would be biased against galaxies with low SSFRs. 
The sample analyzed by \citet{Bothwell2009} is heterogenous, making use of existing compilations
of HI and H$\alpha$ data derived for other purposes. For example, the local low SSFR galaxies 
with low $M_*$ were particularly targeted for the 11HUGS program which contributes significantly
to the H$\alpha$ data used by \citet{Bothwell2009}. Similarly, the HI data analyzed in
\citet{Bothwell2009} includes a large population of massive cluster galaxies 
which have low SSFRs and were included in the targeted but deep surveys of
nearby rich clusters \citep{Springob2005}.
ALFALFA probes to a slightly lower 
stellar mass range than the sample examined by \citet{Bothwell2009} and reveals 
outliers from star-forming sequence at both mass ends similar to those noticed 
by those authors. Thus, the break-down of the star-forming sequence at 
the low mass end is real and indicates that the stellar mass is no longer the dominant regulator 
of SF in dwarfs. It is possible that dwarfs are more vulnerable to environmental effects 
so that their gas supply can be easily disturbed, and thus star formation quenched. 
In the intermediate $M_*$ range, many outliers with extremely low SSFRs 
exist within the $\alpha.40$ parent sample, but their number density relative to the contoured region
is so low that the dispersion of SSFRs remains low. However, while ALFALFA reveals
that outliers with anomalously low SSFR for their stellar mass exist at all masses,
intermediate mass outliers are missing in similar plots (e.g. Figure 8) from \citet{Bothwell2009}. 
Given the heterogenous sample which the \citet{Bothwell2009} analysis is based on, 
the absence of low SSFR galaxies in the intermediate mass range could be due to some combined bias in the 
sample selection criteria by the dual requirement of H$\alpha$ 
and HI (from the Cornell digital HI archive, selected typically by luminosity, optical size and/or cluster
membership) measures. 

The contours in the SSFR versus $u-r$ diagram (middle column)
show that the SSFRs of $\alpha.40$ galaxies are tightly correlated with the
$u-r$ color: for HI-selected galaxies nearly all of which are blue, the SSFRs, 
and thus the recent star-formation histories, are well constrained by 
the $u-r$ color itself. 
Given the fact that the NUV band better traces star formation than the $u$-band, 
we would have expected that the $m_{NUV}-r$ color would be better correlated with SSFR 
than is the $u-r$ color. However, such an improvement is not obvious in the right panel:
the larger attenuation correction introduces a greater dispersion in the observed NUV, 
so that a better correlation between SSFRs and $m_{NUV}-r$ colors is in fact
not clearly evident among the {\it s-sed} dwarf galaxies. 

Finally, we consider another quantity related to the SSFR, namely, the birthrate or $b$-parameter, 
$b = SFR/<SFR>=(SFR/M_*) T R$, where $T$ is the age of the galaxy and $R$ is the fraction of 
the mass formed over the galaxy's lifetime that does not eventually get returned to the ISM or 
IGM \citep{Brinchmann2004}. A typical value of $R$ is $\sim 0.5$ \citep{Brinchmann2004}, and on the
assumption that all galaxies started forming stars shortly after the Big Bang, $T\sim13$ Gyr. 
For these choices of $T$ and $R$, a galaxy with $b=1$, i.e. a constant SFR, would have a SSFR of 
$10^{-9.8}$ yr $^{-1}$. Thus, galaxies with SSFRs larger than this value appear to have current SFRs 
above their lifetime average. With the caveat that SFR measures based on H$\alpha$ emission 
in the low SFR regime may be systematically underestimated, \citet{Lee2009b} found that 
only $\sim$6\% of low mass 11HUGS galaxies are currently experiencing 
massive global bursts (H$\alpha$ EW above 100~$\rm \AA$ or $b\gtrsim2.5$). 
Here, we find that 13 galaxies (18\%) of the ALFALFA {\it s-sed} sample
have $b>1$ and 5 galaxies (7\%) even have $b>2.5$. 
Because of the relationship between the SSFR and the $b$-parameter, the distributions of $\log b$ vs. 
stellar mass, $u-r$ and $m_{NUV}-r$ colors, shown in the bottom row 
of Figure \ref{fig:SFS}, resemble the corresponding 
distributions in the second row, suggesting that dwarfs have both higher SSFRs and $b$-parameters, 
on average, compared to higher mass galaxies.  For this reason, it is even more essential to take 
bursty behavior into account when modeling the SEDs of HI-selected dwarfs. At the same time,
the equivalent widths of emission lines, 
e.g. H$\alpha$, which also characterize the SSFRs or $b$-parameters \citep{Lee2009b}, 
could be large enough to modify the continuum emitted by the underlying stellar population, 
leading to large $\chi^2$ in the SED fitting (\S\ref{sed}). Despite the caveats,
the wide spread in $b$ values evident among the {\it s-sed} sample suggests
that the SFHs of the ALFALFA dwarfs are highly variable and no longer a strong function of $M_*$. 
This result suggests further that
other factors including environmental effects may play an important role in the regulation 
of star formation in HI-selected dwarfs. 

\subsection{Star formation efficiency and the gas depletion timescale}
\label{SFE}

In addition to normalizing the SFRs by stellar mass to infer the level of 
the current SFR relative to its past average, we can also normalize the SFR by the HI mass 
to infer the ratio relative to potential future 
star formation, as inferred from the present amount of HI gas. 
Following \citet{Schimin2010}, we define the star formation efficiency 
as $SFE=SFR/M_{HI}$, and its reciprocal, as the gas consumption timescale $T_{cons}=M_{HI}/SFR$, 
also known as the Roberts time \citep{Roberts1963}. 
Observationally, low SFEs have been measured for low-mass galaxies and LSB galaxies 
\citep{Boissier2008}, and conversely, high efficiencies have been measured in starburst galaxies 
and interacting systems. 

Based on a sample of $\sim$1000 local galaxies $10^7 M_\odot \lesssim M_* \lesssim 10^{11} M_\odot$, 
with HI and H$\alpha$ fluxes compiled from the literature,
\citet{Bothwell2009} found that, with increasing luminosity, the HI gas fraction $M_{HI}/M_*$
drops off faster than the H$\alpha$-derived SFR, resulting in shorter $T_{cons}$ for more luminous galaxies. 
Those authors report $T_{cons}$ in excess of several Hubble times ($t_H$) for samples of dwarf irregular galaxies. 
In contrast, using the FUV luminosity instead of H$\alpha$ to measure the SFR, \citet{Lee2011} found that dwarf 
galaxies may not be as drastically inefficient in converting gas into stars as suggested by H$\alpha$ 
estimates, with SFEs exceeding $t_{H}^{-1}$. 
Although the \citet{Bothwell2009} sample also exploits HI measures for a large sample to derive
gas fractions, it is based on optical, rather than HI, selection, deriving most of its
HI properties from the extensive compilation of HI spectra contained in the Cornell digital HI archive
\citep{Springob2005}. In fact, it is important to note that more than $2/3$ of ALFALFA HI detections
are new and were not included in any of the extensive pointed observations which contributed to
the digital archive dataset \citep{Haynes2011}. In terms of the availability of HI measures, the
$\alpha.40$ parent sample is more homogeneous and its size is more than 10 times larger than the
\citet{Bothwell2009} sample. Moreover, we rely on SED fitting to derive physical quantities, rather 
than deriving stellar masses based on the
$M/L$ versus color relation (see \S\ref{smass}), or the H$\alpha$ luminosity for the SFR 
(see \S\ref{SFR}). 
In addition, the ALFALFA dwarf population used here extends to lower HI masses compared to those in \citet{Lee2011}. 
Therefore, the ALFALFA-based galaxy sample is more comprehensive than either of those two previous studies 
both in terms of probing the general trends exhibited by HI-selected galaxies as well as exploring the 
behavior of the lowest mass dwarfs. 

\begin{figure*}
\center{
\includegraphics[scale=0.85]{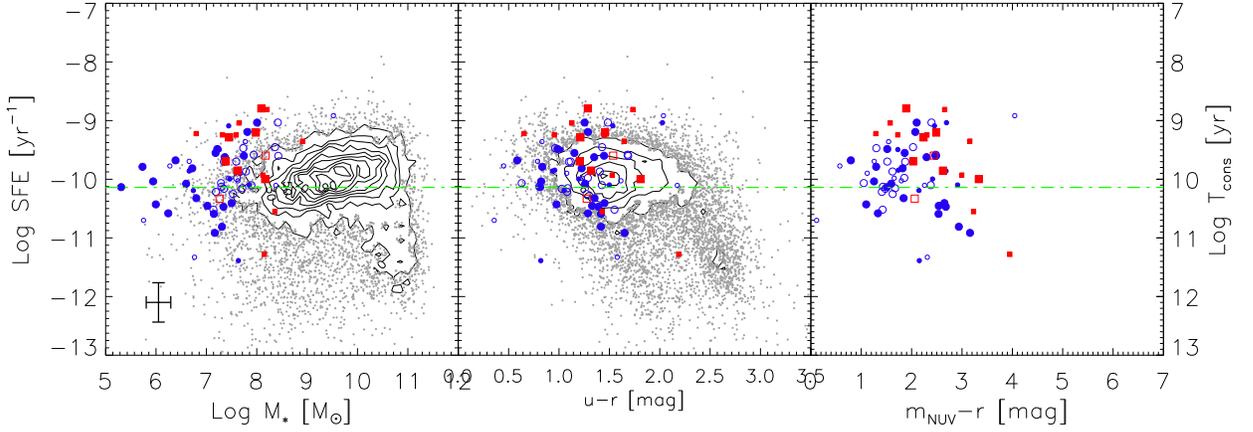}
}
\caption[]{Star formation efficiency ($SFE=SFR/M_{HI}$) and gas depletion timescale ($T_{cons}=SFE^{-1}$). 
Symbol definition follows Figure \ref{fig:SFS}. 
The green dash-dotted line corresponds to $SFE=t_{H}^{-1}$, where $t_{H}$ is the Hubble time. 
{\it Left:} $SFE$ against stellar mass. 
	The SFEs of ALFALFA detections are on average lower than optically selected galaxies. 
	The averaged SFEs in each stellar mass bin increase 
	slowly with stellar mass, with the $SFE \sim t_{H}^{-1}$ at the low mass end. 
	Selected to have lower $f_{gas}$ compared to the $\alpha.40$-SDSS galaxies in this stellar mass range, 
	the {\it s-sed} galaxies have 
	higher SFEs
	compared to the general trend if extrapolated into this 
	stellar mass range. A large portion of them even have $T_{cons} \lesssim t_{H}$. 
	In particular, the majority of Virgo members in the {\it s-sed} sample have $T_{cons} \lesssim t_{H}$.  
{\it Middle\ and\ right:} The $SFE$ is seen to vary very little with color in either the optical or optical/UV bands. 
}
\label{fig:SFE}
\end{figure*}

Figure \ref{fig:SFE} shows the distribution of 
$SFE$ versus $M_*$, $u-r$ and $m_{NUV}-r$ colors for the ALFALFA samples. 
Contours and grey small points in the left two panels depict the distribution of the full 
$\alpha.40$-SDSS dataset, 
and the symbol definition follows the same convention as in Figure \ref{fig:SFS}. The superposed  
dash-dotted line corresponds to $SFE = t_H^{-1}$.  Three main conclusions can be
drawn from the general distribution. First, as discussed more fully elsewhere
\citep{Huang2012}, ALFALFA galaxies have, on average, higher SFRs but lower SFEs 
than optically-selected ones, due to their higher than average HI masses.  
At the same time, the SFEs are barely dependent on either the $u-r$ or $m_{NUV}-r$ color as shown in the center and 
right panels of Figure \ref{fig:SFE}. 
Secondly, although the variation in the average SFR is mild across the 
stellar mass range of $10^8 - 10^{11} {\rm M_\odot}$ ($\delta \log <SFE> \sim 0.5$ dex), the averaged SFE in each stellar mass bin
increases with stellar mass. At the low mass end, the $SFE \sim t_H^{-1}$, albeit with considerable scatter.
Thirdly, a population of galaxies with very low SFE, arising mainly because of their 
low SFR (see Figure \ref{fig:SFS}), stands distinct from the general distribution at the 
high mass end ($M_* \gtrsim 2 \times 10^{10} {\rm M_\odot}$). 

Notably, despite the fact that they are selected to have the lowest HI masses 
in the $\alpha.40$ catalog, the {\it s-sed} dwarfs (circles and squares) typically have higher SFEs 
regardless of their low SFRs, compared to an extrapolation of the general trend into this stellar mass range. 
In contrast to the extremely low SFEs derived from H$\alpha$ emission for the dwarfs studied by \citet{Bothwell2009}, the ALFALFA dwarfs studied here are forming stars efficiently, and a large 
portion of them even have $T_{cons} \lesssim t_H$.
In particular, 14 out of 17 Virgo dwarfs (red squares) detected by ALFALFA have $T_{cons} < t_H$, 
suggesting that the cluster environment may actually play a role in enhancing their SFEs by either 
accelerating their evolution and thus driving them to the gas-poor side, or simply stripping part of their gas away. 
In contrast, AGC~223913 and AGC~220819, both dE/dSphs in Virgo, have low SFEs and $T_{cons} > t_H$. 

\subsection{HI mass and gas fraction}
\label{fgas}
As specified in \S \ref{cmd}, we define the gas fraction here simply
as the ratio of the HI mass to the stellar mass ($f_{gas}=M_{HI}/M_*$), without a 
correction for the presence of helium or for other phases of hydrogen. Little is
actually known about the molecular content of dwarf galaxies other than that it is low.
Numerous other studies, e.g. \citet{Walter2007}, have shown that dwarf galaxies, including 
ones detected in HI, are notoriously difficult to detect in CO and contain only small
masses of dust. Moreover, the relationship between CO and H$_2$ may break down in 
galaxies of low metallicity. Since the fraction of the ISM contributed by very cold 
molecular material is likely to be insignificant \citep[e.g.][]{Leroy2007}, and the uncertainty 
in it, large, we assume that the ALFALFA dwarfs are likely dominated by HI at least on global scales
and equate the HI mass with their total gas mass.

\begin{figure*}
\center{
\includegraphics[scale=0.8]{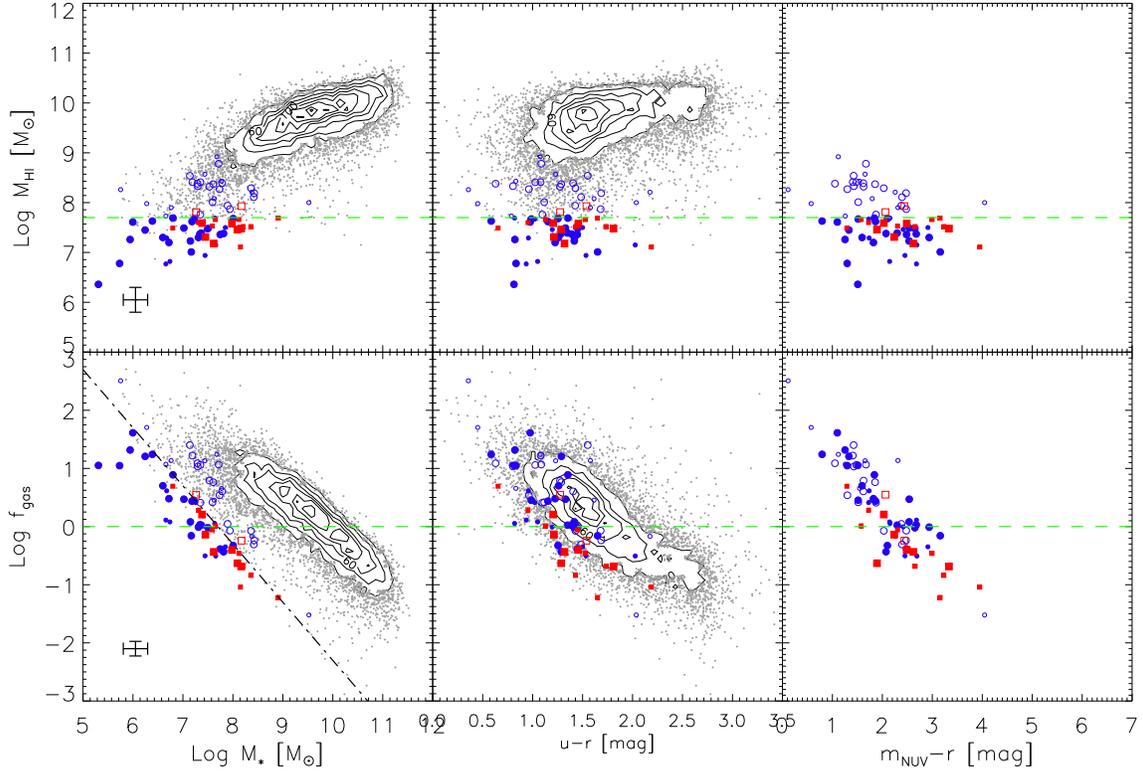}
}
\caption[]{HI mass and gas fraction, $f_{gas}=M_{HI}/M_{*}$. Symbol definition follows Figure \ref{fig:SFS}. 
{\it Upper\ row:} $M_{HI}$ as a function of stellar mass and colors.  The dashed line corresponds to 
	the cutoff in $M_{HI}$ of the {\it s-com} sample.
        The $M_{HI}$ increases with $M_*$, with the {\it s-sed} lying on the low mass tail. 
	At a given $M_{HI}$, Virgo members (red squares) have on average 
	lower $f_{gas}$. 
{\it Bottom\ row:} $f_{gas}$ as a function of the three. The dashed line corresponds to 
	$M_{HI}=M_*$, while the black dash-dotted line shows again $M_{HI}=10^{7.7}M_\odot$. 
	Because ALFALFA is biased against gas-poor galaxies, the correlation that $f_{gas}$ 
	decreases with increasing $M_*$ is clean. 
	Specifically, the HI mass selection of the ALFALFA dwarf sample results in low $f_{gas}$, relative to the 
	$\alpha.40$-SDSS in this {\it stellar mass} range. 
	However,
	46 of the 74 {\it s-sed} ALFALFA dwarfs 
	still have $f_{gas} > 1$. 
	For galaxies bluewards of $u-r=2$, $f_{gas}$ is lower in the redder ones.
}
\label{fig:fgas}
\end{figure*}

Figure \ref{fig:fgas} presents the distribution of HI mass (upper panels) and gas fraction $f_{gas}$ (lower panels) 
as a function of stellar mass, $u-r$ and $m_{NUV}-r$ colors, respectively. As before, in the left two
panels, the ALFALFA dwarf galaxies
in the {\it s-sed} sample are shown as circles and squares superposed on the parent $\alpha.40$-SDSS sample 
(black contours and grey points). The horizontal dashed line in the upper panels marks the HI mass cutoff
$M_{HI} < 10^{7.7} M_\odot$
imposed on the {\it s-com} sample. In the lower plots, horizontal dashed line 
traces $M_{HI} = M_*$ while in the lower left panel, the 
diagonal dash-dotted line traces the
the locus where $M_{HI} = 10^{7.7} M_\odot$. 
The distributions for the full ALFALFA population are smooth, 
with the mean $M_{HI}$ increasing with  $M_*$, and the mean $f_{gas}$ 
decreasing monotonically as $M_*$ grows. Unsurprisingly, $f_{gas}$ is tightly correlated with both the
$u-r$ and $m_{NUV}-r$ color: bluer galaxies have higher gas fractions. 
Note again that ALFALFA probes significantly lower masses, both stellar
and HI, than other surveys. Imposing a limit in $M_{HI}$ confines the ALFALFA {\it s-com} sample to 
fall below the horizontal dashed line in the upper panels and to the left of the 
the dashed-dotted line in the bottom left panel. 

As evident in the bottom row of panels, some of the lowest mass galaxies
have extremely high gas fractions ($f_{gas}>100$). 
Visual inspection of the most egregious cases shows that these values are corrupted by 
poor SDSS photometry or, possibly, by the assignment of the wrong OC. 
Although we have checked the photometry for the
dwarf sample individually, only a portion of the OCs in the $\alpha.40$-SDSS sample
have been inspected individually to set a SDSS photometric quality 
flag \citep{Haynes2011}. As a result, it is not surprising that
some of the $\alpha.40$ galaxies with $f_{gas}>100$ suffer from 
shredding, with a fainter child assigned as the SDSS cross-match and thus producing
an underestimate of $M_*$. In other cases, confusion within the ALFA beam may result
in a match to the wrong object albeit the ``most probable'' OC.  The complexities and caveats 
involved in assigning OCs to the ALFALFA dataset are discussed in \citet{Haynes2011}. 
However, some dwarfs really do appear to be highly HI dominated. For example,
AGC~174514 in the supplementary {\it s-sup} sample also has a $f_{gas}>100$; 
its images are shown in Figure \ref{fig:img}. 
It is the bluest galaxy among the {\it s-sed} sample, suggesting that it is truly a galaxy 
with lots of gas and not many stars. However, it may not be a single object and
the $f_{gas}$ could be a slight overestimate due to the uncertainty of the distribution of the HI; 
there are several additional small blue LSB clumps within the ALFALFA beam which may contribute comparable
optical luminosities, thereby raising the stellar mass by some factor, 
but certainly not $10^2$; this galaxy remains an unusual object with a very high gas fraction. 

Using a similar $f_{gas}$ vs. $M_B$ diagram,
\citet{Bothwell2009} found a population of outliers, namely early type galaxies 
with extremely low $f_{gas}< 0.05$, extending below the main distribution (Figure 3
of that paper) at the bright end ($M_B < -18$). They further 
identified this phenomenon with the large spread in the SSFRs of massive 
galaxies, suggesting that a dearth of fuel for star formation causes the extremely low SSFRs. 
In comparison with Figure 3 of \citet{Bothwell2009}, Figure \ref{fig:fgas} reveals no concentration of galaxies 
with low $f_{gas}$ below the main distribution at the high mass end, despite the existence of 
many low SSFR massive galaxies. The absence of very low $f_{gas}$ massive galaxies in ALFALFA is not
surprising given its well-characterized flux completeness and near-constant integration time. In contrast, 
some of the HI measures contributed by the \citet{Springob2005} dataset and used by \citet{Bothwell2009} 
made use of very deep targeted observations designed specifically to probe to very low gas fractions
in massive early type and cluster galaxies. The desire to probe the true range of gas fractions
representative of the most massive galaxies likewise motivated the targeted deep observations
of the GASS program \citep{Catinella2010}.

At the same time, they found that the population of low mass galaxies with a large spread in SSFRs is not 
matched by an equivalent shift to a larger range of $f_{gas}$. 
On the other hand, the broadening of the $f_{gas}$ distribution at the low mass end is more evident 
in the ALFALFA dataset, compared to that seen by \citet{Bothwell2009}.  
Among the 176 ALFALFA dwarfs in the low HI mass {\it s-com} set, 
only 2 are included in the \citet{Springob2005} catalog. Hence, it is not surprising that
\citet{Bothwell2009} see no broadening of the $f_{gas}$ distribution at the low mass end. 
HI selection yields a high overall gas fraction across the full range of sampled masses,
and at the same time, still a larger dispersion in the$SSFR$ distribution at both the high and low 
stellar mass ends (see \S\ref{SFS}).

In Figure \ref{fig:fgas}, the ALFALFA dwarfs in the {\it s-sed} sample (circles and squares) lie on the 
low mass tail of both the $M_{HI}$ and $f_{gas}$ distributions: at low stellar masses,
the ALFALFA dwarfs probe a relatively low gas fraction population. This departure from the
norm for the majority of the ALFALFA detections results from the imposition of the low HI mass cutoff. 
Among the low mass {\it s-com} galaxies (filled symbols), at a given $M_{HI}$, the Virgo members (red squares) 
have, on average, higher $M_*$ and thus lower gas fractions, than the non-Virgo ones (blue circles). 
Given the general trend between $f_{gas}$ and $M_*$, the most gas-rich galaxies have very low $M_*$.
Even if their $f_{gas} > 1$, the least massive may lie below the HI sensitivity of ALFALFA 
at the Virgo distance (log $M_{HI} \sim 7.3$). Alternatively, the low gas fractions may reflect a
possible connection between the cluster environment and relative gas poorness; in fact,
\citet{Hallenbeck2012} show that some morphologically early Virgo dwarfs contain
``normal'' gas fractions and argue that they must have recently accreted their HI. 
Here, as is seen in Figure \ref{fig:fgas}, more than half (46) 
of the {\it s-sed} galaxies have $f_{gas} > 1$, meaning that their baryonic mass is dominated 
by their atomic gas, rather than by their stars. 
 
\subsection{The star formation law and the global interplay of gas and stars}

It is well known that the local surface density of the star formation rate is 
strongly correlated with the local surface density of gas, $\Sigma_{SFR} \propto \Sigma_{gas}^{\alpha}$, 
at least when averaged over $\sim$ kpc scales.
A precise measurement of its exponent is the `Kennicutt-Schmidt law' \citep{Kenn1998b}, 
\begin{equation}
\label{KSL}
	\log (\Sigma_{SFR} [{\rm M_\odot\ yr ^{-1}\ kpc^{-2}}]) 
	= \alpha \log (\Sigma_{gas} [{\rm M_\odot\ pc^{-2}}]) + C, 
\end{equation}
where $C=\log \left[ \frac{(2.5\pm0.7)\times10^{-4}}{1.5} \right]$, 
 $\alpha = 1.4 \pm 0.15$ and $\Sigma_{gas} = \Sigma_{HI} + \Sigma_{H_2}$. 
The denominator of $C$, 1.5, results from the Chabrier IMF adopted in this work.
Such a super-linear Schmidt law ($\alpha > 1$) suggests a star formation efficiency 
which rises with gas surface density. 
Measurements of azimuthally averaged gas and SFR profiles show that the SFR correlates better with the molecular 
hydrogen component than with the total gas density at least within the optical disk \citep{Bigiel2008}. 
Yet, on global scales averaged over the whole galaxy, SFRs appear to correlate better with the total gas (HI + H$_2$), 
rather than the molecular gas, surface density \citep{Kenn1998b}. Global quantities such as those available to us
sample a range of gas surface densities, timescales and conditions within the ISM, and since we have only
global HI measures, we can only 
estimate $\Sigma_{gas}$ by proxy, using the global HI mass and the optical size of the 
stellar disk. 

\begin{figure*}
\center{
\includegraphics[scale=0.75]{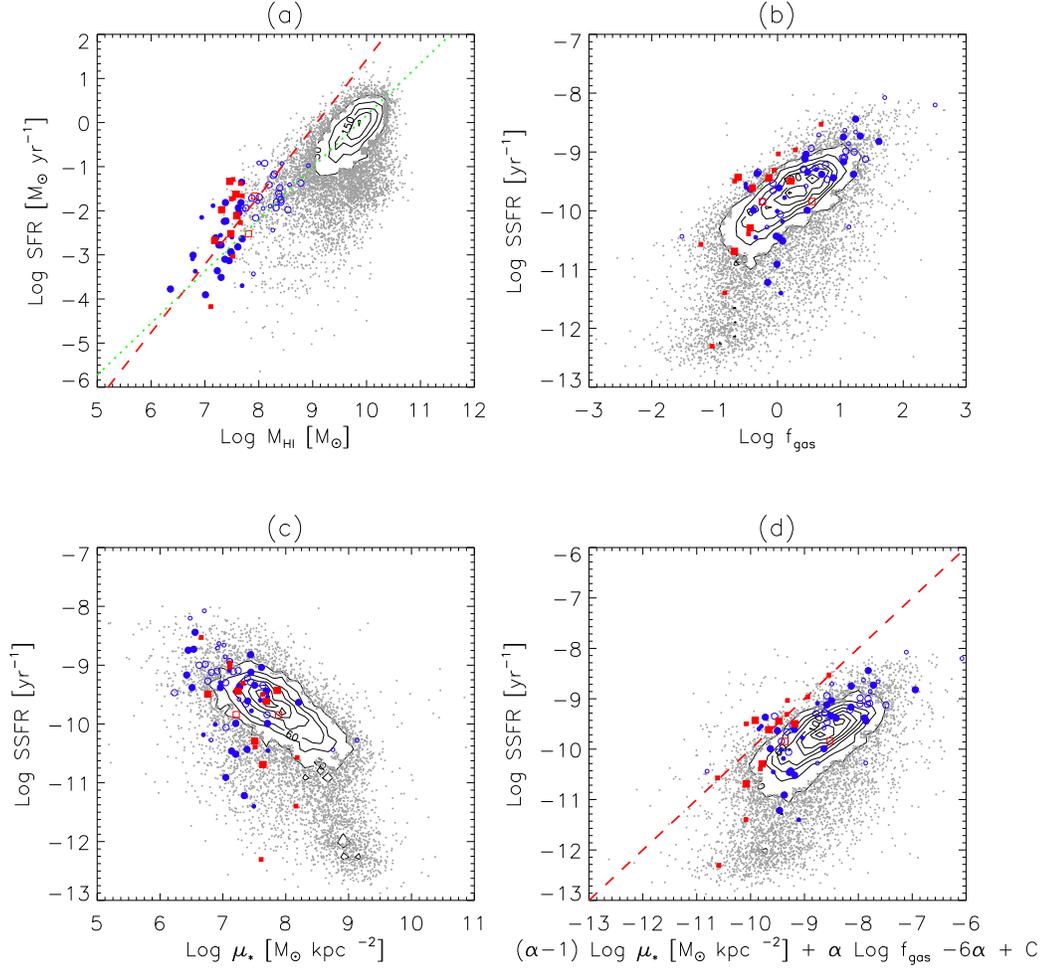}
}
\caption[]{
Gas content and star formation. Symbol definition follows Figure \ref{fig:SFS}. 
{\it Panel~(a)}: The SFR increases with the HI mass. 
	The dotted line has a slope of 
	1.2, corresponding to the linear fit to all the $\alpha.40$-SDSS galaxies. 
	The dashed line, of slightly steeper slope 
	1.5 is the linear fit to the {\it s-sed} only.
{\it Panel~(b)}: 
	Both axes in panel (a) are normalized by the stellar mass.  
        Besides the general trend that the SSFR increases with the gas fraction, 
	the SSFRs of a subset of galaxies drop dramatically once the gas fraction is below $\sim 0.3$. 
{\it Panel~(c)}: SSFRs decrease with increasing stellar surface mass density, $\mu_*$. 
	{\it s-sed} have relatively low $\mu_*$. 
{\it Panel~(d)}: By assuming that the aperture with a radius of $r_{50, z}$ 
	also contains half of the stellar mass, SFR and HI mass, the star formation law 
	\citet{Kenn1998b} is substituted into the expression of SSFR, resulting in the quantity on x-axis. 
	The adopted values of the scaling factor $C$ and exponent $\alpha$ reflect the \citet{Kenn1998b} result. 
	Although a clear correlation appears, the main distribution is shifted to the right of the dashed 
	one-to-one line. 
}
\label{fig:SFL}
\end{figure*}

Albeit crude in comparison to more detailed studies of spatial resolved atomic and molecular distributions, 
we examine via Figure \ref{fig:SFL} the correlations of global quantities related to the Schmidt law 
to demonstrate the constraints placed by the global HI measures on global SFRs. As before, both the
$\alpha.40$ and the dwarf {\it s-sed} samples are shown, with symbol definitions following Figure \ref{fig:SFS}.
Panel (a) of Figure \ref{fig:SFL}, shows the distribution of the SFR as a function of $M_{HI}$; this can
be thought of as the global atomic and volumetric star formation law. 
The green dotted line with a slope of
1.2 is the linear fit to all the $\alpha.40$-SDSS galaxies. 
The dashed line, of slightly steeper slope 
1.5, is the linear fit to the
galaxies in the dwarf {\it s-sed} sample only.
While it is natural to expect higher SFRs in galaxies with more gas available
to form stars, the ALFALFA {\it s-sed} dwarf galaxies define the low HI mass tail
of the distribution of star forming galaxies. 
Note also the presence of a second tail exhibited by the parent $\alpha.40$-SDSS population 
at very low SFRs; these galaxies may include a population of objects whose HI distributions
extend beyond their star-forming disks. 
In \S\ref{SFE}, we have already presented the result that the $SFE$ increases slowly with $M_*$. 
In fact, other recent works demonstrate that the $\Sigma_{SFR} - \Sigma_{gas}$ relation steepens at low 
gas surface densities due to the transition from atomic to molecular hydrogen \citep[e.g., ][]{Bigiel2008}. 
The dependence of SFR on $M_{HI}$ for the ALFALFA dwarfs shown in
panel (a) shows a suggestive mild steepening at low $M_{HI}$. 
Similarly, \citet{Roy2009} found that the SFR is generally lower in gas rich dwarfs 
than predicted by the `Kennicutt-Schmidt law'. 
Again, we note that the linkage of the scaling between SFR and $M_{HI}$ with the power law 
derived for $\Sigma_{SFR}$ and $\Sigma_{gas}$ is subject to various uncertainties, including the 
contribution from molecular gas, and the relative extent of the star forming and HI disks. 
Therefore, this comparison of global quantities yields only a suggestive empirical result. 

In panel (b), we normalize both axes by $M_*$, giving the relationship between
SSFR and $f_{gas}$. Normalization by $M_*$ moves the galaxies in the ALFALFA dwarf {\it s-sed} 
sample from the tail seen in panel (a)
back to within the main distribution. For the majority of ALFALFA galaxies, except those with extremely
low SSFR, a clear trend exists: the more gas-rich (higher gas fraction), the higher is the SFR 
relative to the accumulated stellar mass (higher SSFR). 

Given the fact that the majority of HI disks are un-resolved by ALFALFA, we cannot quantify $\Sigma_{gas}$ 
directly. However, we can characterize the surface densities by introducing an area-related quantity: 
the stellar surface mass density, $\mu_*$. Following \citet{Schimin2010}, we define the stellar surface mass density:
\begin{eqnarray*}
	\mu_* [{\rm M_\odot\ kpc^{-2}}] = \frac{0.5 M_* [{\rm M_\odot]}}{\pi(r_{50, z} [{\rm kpc}] )^2}, 
\end{eqnarray*}
where $r_{50, z}$ is the Petrosian half-light radius according to the SDSS pipeline measurements. 
Panel (c) of Figure \ref{fig:SFL} shows that SSFR decreases with increasing $\mu_*$ in general, 
and that the dwarf {\it s-sed} galaxies, on average, have lower $\mu_*$ than the full ALFALFA population. 
For comparison with other studies, we adopt an aperture of radius $r_{50, z}$ and assume that, 
by definition, it encloses also one half of the stellar mass. Although it is likely not a correct
assumption, we assume first that this aperture also 
contains exactly half of the global SFR, as well as half of the HI mass, and thus 
\begin{eqnarray*}
	\log (\Sigma_{SFR} [{\rm M_\odot\ yr ^{-1}\ kpc^{-2}}]) 
		= \log \frac{0.5 SFR [{\rm M_\odot yr^{-1}}]}{\pi(r_{50, z} [{\rm kpc}] )^2}  \\
	\log (\Sigma_{gas} [{\rm M_\odot\ pc^{-2}}]) 
	 	= \log (\Sigma_{gas} [{\rm M_\odot\ kpc^{-2}}]) - 6
		= \log \frac{0.5 M_{HI} [{\rm M_\odot}]}{\pi(r_{50, z} [{\rm kpc}] )^2} - 6
\end{eqnarray*}
The star formation law averaged within this aperture can be re-written from equation (\ref{KSL}) as: 
\begin{equation}
\label{KSL2}
	\log (\Sigma_{SFR} [{\rm M_\odot\ yr ^{-1}\ kpc^{-2}}]) 
		= \alpha (\log \mu_* [{\rm M_\odot\ kpc^{-2}}]-6) 
		+ \alpha \log f_{gas} + C
\end{equation}

Certainly the adoption of the right hand side of equation (\ref{KSL2}) to infer $\Sigma_{SFR}$, 
is crude at best. It is true or, at least, likely that 
(i) star formation may be more concentrated than the older population (see \S\ref{Gdata}), 
so that $\Sigma_{SFR}$ is underestimated; (ii) the HI disk is more extended than the stellar one so 
that $\Sigma_{gas}$ and thus $\Sigma_{SFR}$ are both overestimated; (iii) ignoring the contribution of H$_2$ to 
the $\Sigma_{gas}$ also leads to an underestimate of $\Sigma_{SFR}$. 

With the above caveats, equation (\ref{KSL2}) gives a crude relation between the SSFR, $\mu_*$ and $f_{gas}$: 
\begin{eqnarray}
\label{SFmulaw}
	\log SSFR [{\rm yr^{-1}}]
	= \log \frac{0.5 SFR [{\rm M_\odot yr^{-1}}]}{\pi(r_{50, z} [{\rm kpc}] )^2} 
		- \log \frac{0.5 M_* [{\rm M_\odot]}}{\pi(r_{50, z} [{\rm kpc}] )^2} \nonumber\\
	= (\alpha-1) \log \mu_* [{\rm M_\odot\ kpc^{-2}}] 
		+ \alpha \log f_{gas} - 6\alpha + C. 
\end{eqnarray}

Adopting the values of $C$ and $\alpha$ from equation \ref{KSL}, we plot the right hand side of 
this equation on the x axis of Figure \ref{fig:SFL} panel (d) and the left hand side ($\log SSFR$)
on the y axis. 
Although these two quantities are correlated as expected with a slope of $\sim 1$, 
the main distribution is shifted from the one-to-one correlation, represented by the red dashed line, 
in the sense that the right hand side overestimates the SSFR, i.e. $\Sigma_{SFR}$. 
Among the three possible caveats to the adoption of the right hand side of equation (\ref{KSL2}) to 
infer $\Sigma_{SFR}$ which might lead to systematic errors in the $\Sigma_{SFR}$ estimates, 
only the second one, that the HI gas extends beyond the stellar disk, would yield a shift
in the correct direction. Additional factors that might cause such a shift include: 
(iv) The ALFALFA galaxies have sub-solar metallicity on average, requiring an even smaller $C$ (see \S\ref{Z}); 
(v) because they are gas-rich in general, they also have lower SFEs relative to the typical SDSS galaxies 
(see \S\ref{SFE}), and thus smaller $C$ values. Both these interpretations
are consistent with what is evident in Figure \ref{fig:SFL} panel (d), i.e., the standard $C$ \citep{Kenn1998b} 
may be an overestimate for an HI selected population. 
Furthermore, a plot with the same x and y axes as in Figure \ref{fig:SFL}(d) but color coded by the mean SFEs of galaxies 
within each grid cell clearly shows that the offset from the one-to-one line increases with decreasing SFE. 

Rather than adopting $C$ and $\alpha$ values from the Kennicutt-Schmidt law, 
\citet{Zhang2009} calibrated an empirical estimator of $f_{gas}$ from $SSFR$ and $\mu_*$, reformulated as, 
\begin{eqnarray*}
	\log SSFR = 2.96 \log \mu_* + 3.85 \log f_{gas} - 32.81, 
\end{eqnarray*}
corresponding to larger $\alpha$ and smaller $C$ values than given in equation \ref{KSL}. 
Their best estimator of $f_{gas}$ had $SSFR$ substituted by $g-r$ color, 
since optical colors can be good proxies for SSFRs (see \S\ref{SFS}). 
In agreement with \citet{Zhang2009}, we also prefer a smaller SFE scaling factor than the 
\citet{Kenn1998b} value (see Eqn \ref{KSL}). However, \citet{Zhang2009} calibrate their $f_{gas}$ estimator 
from an SDSS selected sample with existing HI data and therefore relatively gas poor compared to the 
ALFALFA HI-selected population. For this reason, their
estimator systematically underpredicts the $f_{gas}$ of the $\alpha.40$-SDSS galaxies by $\sim$0.3 dex.  

Panel (d) of Figure \ref{fig:SFL} shows that, compared to the full 
$\alpha.40$ distribution, the ALFALFA dwarf galaxies 
appear slightly closer to the one-to-one line given by equation (\ref{SFmulaw}). 
Again, this result may arise from
their selection as the lowest $M_{HI}$ systems. As we have seen, the dwarfs studied here
are relatively gas-poor (see \S\ref{fgas}) and have higher SFEs (see \S\ref{SFE}) 
compared to those estimated for the full $\alpha.40$ population in the same {\it stellar mass} 
range. Therefore the fact that the dwarfs lie closer to the one-to-one relation 
traced by equation (\ref{SFmulaw}) is consistent 
with the finding above that lower than typical SFEs drive the overall ALFALFA distribution 
to lower SSFRs for their $\mu_*$ and $f_{gas}$. Comparisons of the star forming properties of samples selected 
either by stellar mass or by HI mass must account for the biases imposed on those properties
by the various selection criteria.

\section{Summary}
\label{summary}
With unprecedented sensitivity, areal coverage, angular and velocity resolution, the
ALFALFA extragalactic HI survey provides us with a homogeneous parent sample of gas-bearing galaxies 
which approaches a full census of star-forming galaxies. More than 2/3 of ALFALFA sources 
are new HI detections and thus would not have been included in previous star formation studies of 
large HI samples based on optical target selection, e.g. \citet{Bothwell2009}. 
In this paper, we have examined the population of galaxies detected by ALFALFA 
with the lowest HI masses, combining photometric and spectroscopic data from the SDSS
and new UV images from GALEX to probe the relationship between their HI, stellar and star-forming 
components. Compared to other dwarf studies such as 
FIGGS \citep{Begum2008} or 11HUGS \citep{Lee2007}, the ALFALFA dwarf sample includes a much larger 
number of very low HI mass objects ($\log M_{HI} < 7.7$). 
Moreover, only 56\% of the ALFALFA dwarfs within the SDSS footprint have a 
counterpart in the SDSS spectroscopic survey confirming that the ALFALFA 
survey is providing an important contribution to the
census of dwarf galaxies in the nearby universe.

Being sensitive to very low levels of star formation activity (below $10^{-2} {\rm M_\odot\ yr^{-1}}$),
the FUV luminosity is a more robust tracer of the SFR in metal- and dust-poor dwarfs than the 
H$\alpha$ or IR predictors. We have reprocessed all of the available MIS-level 
GALEX FUV images for 118 ALFALFA dwarfs. 
Only $\it one$ out of those 118 dwarfs is not detected in the FUV: 
virtually all HI-selected gas-rich dwarfs show some evidence of recent star formation.
Examinations of the CMD, UV colors, BPT diagram and $D_n(4000)$ 
all confirm the general notion that HI-selected dwarf galaxy samples are dominated by faint blue cloud
galaxies and are more likely to be star-forming than galaxies typically included
in optical-UV selected samples
\citep[e.g.][]{Brinchmann2004, Salim2007, Wyder2007, Catinella2010}. 

SED fitting to UV/optical bands is improved via trimming out the models with unrealistically high extinction. 
We demonstrate that estimates of the $SFR_{FUV}$ still suffer from 
internal extinction in the ALFALFA dwarfs. Even in such low metallicity, low 
luminosity systems, 
neglecting the extinction at FUV wavelengths $A_{FUV}$, 
as has been the practice of some authors, e.g. \citet{Roy2009}, 
will lead to the systematic underestimate of FUV luminosity. 
At the same time, application of the standard $IRX-\beta$ relation
given their UV colors would over-predict $A_{FUV}$
in many of the lowest HI mass galaxies, possibly because of their 
different SFH and/or dust geometry.

Although limited to the SDSS fiber apertures, the SFRs derived from SDSS H$\alpha$ 
line flux are compared to those derived from SED-fitting and we cannot rule out
systematic deficiency in the SFR values derived from H$\alpha$ as might be
consistent with 
\citet{Lee2009}. Because the
ALFALFA dwarfs have low SFRs on average, they provide an ideal sample 
for H$\alpha$ imaging to probe this question; this work is in progress.
 
Many ALFALFA dwarfs have high SSFRs and $b$-parameters greater than 1, 
suggesting that they are currently forming stars in a bursty manner. As a result, it is essential 
to take the possibility of bursty star formation into full account when modeling their SEDs. 
If a constant SFH is assumed, the calibrations used most often will
over-predict the derived properties, e.g. the stellar mass as derived from 
the mass to light ratio versus color relation \citep{Bell2003}, 
or the SFR obtained from the FUV luminosity \citep{Kenn1998a, Lee2009}.

In the SSFR vs. $M_*$ diagram, 
the red sequence galaxies possess low SSFRs.
The dispersion in the SSFR distribution increases at both the high and low mass ends. 
Factors other than $M_*$ such as environmental effects likely 
play important roles in the regulation and quenching of star formation in low mass galaxies. 
In the $f_{gas}$ vs. $M_*$ plot, 
although \citet{Bothwell2009} reported broadening in the $f_{gas}$ 
distribution only at the high mass end, we find a mild increase in 
the dispersion of $f_{gas}$ only at the low mass end.  
Within the ALFALFA dwarf {\it s-sed} sample, Virgo cluster members have lower gas fractions
at a given $M_{HI}$ with a wide spread also in the distributions
of SSFR or the $b$-parameter. A study of the
gas and stars in a larger sample of VCC Virgo dwarfs selected by optical criteria will be
presented elsewhere \citep{Hallenbeck2012}.

A clear decrease of the $f_{gas}$ with increasing $M_*$ is seen for the whole distribution, 
i.e. on average, the low mass galaxies are extremely gas rich for their $M_*$ and 
relatively unevolved. However, because the ALFALFA dwarfs are selected to be the sources with 
lowest HI masses, the imposition of an HI mass cutoff leads to the selection of
a dwarf sample with lower gas fractions for their stellar mass than is characteristic
of the $\alpha.40$ sample overall. Therefore, 
a large portion of the ALFALFA dwarfs studied here still have $T_{cons} \lesssim t_H$, 
in particular the more gas poor Virgo members. 

The HI mass correlates well with the SFR on global scales. 
Under the simple assumption that the aperture subtended by the half light radius also contains exactly half of the 
stellar mass, HI mass and star formation, we substitute the general ``Kennicutt-Schmidt'' 
star-formation law \citep{Kenn1998b} into the expression of the SSFR, leading to a predicted relation
between SSFR, stellar surface mass density and gas fraction as given by equation (\ref{SFmulaw}); this relation 
also depends on the exponent $\alpha$ and scaling factor $C$ in the star formation law. 
The observed trend is consistent with the assumption that HI disks are in general more extended than the corresponding 
stellar ones. Additional possible contributions are that ALFALFA galaxies have generally 
sub-solar metallicity and, probably more importantly, lower SFEs, both of which may lead to lower $C$ values. 

A statistical sample such as that provided by ALFALFA allows the study only of global relations between
the constituent populations within the galaxies. 
Clearly, a full understanding of the interaction between gas and stars and the physical mechanisms
responsible for laws of star formation requires resolved maps of the distribution of HI, and where 
feasible, the molecular component as well as deeper and better probes of the stellar population. 
ALFALFA is already providing the fundamental target list for an on-going EVLA/Spitzer/HST project exploring
the lowest HI mass galaxies detected by ALFALFA: SHIELD 
\citep[Survey of HI in Extremely Low Mass Dwarfs,][]{Cannon2011}. The completion of the ALFALFA survey 
over then next few years will 
yield an even richer sample of dwarf galaxies than that studied here.

\acknowledgements
The authors would like to acknowledge the work of the entire ALFALFA collaboration team 
in observing, flagging, and extracting the catalog of galaxies used in this work. 
The ALFALFA team at Cornell is supported by NSF grant AST-0607007 and AST-1107390
and by grants from the Brinson Foundation. 

GALEX is a NASA Small Explorer, launched in 2003 
April. We gratefully acknowledge NASA's support for construction, operation and science 
analysis for the GALEX mission, developed in cooperation with the Centre National 
d'Etudes Spatiales of France and the Korean Ministry of Science and Technology. 
SH, SS and MPH acknowledge support for this work 
from the GALEX Guest Investigator program under NASA grants NNX07AJ12G, NNX08AL67G
and NNX09AF79G. 

Funding for the SDSS and SDSS-II has been provided by the Alfred P. Sloan Foundation, 
the participating institutions, the National Science Foundation, the US Department 
of Energy, the NASA, the Japanese Monbukagakusho, the Max Planck Society and 
the Higher Education Funding Council for England. The SDSS Web Site is 
http://www.sdss.org/.
The SDSS is managed by the Astrophysical Research Consortium for the 
participating institutions. The participating institutions are the American 
Museum of Natural History, Astrophysical Institute Potsdam, University of Basel, 
University of Cambridge, Case Western Reserve University, University of Chicago, 
Drexel University, Fermilab, the Institute for Advanced Study, the Japan 
Participation Group, Johns Hopkins University, the Joint Institute for Nuclear 
Astrophysics, the Kavli Institute for Particle Astrophysics and Cosmology, the
 Korean Scientist Group, the Chinese Academy of Sciences (LAMOST), Los Alamos 
National Laboratory, the Max Planck Institute for Astronomy, the MPA, New Mexico 
State University, Ohio State University, University of Pittsburgh, University 
of Portsmouth, Princeton University, the United States Naval Observatory and 
the University of Washington.

\begin{deluxetable}{rrrrrrrrrrrr}
\rotate
\tablecolumns{12}
\tablewidth{0pt}
\tabletypesize{\scriptsize}
\tablecaption{Properties of dwarf sample \label{table:dwarf}}
\tablehead{
\colhead{AGC}	&\colhead{$RA$(J2000)}	&\colhead{$Dec$(J2000)}	&
\colhead{$m_{FUV}$}	&\colhead{$m_{NUV}$}	&
\colhead{$r$}	&\colhead{$u-r$}&\colhead{sFlag}&
\colhead{$D$}	&\colhead{$\log M_{HI}$}	&
\colhead{$\log M_*$}	&\colhead{$\log SFR$}	\\
\colhead{}	&\colhead{[$^{h}\ ^{m}\ ^{s}$]}	&\colhead{[$^\circ\ '\ ''$]} &
\colhead{[mag]} 	&\colhead{[mag]}		&
\colhead{[mag]}	&\colhead{[mag]}	&\colhead{}&
\colhead{[Mpc]}	&\colhead{[M$_\odot$]}	&
\colhead{[M$_\odot$]}	&\colhead{[M$_\odot$ yr$^{-1}$]}\\
\colhead{1} & \colhead{2} & \colhead{3} & \colhead{4} &
\colhead{5} & \colhead{6} & \colhead{7} & \colhead{8} &
\colhead{9} & \colhead{10} & \colhead{11} & \colhead{12}
}
\startdata
102728*	&00 00 21.4	&+31 01 19	&20.42(0.09)	&20.00(0.05)	&18.70(0.04)	&0.83(0.16)	&par	&9.1(2.3)	&6.78(0.52)	&5.73(0.53)	&-3.01( 0.55)\\
748778*	&00 06 34.3	&+15 30 39	&19.78(0.09)	&19.65(0.05)	&18.14(0.05)	&0.81(0.19)	&par	&4.6(2.3)	&6.36(1.01)	&5.31(1.01)	&-3.78( 1.12)\\
102558	&00 07 04.6	&+27 01 28	&20.42(0.10)	&20.10(0.06)	&17.82(0.02)	&0.96(1.31)	&pbphot	&41.7(2.3)	&8.27(0.14)	&...	&...\\
748779	&00 07 51.8	&+15 45 18	&...	&...	&18.86(0.05)	&1.08(0.29)	&par	&12.4(2.3)	&7.09(0.38)	&...	&...\\
101772	&00 11 08.6	&+14 14 23	&...	&...	&16.70(0.01)	&0.98(0.06)	&par	&11.7(2.3)	&7.54(0.40)	&...	&...\\
102655	&00 30 13.6	&+24 17 59	&...	&18.90(0.02)	&17.32(0.01)	&1.02(0.06)	&oly	&23.1(2.3)	&7.81(0.22)	&...	&...\\
113753*	&01 21 02.1	&+26 05 20	&19.68(0.07)	&19.33(0.04)	&17.90(0.01)	&0.80(0.05)	&oly	&42.2(2.3)	&8.33(0.14)	&7.30(0.13)	&-1.80( 0.72)\\
114027	&01 34 41.7	&+14 38 36	&...	&...	&17.52(0.02)	&0.33(0.06)	&domi	&10.1(2.3)	&7.40(0.46)	&...	&...\\
112503	&01 38 00.3	&+14 58 58	&...	&...	&16.29(0.01)	&1.24(0.04)	&par	&10.2(2.3)	&7.14(0.46)	&...	&...\\
1171	&01 39 44.8	&+15 53 58	&18.21(0.03)	&17.74(0.01)	&...	&...	&...	&7.3(1.5)	&7.42(0.41)	&...	&...\\
112505	&01 40 09.6	&+15 56 24	&20.03(0.09)	&19.84(0.04)	&...	&...	&...	&10.3(2.3)	&7.12(0.46)	&...	&...\\
112521	&01 41 08.0	&+27 19 20	&20.71(0.13)	&20.00(0.07)	&...	&...	&...	&4.6(2.3)	&6.53(1.02)	&...	&...\\
110482	&01 42 17.3	&+26 22 00	&18.23(0.04)	&17.95(0.02)	&...	&...	&...	&5.6(2.3)	&6.99(0.84)	&...	&...\\
111945	&01 44 42.7	&+27 17 18	&...	&...	&...	&...	&...	&6.4(2.3)	&7.48(0.73)	&...	&...\\
111946	&01 46 42.2	&+26 48 05	&...	&...	&...	&...	&...	&5.7(2.3)	&6.76(0.82)	&...	&...\\
111977	&01 55 20.2	&+27 57 14	&18.10(0.03)	&17.74(0.02)	&...	&...	&...	&5.5(0.3)	&6.78(0.13)	&...	&...\\
111164	&02 00 10.1	&+28 49 52	&...	&...	&...	&...	&...	&4.9(0.3)	&6.57(0.14)	&...	&...\\
122401*	&02 28 19.5	&+26 07 31	&17.76(0.03)	&17.39(0.02)	&14.92(0.01)	&1.69(0.05)	&domi	&23.0(2.3)	&8.19(0.21)	&8.43(0.22)	&-1.41( 0.26)\\
122206	&02 31 00.3	&+27 57 30	&18.05(0.05)	&17.75(0.03)	&...	&...	&...	&21.2(2.3)	&8.60(0.22)	&...	&...\\
122400	&02 31 22.1	&+25 42 45	&...	&...	&18.71(0.10)	&0.59(0.42)	&pbphot	&12.7(2.3)	&7.53(0.37)	&...	&...\\
122212*	&02 31 39.4	&+27 10 46	&18.06(0.03)	&17.81(0.02)	&15.32(0.00)	&1.45(0.03)	&domi	&13.3(2.3)	&7.36(0.36)	&7.74(0.36)	&-2.24( 0.46)\\
123162	&02 32 39.7	&+29 26 21	&...	&...	&...	&...	&...	&13.3(2.3)	&7.30(0.36)	&...	&...\\
122397	&02 38 05.9	&+30 40 16	&...	&...	&...	&...	&...	&11.7(2.3)	&7.60(0.40)	&...	&...\\
122219	&02 40 55.6	&+26 40 05	&...	&...	&16.62(0.02)	&1.13(0.07)	&par	&19.7(2.3)	&7.66(0.26)	&...	&...\\
123170	&02 44 03.2	&+29 17 19	&...	&...	&17.34(0.02)	&1.67(0.24)	&pbphot	&12.1(2.3)	&7.68(0.39)	&...	&...\\
122424	&02 45 07.1	&+25 56 10	&...	&...	&...	&...	&...	&20.4(2.3)	&7.62(0.25)	&...	&...\\
122226	&02 46 38.9	&+27 43 35	&...	&...	&...	&...	&...	&6.8(2.3)	&7.39(0.68)	&...	&...\\
122900	&02 50 27.3	&+24 18 34	&...	&18.56(0.04)	&...	&...	&...	&18.5(2.3)	&7.94(0.25)	&...	&...\\
174585	&07 36 10.3	&+09 59 11	&19.16(0.04)	&18.76(0.02)	&...	&...	&...	&5.0(2.3)	&6.50(0.94)	&...	&...\\
174605	&07 50 21.6	&+07 47 40	&...	&...	&...	&...	&...	&4.8(2.3)	&6.55(0.97)	&...	&...\\
174514	&07 52 30.9	&+11 49 40	&20.12(0.07)	&20.09(0.04)	&19.97(0.04)	&0.36(0.13)	&oly	&30.4(2.3)	&8.26(0.16)	&5.75(0.19)	&-2.44( 0.17)\\
4115	&07 57 01.9	&+14 23 29	&15.48(0.01)	&15.48(0.01)	&15.14(0.01)	&1.18(0.03)	&pbphot	&7.7(0.5)	&8.51(0.12)	&...	&...\\
181471	&08 03 24.6	&+15 08 28	&18.67(0.03)	&18.27(0.02)	&17.56(0.02)	&1.36(0.11)	&pbphot	&30.3(2.3)	&8.36(0.16)	&...	&...\\
182460	&08 03 43.9	&+10 08 58	&18.23(0.02)	&18.00(0.01)	&17.23(0.01)	&0.90(0.03)	&pbphot	&39.1(2.3)	&8.60(0.14)	&...	&...\\
188862	&08 09 17.8	&+08 43 39	&...	&...	&19.47(0.02)	&1.47(0.12)	&oly	&17.2(2.3)	&7.54(0.29)	&...	&...\\
188955	&08 21 37.0	&+04 19 01	&...	&...	&18.35(0.03)	&0.47(0.07)	&pbphot	&11.8(2.3)	&7.30(0.40)	&...	&...\\
188762*	&08 23 31.3	&+15 09 05	&20.01(0.09)	&19.65(0.06)	&18.60(0.04)	&0.63(0.13)	&domi	&38.7(2.3)	&8.38(0.14)	&7.29(0.23)	&-1.69( 0.20)\\
188875*	&08 26 30.6	&+11 47 12	&19.12(0.04)	&18.79(0.02)	&16.41(0.01)	&1.43(0.05)	&par	&29.3(2.3)	&7.94(0.18)	&7.89(0.18)	&-2.16( 0.55)\\
182595	&08 51 12.1	&+27 52 48	&...	&...	&16.47(0.02)	&1.27(0.05)	&domi	&5.9(2.3)	&6.53(0.80)	&...	&...\\
182462	&08 52 33.8	&+13 50 28	&18.84(0.06)	&18.26(0.03)	&16.60(0.01)	&1.23(0.06)	&domi	&23.9(2.3)	&8.57(0.20)	&7.44(0.20)	&-1.41( 0.24)\\
191791	&09 08 53.8	&+14 35 02	&20.72(0.10)	&20.04(0.06)	&17.36(0.03)	&0.94(0.09)	&par	&9.5(0.8)	&6.77(0.36)	&6.66(0.19)	&-3.09( 0.60)\\
198507	&09 15 25.8	&+25 25 10	&...	&...	&18.48(0.03)	&0.52(0.07)	&oly	&7.4(2.3)	&6.95(0.63)	&...	&...\\
198354	&09 16 30.9	&+09 10 24	&...	&...	&18.05(0.03)	&1.91(0.25)	&oly	&20.7(2.3)	&7.47(0.26)	&...	&...\\
191894*	&09 21 15.0	&+09 43 52	&19.89(0.06)	&19.50(0.03)	&17.21(0.01)	&1.35(0.04)	&oly	&21.9(2.3)	&7.39(0.25)	&7.37(0.23)	&-2.23( 0.37)\\
198508	&09 22 57.0	&+24 56 48	&...	&...	&17.47(0.02)	&1.15(0.08)	&domi	&7.7(2.3)	&6.71(0.62)	&...	&...\\
192039*	&09 47 31.4	&+10 29 32	&19.65(0.06)	&19.23(0.04)	&17.61(0.02)	&1.40(0.08)	&oly	&47.4(2.3)	&8.39(0.15)	&7.78(0.14)	&-1.48( 0.17)\\
191803	&09 48 05.9	&+07 07 43	&...	&...	&15.99(0.01)	&1.27(0.04)	&domi	&7.2(2.3)	&7.33(0.65)	&...	&...\\
193921	&09 49 14.9	&+15 48 27	&20.63(0.10)	&20.67(0.07)	&18.36(0.04)	&1.58(0.24)	&oly	&23.3(2.3)	&7.90(0.22)	&6.76(0.22)	&-3.43( 0.67)\\
731430	&09 57 29.4	&+27 45 24	&...	&...	&17.44(0.02)	&1.11(0.06)	&par	&19.3(2.3)	&7.55(0.26)	&...	&...\\
5373	&10 00 00.0	&+05 19 56	&13.66(0.00)	&13.44(0.00)	&20.60(0.04)	&0.30(0.10)	&pbphot	&1.3(0.6)	&7.51(0.91)	&...	&...\\
205097	&10 00 02.4	&+15 46 07	&20.46(0.09)	&19.78(0.05)	&17.52(0.02)	&2.17(0.19)	&domi	&35.9(2.3)	&8.21(0.16)	&8.38(0.14)	&-1.89( 0.25)\\
205104*	&10 01 55.2	&+15 45 37	&19.50(0.05)	&19.34(0.03)	&17.72(0.02)	&1.27(0.11)	&par	&34.8(2.3)	&8.32(0.15)	&7.53(0.16)	&-1.93( 0.19)\\
205105*	&10 02 51.0	&+14 33 12	&18.35(0.03)	&18.17(0.02)	&16.87(0.01)	&0.96(0.03)	&domi	&42.8(2.3)	&8.27(0.14)	&7.73(0.13)	&-1.20( 0.17)\\
202240	&10 21 20.2	&+12 10 37	&17.68(0.02)	&17.23(0.01)	&15.73(0.02)	&1.52(0.03)	&pbphot	&43.3(2.3)	&8.83(0.11)	&...	&...\\
731448	&10 23 45.0	&+27 06 39	&...	&...	&16.02(0.01)	&1.01(0.02)	&oly	&7.5(2.3)	&7.10(0.62)	&...	&...\\
202243*	&10 26 41.8	&+11 53 51	&18.75(0.03)	&18.44(0.02)	&16.77(0.02)	&1.09(0.04)	&par	&35.3(2.3)	&8.78(0.13)	&7.71(0.15)	&-1.37( 0.23)\\
208394	&10 28 43.8	&+04 44 04	&...	&...	&21.54(0.12)	&0.58(0.38)	&pbphot	&19.2(2.3)	&7.69(0.26)	&...	&...\\
731454	&10 28 58.6	&+25 17 13	&...	&...	&16.90(0.02)	&1.31(0.04)	&domi	&20.7(2.3)	&7.31(0.27)	&...	&...\\
749315	&10 29 06.4	&+26 54 38	&...	&...	&18.64(0.02)	&0.67(0.06)	&oly	&9.2(2.3)	&6.83(0.52)	&...	&...\\
749316	&10 30 09.6	&+27 23 19	&...	&...	&18.03(0.02)	&0.91(0.09)	&pbphot	&21.7(2.3)	&7.69(0.23)	&...	&...\\
203709	&10 30 44.3	&+06 07 31	&...	&...	&15.90(0.02)	&1.21(0.04)	&domi	&8.2(2.3)	&7.25(0.57)	&...	&...\\
205156	&10 30 52.9	&+12 26 48	&...	&...	&18.17(0.02)	&1.29(0.06)	&oly	&11.1(0.7)	&6.97(0.18)	&...	&...\\
202015	&10 31 54.1	&+12 55 38	&20.16(0.08)	&19.70(0.05)	&20.69(0.07)	&2.07(0.80)	&pbphot	&43.1(2.3)	&8.28(0.16)	&...	&...\\
731457	&10 31 55.8	&+28 01 33	&17.80(0.02)	&17.62(0.01)	&16.11(0.01)	&1.23(0.03)	&pbphot	&6.1(2.3)	&6.74(0.76)	&...	&...\\
204139	&10 32 01.3	&+04 20 46	&...	&...	&17.51(0.02)	&1.11(0.06)	&par	&18.7(2.3)	&7.58(0.27)	&...	&...\\
202248	&10 34 56.1	&+11 29 32	&18.87(0.04)	&18.63(0.02)	&16.89(0.01)	&0.96(0.05)	&par	&11.1(0.7)	&7.29(0.15)	&6.68(0.15)	&-2.56( 0.49)\\
205165	&10 37 04.8	&+15 20 15	&18.51(0.03)	&18.01(0.02)	&15.56(0.01)	&1.53(0.04)	&domi	&11.1(0.7)	&6.94(0.16)	&7.44(0.15)	&-2.15( 0.20)\\
208397	&10 38 58.1	&+03 52 27	&...	&...	&20.01(0.05)	&0.72(0.21)	&oly	&10.2(2.3)	&7.01(0.47)	&...	&...\\
200512	&10 39 55.6	&+13 54 34	&20.04(0.13)	&19.10(0.06)	&18.85(0.10)	&1.25(0.56)	&pbphot	&11.1(0.7)	&6.95(0.16)	&...	&...\\
208399	&10 40 10.7	&+04 54 32	&...	&...	&19.22(0.06)	&1.40(0.38)	&pbphot	&9.9(2.3)	&7.39(0.47)	&...	&...\\
205078	&10 41 26.1	&+07 02 16	&...	&...	&18.42(0.05)	&0.90(0.19)	&par	&19.4(2.3)	&7.58(0.26)	&...	&...\\
200532*	&10 42 00.3	&+12 20 07	&18.38(0.03)	&17.85(0.01)	&15.20(0.00)	&1.45(0.02)	&domi	&11.1(0.7)	&7.48(0.14)	&7.50(0.14)	&-2.93( 0.51)\\
203082	&10 42 26.5	&+13 57 26	&...	&...	&16.57(0.02)	&1.58(0.07)	&domi	&17.5(1.1)	&7.59(0.15)	&...	&...\\
202024	&10 44 57.5	&+11 54 58	&19.49(0.04)	&19.24(0.03)	&17.09(0.02)	&1.05(0.06)	&par	&11.1(0.7)	&6.82(0.18)	&6.74(0.16)	&-3.37( 0.71)\\
205270	&10 45 09.7	&+15 27 00	&...	&...	&16.23(0.00)	&1.38(0.03)	&oly	&17.5(1.1)	&7.49(0.17)	&...	&...\\
201970	&10 46 53.2	&+12 44 40	&20.84(0.24)	&20.13(0.12)	&18.63(0.10)	&1.24(0.49)	&pbphot	&11.1(0.7)	&7.27(0.13)	&...	&...\\
201959*	&10 47 27.5	&+13 53 22	&19.41(0.04)	&18.74(0.02)	&16.35(0.01)	&1.48(0.03)	&domi	&45.6(2.3)	&8.11(0.15)	&8.42(0.12)	&-0.92( 0.18)\\
200603	&10 49 17.1	&+12 25 20	&17.72(0.03)	&17.21(0.01)	&16.04(0.01)	&1.15(0.03)	&pbphot	&17.5(1.1)	&8.48(0.13)	&...	&...\\
205197*	&10 49 42.8	&+13 49 41	&20.76(0.08)	&20.64(0.06)	&19.31(0.04)	&1.29(0.16)	&oly	&17.5(1.1)	&7.45(0.15)	&6.24(0.17)	&-3.13( 0.18)\\
205198*	&10 50 01.8	&+13 47 05	&18.35(0.02)	&18.05(0.01)	&16.53(0.00)	&0.99(0.02)	&oly	&17.5(1.1)	&7.67(0.15)	&7.22(0.14)	&-1.82( 0.16)\\
205327*	&10 53 35.5	&+11 00 21	&19.53(0.05)	&19.49(0.03)	&18.01(0.01)	&1.03(0.05)	&oly	&44.7(2.3)	&8.41(0.16)	&7.35(0.12)	&-1.75( 0.17)\\
6014	&10 53 42.7	&+09 43 39	&16.92(0.01)	&16.77(0.01)	&15.09(0.01)	&1.48(0.05)	&pbphot	&11.1(0.7)	&7.97(0.13)	&...	&...\\
202035	&10 56 13.9	&+12 00 37	&17.76(0.02)	&17.55(0.01)	&16.29(0.01)	&0.79(0.04)	&domi	&11.1(0.7)	&7.73(0.13)	&6.66(0.14)	&-1.97( 0.23)\\
205278*	&10 58 52.2	&+14 07 46	&20.57(0.08)	&19.55(0.03)	&16.39(0.01)	&1.65(0.06)	&par	&11.1(0.7)	&7.01(0.19)	&7.17(0.15)	&-3.90( 0.74)\\
215256*	&11 03 26.3	&+16 01 00	&19.11(0.04)	&18.56(0.03)	&16.08(0.03)	&1.68(0.07)	&domi	&21.0(2.3)	&7.87(0.23)	&7.94(0.24)	&-1.71( 0.27)\\
219117*	&11 03 46.7	&+08 34 19	&20.12(0.07)	&19.85(0.04)	&18.01(0.02)	&1.35(0.08)	&par	&17.5(1.1)	&7.69(0.15)	&6.80(0.15)	&-2.63( 0.18)\\
210023*	&11 04 26.3	&+11 45 21	&17.50(0.02)	&17.21(0.01)	&15.34(0.01)	&1.10(0.02)	&par	&11.1(0.7)	&7.76(0.13)	&7.35(0.15)	&-1.94( 0.29)\\
213757	&11 05 59.6	&+07 22 25	&...	&...	&16.77(0.01)	&1.34(0.03)	&oly	&17.5(1.1)	&7.64(0.15)	&...	&...\\
215262	&11 06 35.3	&+12 13 48	&...	&...	&17.93(0.03)	&1.63(0.22)	&pbphot	&17.5(1.1)	&7.61(0.14)	&...	&...\\
731550	&11 07 07.7	&+28 03 23	&...	&...	&16.76(0.01)	&1.12(0.05)	&par	&24.6(2.3)	&7.62(0.26)	&...	&...\\
210082*	&11 09 23.2	&+10 50 03	&16.88(0.02)	&16.47(0.01)	&14.44(0.00)	&1.43(0.01)	&par	&17.5(1.1)	&8.29(0.13)	&8.36(0.14)	&-1.16( 0.21)\\
210111	&11 10 25.1	&+10 07 34	&16.89(0.02)	&16.67(0.01)	&15.57(0.01)	&1.32(0.03)	&pbphot	&17.5(1.1)	&8.34(0.13)	&...	&...\\
219368	&11 12 21.6	&+24 04 39	&...	&...	&21.20(0.12)	&1.57(0.94)	&pbphot	&10.2(2.3)	&7.31(0.46)	&...	&...\\
6245	&11 12 39.8	&+09 03 21	&17.85(0.02)	&16.59(0.01)	&12.54(0.00)	&2.04(0.01)	&domi	&17.5(1.1)	&8.00(0.14)	&9.52(0.16)	&-0.92( 0.28)\\
213796*	&11 12 52.7	&+07 55 19	&19.12(0.04)	&18.69(0.02)	&16.83(0.01)	&1.15(0.03)	&oly	&17.5(1.1)	&7.62(0.15)	&7.18(0.14)	&-1.93( 0.17)\\
215240	&11 13 50.8	&+09 57 39	&...	&...	&18.04(0.02)	&0.95(0.06)	&oly	&17.5(1.1)	&7.52(0.15)	&...	&...\\
212824	&11 13 59.4	&+11 19 49	&18.11(0.02)	&17.91(0.02)	&20.39(0.07)	&1.07(0.38)	&pbphot	&44.5(2.3)	&9.04(0.11)	&...	&...\\
215282	&11 14 25.2	&+15 32 02	&...	&...	&16.07(0.00)	&1.28(0.02)	&pbphot	&11.3(2.3)	&6.94(0.43)	&...	&...\\
202256*	&11 14 45.0	&+12 38 51	&18.98(0.04)	&18.70(0.02)	&16.87(0.01)	&1.22(0.04)	&par	&10.0(0.6)	&7.20(0.13)	&6.72(0.14)	&-2.61( 0.16)\\
215284	&11 15 32.4	&+14 34 38	&...	&...	&17.25(0.02)	&1.17(0.05)	&oly	&19.7(2.3)	&7.59(0.25)	&...	&...\\
215186	&11 17 01.2	&+04 39 44	&...	&...	&17.75(0.01)	&1.41(0.07)	&oly	&24.0(2.3)	&7.56(0.24)	&...	&...\\
210220	&11 17 01.1	&+13 05 55	&19.08(0.04)	&18.18(0.02)	&15.49(0.01)	&2.03(0.06)	&domi	&10.0(0.6)	&7.15(0.14)	&7.66(0.13)	&-1.89( 0.16)\\
215286	&11 19 12.7	&+14 19 40	&...	&...	&20.32(0.07)	&1.68(0.44)	&pbphot	&10.0(0.6)	&7.12(0.13)	&...	&...\\
202257	&11 19 14.4	&+11 57 07	&17.75(0.02)	&17.54(0.01)	&16.32(0.02)	&1.32(0.06)	&pbphot	&10.7(2.3)	&7.95(0.44)	&...	&...\\
213074	&11 19 28.1	&+09 35 44	&17.80(0.02)	&17.58(0.01)	&17.01(0.01)	&0.45(0.02)	&domi	&13.7(2.3)	&7.98(0.34)	&6.28(0.35)	&-1.79( 0.34)\\
213511	&11 22 23.4	&+11 47 38	&...	&...	&17.08(0.01)	&1.15(0.03)	&oly	&17.5(1.1)	&7.47(0.18)	&...	&...\\
213440	&11 23 37.6	&+12 53 45	&...	&19.18(0.03)	&16.44(0.01)	&1.42(0.05)	&oly	&10.0(0.6)	&6.77(0.17)	&...	&...\\
215142*	&11 24 44.5	&+15 16 32	&18.38(0.04)	&17.95(0.02)	&16.10(0.01)	&1.29(0.03)	&par	&20.0(2.3)	&8.37(0.23)	&7.61(0.25)	&-1.61( 0.25)\\
215296*	&11 26 55.1	&+14 50 03	&19.94(0.08)	&19.58(0.04)	&18.33(0.03)	&0.82(0.07)	&par	&11.5(2.3)	&7.26(0.41)	&5.94(0.41)	&-2.78( 0.59)\\
219203	&11 27 28.9	&+05 37 02	&...	&...	&18.03(0.03)	&0.95(0.09)	&par	&25.0(2.3)	&7.67(0.21)	&...	&...\\
212837	&11 30 53.4	&+14 08 46	&...	&17.74(0.02)	&16.68(0.01)	&1.17(0.07)	&pbphot	&10.7(2.3)	&7.72(0.44)	&...	&...\\
215303	&11 31 08.8	&+13 34 14	&...	&19.04(0.03)	&17.32(0.03)	&1.12(0.05)	&domi	&15.0(2.3)	&7.48(0.32)	&...	&...\\
215306	&11 33 50.1	&+14 49 28	&...	&18.67(0.02)	&16.48(0.02)	&1.29(0.04)	&domi	&20.4(2.3)	&7.66(0.24)	&...	&...\\
215248	&11 33 50.9	&+14 03 15	&...	&18.85(0.02)	&16.55(0.01)	&1.31(0.04)	&par	&11.3(2.3)	&6.80(0.44)	&...	&...\\
212838	&11 34 53.4	&+11 01 10	&...	&17.70(0.02)	&17.35(0.02)	&0.32(0.05)	&pbphot	&10.3(2.3)	&7.60(0.45)	&...	&...\\
213155	&11 37 08.6	&+13 15 03	&...	&...	&17.09(0.01)	&0.62(0.04)	&par	&12.0(2.3)	&7.64(0.39)	&...	&...\\
6655*	&11 41 50.5	&+15 58 24	&16.43(0.01)	&16.23(0.01)	&14.15(0.00)	&1.29(0.01)	&oly	&8.7(2.3)	&7.38(0.54)	&7.81(0.54)	&-1.81( 0.54)\\
213333	&11 43 27.0	&+11 23 54	&...	&...	&15.86(0.00)	&0.97(0.02)	&oly	&10.3(2.3)	&7.31(0.46)	&...	&...\\
731804	&11 49 25.8	&+25 37 00	&...	&...	&17.68(0.02)	&1.33(0.06)	&oly	&29.2(2.3)	&7.65(0.28)	&...	&...\\
210822	&11 50 02.7	&+15 01 24	&...	&...	&14.95(0.00)	&0.45(0.01)	&pbphot	&8.6(2.3)	&7.45(0.54)	&...	&...\\
213174	&11 51 04.8	&+05 14 46	&...	&18.54(0.02)	&17.01(0.01)	&0.46(0.03)	&par	&25.0(2.3)	&7.84(0.21)	&...	&...\\
215213	&11 52 20.2	&+15 27 36	&...	&...	&17.61(0.16)	&0.96(0.18)	&pbphot	&9.0(2.3)	&7.15(0.52)	&...	&...\\
215145	&11 54 12.4	&+12 26 06	&...	&18.86(0.03)	&19.01(0.08)	&0.57(0.30)	&par	&16.7(4.2)	&8.00(0.51)	&...	&...\\
224237	&12 04 47.1	&+10 37 35	&18.73(0.04)	&18.32(0.02)	&16.78(0.01)	&0.83(0.03)	&par	&39.3(2.3)	&8.43(0.15)	&7.77(0.16)	&-0.93( 0.20)\\
226606	&12 09 21.2	&+25 12 03	&...	&...	&16.29(0.01)	&1.29(0.03)	&domi	&8.7(2.3)	&7.01(0.54)	&...	&...\\
224231	&12 11 59.5	&+05 55 02	&...	&...	&16.78(0.01)	&1.24(0.05)	&par	&8.1(2.3)	&6.78(0.58)	&...	&...\\
220195	&12 12 22.7	&+06 58 46	&19.76(0.05)	&19.55(0.02)	&17.13(0.01)	&1.20(0.08)	&par	&24.2(4.8)	&7.46(0.44)	&7.47(0.40)	&-2.54( 0.55)\\
223286	&12 13 48.1	&+12 41 26	&...	&...	&16.33(0.01)	&1.44(0.05)	&par	&17.5(5.1)	&7.60(0.59)	&...	&...\\
7285	&12 15 56.3	&+14 25 57	&...	&...	&14.42(0.01)	&1.96(0.02)	&domi	&16.7(1.7)	&7.48(0.22)	&...	&...\\
220257	&12 15 53.7	&+14 01 30	&...	&...	&16.68(0.02)	&3.78(0.62)	&pbphot	&16.7(1.7)	&7.51(0.21)	&...	&...\\
222297	&12 16 13.0	&+07 55 45	&...	&18.37(0.02)	&14.61(0.00)	&1.65(0.02)	&domi	&24.2(4.8)	&7.57(0.45)	&...	&...\\
220261	&12 16 11.8	&+08 22 24	&19.56(0.05)	&19.22(0.03)	&17.53(0.03)	&2.40(0.47)	&pbphot	&16.6(2.0)	&7.36(0.27)	&...	&...\\
220282	&12 16 52.4	&+14 30 55	&...	&...	&14.34(0.01)	&1.89(0.03)	&domi	&16.7(1.7)	&7.57(0.22)	&...	&...\\
220289	&12 17 10.7	&+06 25 54	&...	&...	&15.91(0.00)	&1.31(0.02)	&domi	&24.2(4.8)	&7.55(0.43)	&...	&...\\
732041	&12 17 42.4	&+27 29 03	&...	&...	&16.18(0.01)	&1.32(0.04)	&par	&21.6(2.3)	&7.52(0.24)	&...	&...\\
223390	&12 18 07.7	&+05 55 47	&19.09(0.04)	&18.61(0.02)	&16.46(0.01)	&0.82(0.03)	&oly	&29.6(2.3)	&7.69(0.23)	&7.64(0.17)	&-3.70( 0.91)\\
229053	&12 18 15.5	&+25 34 05	&...	&...	&17.32(0.02)	&0.89(0.08)	&par	&8.1(2.0)	&7.08(0.50)	&...	&...\\
220321	&12 18 15.3	&+13 44 56	&...	&...	&15.29(0.00)	&1.46(0.02)	&oly	&16.7(1.7)	&7.50(0.22)	&...	&...\\
223407	&12 18 43.8	&+12 23 08	&19.09(0.05)	&18.41(0.02)	&16.11(0.01)	&1.45(0.05)	&domi	&16.7(0.6)	&7.53(0.11)	&7.59(0.09)	&-1.72( 0.17)\\
220336	&12 18 51.3	&+12 35 50	&18.19(0.03)	&17.83(0.02)	&16.10(0.01)	&0.95(0.04)	&par	&16.7(1.7)	&7.60(0.21)	&7.32(0.21)	&-1.64( 0.26)\\
220354*	&12 19 15.6	&+06 17 37	&18.21(0.02)	&17.60(0.01)	&15.11(0.01)	&1.46(0.02)	&domi	&16.6(2.0)	&7.58(0.26)	&7.98(0.25)	&-1.62( 0.42)\\
223449*	&12 20 43.8	&+14 37 51	&19.29(0.07)	&18.65(0.03)	&16.41(0.01)	&1.21(0.03)	&oly	&16.7(0.6)	&7.31(0.15)	&7.45(0.12)	&-1.98( 0.95)\\
220409	&12 20 40.2	&+13 53 20	&...	&...	&16.27(0.01)	&0.31(0.02)	&oly	&16.7(1.7)	&7.65(0.21)	&...	&...\\
229052	&12 20 41.1	&+24 57 21	&...	&...	&21.29(0.14)	&0.80(0.43)	&pbphot	&9.5(2.3)	&7.04(0.50)	&...	&...\\
224272*	&12 20 38.6	&+05 54 32	&20.60(0.08)	&20.02(0.04)	&17.48(0.01)	&1.33(0.07)	&oly	&16.6(2.0)	&7.49(0.27)	&7.02(0.26)	&-2.96( 0.63)\\
220418	&12 20 57.6	&+06 20 23	&...	&...	&15.33(0.01)	&1.41(0.02)	&par	&16.4(2.5)	&7.49(0.32)	&...	&...\\
220419*	&12 21 00.1	&+12 43 33	&19.02(0.05)	&18.48(0.02)	&16.42(0.01)	&1.27(0.04)	&oly	&16.7(1.7)	&7.81(0.21)	&7.26(0.22)	&-2.52( 0.83)\\
220435	&12 21 27.2	&+15 01 17	&20.10(0.12)	&19.71(0.06)	&20.12(0.09)	&0.82(0.34)	&pbphot	&16.7(1.7)	&7.59(0.21)	&...	&...\\
220450	&12 22 07.6	&+15 47 56	&18.76(0.03)	&17.96(0.01)	&16.72(0.04)	&0.95(0.09)	&pbphot	&16.7(1.7)	&7.62(0.21)	&...	&...\\
220460	&12 22 38.3	&+06 00 53	&...	&...	&16.26(0.01)	&1.57(0.08)	&par	&16.4(2.5)	&7.58(0.32)	&...	&...\\
220483	&12 23 16.1	&+07 41 15	&-999(-999)	&22.04(0.11)	&18.35(0.07)	&1.87(0.80)	&pbphot	&16.4(2.5)	&7.23(0.34)	&...	&...\\
220493	&12 23 28.4	&+05 48 59	&...	&...	&15.62(0.01)	&1.34(0.02)	&par	&16.4(2.5)	&7.57(0.32)	&...	&...\\
226326*	&12 23 58.2	&+07 27 02	&18.60(0.02)	&18.50(0.01)	&17.70(0.01)	&0.59(0.04)	&oly	&16.6(2.0)	&7.63(0.25)	&6.39(0.27)	&-2.05( 0.28)\\
227889*	&12 24 50.9	&+07 53 56	&19.92(0.06)	&19.79(0.03)	&18.20(0.02)	&1.26(0.14)	&par	&16.6(2.0)	&7.30(0.26)	&6.60(0.26)	&-2.78( 0.28)\\
220542*	&12 25 21.4	&+13 04 13	&18.61(0.03)	&18.30(0.01)	&16.27(0.01)	&1.20(0.04)	&par	&16.7(1.7)	&7.59(0.21)	&7.38(0.22)	&-2.11( 0.34)\\
224232*	&12 25 31.5	&+11 09 30	&19.94(0.06)	&19.50(0.02)	&16.82(0.01)	&1.39(0.06)	&domi	&16.6(2.0)	&7.37(0.27)	&7.34(0.26)	&-3.10( 0.59)\\
749236	&12 25 42.4	&+26 48 36	&...	&...	&16.14(0.01)	&0.59(0.03)	&domi	&5.9(2.0)	&7.26(0.68)	&...	&...\\
220555	&12 25 47.4	&+14 57 08	&18.55(0.03)	&17.90(0.01)	&14.90(0.01)	&1.53(0.02)	&domi	&16.7(1.7)	&7.66(0.21)	&8.12(0.22)	&-2.27( 0.52)\\
749237	&12 26 23.4	&+27 44 44	&...	&...	&15.94(0.01)	&1.28(0.03)	&domi	&7.0(2.0)	&7.32(0.57)	&...	&...\\
220609	&12 27 30.1	&+09 20 28	&19.15(0.03)	&18.71(0.02)	&16.56(0.02)	&1.42(0.07)	&pbphot	&16.4(2.5)	&7.22(0.34)	&...	&...\\
220616*	&12 27 33.4	&+10 00 14	&19.28(0.04)	&18.31(0.02)	&14.98(0.01)	&1.81(0.04)	&par	&16.4(2.5)	&7.48(0.32)	&8.17(0.32)	&-2.52( 0.74)\\
223691*	&12 28 27.2	&+06 56 45	&19.98(0.06)	&19.47(0.02)	&16.93(0.01)	&1.42(0.08)	&par	&16.6(2.0)	&7.23(0.27)	&7.15(0.25)	&-3.36( 0.79)\\
7596	&12 28 34.0	&+08 38 22	&18.27(0.03)	&17.22(0.01)	&14.57(0.00)	&1.73(0.04)	&domi	&16.4(2.5)	&7.52(0.31)	&8.20(0.32)	&-1.29( 0.58)\\
222021	&12 28 55.4	&+08 49 00	&20.05(0.06)	&19.96(0.04)	&20.59(0.17)	&0.55(0.44)	&pbphot	&16.6(2.0)	&7.68(0.25)	&...	&...\\
227861	&12 29 59.4	&+08 25 54	&19.26(0.04)	&18.34(0.02)	&15.42(0.00)	&1.50(0.02)	&domi	&16.6(2.0)	&7.50(0.26)	&7.85(0.25)	&-2.60( 0.76)\\
223771	&12 30 32.3	&+10 15 39	&...	&...	&19.62(0.16)	&0.92(0.65)	&pbphot	&16.6(2.0)	&7.40(0.26)	&...	&...\\
724906	&12 30 56.0	&+26 30 41	&...	&...	&16.94(0.01)	&1.04(0.04)	&domi	&11.6(2.3)	&7.40(0.40)	&...	&...\\
220745*	&12 32 22.8	&+16 01 07	&17.57(0.03)	&17.23(0.01)	&14.79(0.00)	&1.54(0.02)	&domi	&16.7(1.7)	&7.93(0.21)	&8.17(0.21)	&-1.67( 0.25)\\
220755*	&12 32 46.4	&+07 47 57	&19.04(0.04)	&18.63(0.02)	&16.01(0.01)	&1.32(0.04)	&par	&16.4(2.5)	&7.18(0.35)	&7.61(0.32)	&-2.67( 0.55)\\
220768	&12 33 09.9	&+11 20 49	&17.89(0.03)	&17.28(0.01)	&15.00(0.01)	&2.88(0.05)	&pbphot	&16.7(1.7)	&7.53(0.26)	&...	&...\\
223873	&12 34 01.5	&+05 57 10	&...	&...	&18.23(0.05)	&2.28(0.66)	&pbphot	&16.6(2.0)	&7.44(0.25)	&...	&...\\
220819	&12 35 30.9	&+06 20 02	&18.22(0.03)	&17.68(0.01)	&14.45(0.00)	&1.43(0.01)	&domi	&15.8(6.4)	&7.52(0.81)	&8.36(0.81)	&-3.03( 0.91)\\
223913	&12 36 02.8	&+07 12 00	&22.09(0.55)	&19.10(0.04)	&15.15(0.01)	&2.19(0.05)	&domi	&16.6(2.0)	&7.11(0.35)	&8.15(0.26)	&-4.17( 0.69)\\
220837	&12 36 34.9	&+08 03 17	&17.64(0.02)	&16.88(0.01)	&14.30(0.00)	&1.82(0.02)	&pbphot	&16.4(2.5)	&7.41(0.31)	&...	&...\\
220856	&12 38 06.8	&+10 09 54	&18.03(0.02)	&17.89(0.01)	&16.60(0.01)	&0.65(0.02)	&oly	&16.7(1.7)	&7.49(0.23)	&6.80(0.20)	&-1.73( 0.21)\\
220860	&12 38 15.5	&+06 59 40	&18.92(0.03)	&18.36(0.02)	&...	&...	&...	&16.4(2.5)	&7.22(0.36)	&...	&...\\
749241	&12 40 01.7	&+26 19 19	&19.54(0.04)	&19.24(0.02)	&18.90(0.04)	&0.83(0.15)	&pbphot	&5.6(2.3)	&6.75(0.83)	&...	&...\\
220903	&12 40 10.4	&+06 50 48	&...	&...	&16.00(0.01)	&1.26(0.05)	&par	&16.6(2.0)	&7.55(0.25)	&...	&...\\
229001	&12 40 49.4	&+27 33 50	&...	&...	&16.57(0.01)	&1.16(0.03)	&domi	&21.3(2.3)	&7.60(0.26)	&...	&...\\
7889	&12 43 07.7	&+12 03 04	&17.53(0.02)	&16.31(0.01)	&13.16(0.00)	&1.65(0.01)	&domi	&16.7(1.7)	&7.69(0.21)	&8.91(0.23)	&-1.66( 0.55)\\
224296	&12 43 22.8	&+05 45 55	&...	&...	&17.04(0.03)	&1.26(0.07)	&par	&15.6(6.9)	&7.60(0.88)	&...	&...\\
225876*	&12 44 57.9	&+12 01 47	&20.26(0.07)	&20.12(0.05)	&19.02(0.04)	&0.98(0.14)	&par	&16.4(6.5)	&7.61(0.80)	&6.00(0.80)	&-2.82( 0.82)\\
221000*	&12 46 04.4	&+08 28 34	&17.04(0.01)	&16.69(0.01)	&14.80(0.00)	&1.29(0.01)	&domi	&16.6(2.0)	&7.46(0.27)	&8.09(0.25)	&-1.33( 0.27)\\
221004	&12 46 15.3	&+10 12 20	&17.53(0.02)	&17.10(0.01)	&15.53(0.01)	&1.13(0.03)	&domi	&16.7(1.7)	&7.66(0.22)	&7.65(0.22)	&-1.38( 0.28)\\
221013	&12 46 55.4	&+26 33 51	&...	&...	&14.21(0.00)	&1.43(0.01)	&domi	&10.4(2.3)	&7.23(0.45)	&...	&...\\
227897	&12 50 04.2	&+06 50 51	&...	&19.26(0.03)	&18.14(0.02)	&0.66(0.09)	&par	&16.6(2.0)	&7.43(0.27)	&...	&...\\
227972	&12 50 24.0	&+04 54 22	&...	&...	&18.20(0.04)	&1.37(0.24)	&par	&16.6(2.0)	&7.41(0.26)	&...	&...\\
227973	&12 50 39.9	&+05 20 52	&...	&...	&19.06(0.05)	&1.12(0.21)	&oly	&16.6(2.0)	&7.39(0.25)	&...	&...\\
224229	&12 53 40.2	&+04 04 32	&...	&...	&16.62(0.01)	&1.50(0.05)	&domi	&16.6(2.0)	&7.48(0.26)	&...	&...\\
224230	&12 53 43.3	&+04 09 14	&...	&...	&16.80(0.02)	&0.62(0.03)	&par	&16.6(2.0)	&7.64(0.25)	&...	&...\\
8030	&12 54 29.1	&+26 18 13	&...	&...	&15.55(0.01)	&1.45(0.04)	&par	&7.8(2.3)	&7.38(0.60)	&...	&...\\
225878	&12 56 03.1	&+12 07 59	&...	&...	&19.74(0.06)	&1.44(0.36)	&oly	&16.0(7.7)	&7.37(0.98)	&...	&...\\
227974	&12 56 03.5	&+04 52 01	&...	&...	&19.48(0.05)	&2.89(1.20)	&pbphot	&16.6(2.0)	&7.68(0.25)	&...	&...\\
227975	&12 57 18.1	&+04 59 29	&...	&...	&18.77(0.05)	&0.83(0.13)	&oly	&16.6(2.0)	&7.53(0.25)	&...	&...\\
8091	&12 58 40.4	&+14 13 02	&...	&...	&15.59(0.01)	&1.43(0.03)	&pbphot	&2.1(0.1)	&7.05(0.12)	&...	&...\\
732418	&12 59 08.2	&+26 22 39	&...	&...	&16.83(0.01)	&1.23(0.04)	&par	&15.0(3.1)	&7.15(0.44)	&...	&...\\
225851	&12 59 42.6	&+11 04 38	&18.24(0.04)	&17.90(0.02)	&16.19(0.01)	&1.10(0.02)	&pbphot	&42.4(2.3)	&8.62(0.13)	&...	&...\\
225852	&12 59 41.9	&+10 43 40	&18.39(0.04)	&18.07(0.02)	&16.35(0.01)	&1.02(0.03)	&domi	&16.6(2.0)	&7.67(0.25)	&7.26(0.25)	&-1.82( 0.28)\\
233575*	&13 02 52.5	&+13 09 23	&20.42(0.18)	&20.17(0.10)	&18.74(0.04)	&1.55(0.20)	&domi	&46.4(2.3)	&8.54(0.12)	&7.14(0.13)	&-1.98( 0.21)\\
238737	&13 13 04.4	&+06 17 07	&...	&...	&17.32(0.02)	&0.90(0.09)	&par	&16.9(5.4)	&7.46(0.65)	&...	&...\\
233627	&13 19 53.0	&+13 48 23	&18.84(0.04)	&18.63(0.02)	&18.69(0.03)	&0.72(0.11)	&pbphot	&16.8(5.2)	&7.98(0.62)	&...	&...\\
732602	&13 21 04.8	&+24 08 36	&...	&...	&16.06(0.01)	&1.19(0.03)	&domi	&11.6(2.3)	&7.41(0.41)	&...	&...\\
238691*	&13 25 17.6	&+05 32 36	&20.42(0.10)	&19.73(0.06)	&16.79(0.02)	&1.42(0.07)	&par	&16.7(4.9)	&7.30(0.60)	&7.31(0.60)	&-3.51( 0.95)\\
233559	&13 28 47.3	&+10 57 41	&19.69(0.08)	&19.89(0.05)	&18.81(0.05)	&1.13(0.24)	&pbphot	&16.0(5.0)	&7.42(0.63)	&...	&...\\
238890	&13 32 30.3	&+25 07 24	&...	&...	&15.96(0.01)	&1.71(0.04)	&domi	&6.6(2.0)	&6.56(0.62)	&...	&...\\
8638	&13 39 19.3	&+24 46 28	&...	&...	&14.43(0.00)	&1.57(0.02)	&pbphot	&4.3(0.4)	&7.27(0.17)	&...	&...\\
238847	&13 45 09.7	&+27 20 11	&...	&...	&19.03(0.05)	&0.60(0.15)	&pbphot	&14.1(2.3)	&7.47(0.34)	&...	&...\\
233681	&13 47 16.0	&+13 10 38	&...	&...	&17.60(0.02)	&1.04(0.10)	&oly	&21.2(2.3)	&7.69(0.24)	&...	&...\\
713655	&13 48 22.8	&+08 12 41	&...	&...	&17.06(0.01)	&0.63(0.03)	&oly	&21.7(2.3)	&7.55(0.23)	&...	&...\\
231980	&13 54 33.5	&+04 14 40	&19.29(0.05)	&18.81(0.03)	&21.05(0.11)	&0.87(0.41)	&pbphot	&2.6(0.2)	&6.11(0.17)	&...	&...\\
238643	&13 55 58.3	&+08 59 36	&19.21(0.04)	&18.88(0.02)	&18.97(0.02)	&0.35(0.07)	&pbphot	&22.3(2.3)	&7.61(0.23)	&...	&...\\
233718*	&13 58 43.4	&+14 15 41	&18.28(0.02)	&18.06(0.01)	&16.54(0.01)	&1.12(0.02)	&domi	&23.1(2.3)	&8.01(0.21)	&7.59(0.22)	&-1.69( 0.24)\\
732832	&13 58 45.0	&+24 09 05	&...	&...	&16.40(0.00)	&1.46(0.02)	&oly	&18.4(2.3)	&7.58(0.27)	&...	&...\\
243852	&14 07 04.5	&+10 42 46	&18.40(0.03)	&18.07(0.01)	&16.59(0.01)	&0.63(0.03)	&pbphot	&21.5(2.3)	&7.90(0.23)	&...	&...\\
244129	&14 18 53.5	&+09 17 29	&...	&...	&17.30(0.02)	&1.12(0.05)	&domi	&21.7(2.3)	&7.56(0.25)	&...	&...\\
241893*	&14 39 44.5	&+05 21 12	&18.66(0.03)	&18.24(0.02)	&16.14(0.01)	&1.25(0.04)	&domi	&28.4(2.3)	&7.68(0.23)	&8.01(0.18)	&-1.35( 0.20)\\
253922*	&15 32 13.0	&+12 01 21	&20.10(0.06)	&19.93(0.04)	&18.52(0.03)	&1.08(0.15)	&oly	&45.8(2.3)	&8.41(0.14)	&7.19(0.14)	&-1.81( 0.22)\\
253923	&15 33 58.9	&+12 03 51	&19.56(0.05)	&19.48(0.03)	&17.68(0.03)	&1.61(0.17)	&domi	&41.3(2.3)	&8.08(0.15)	&7.63(0.15)	&-1.95( 0.18)\\
262734	&16 34 24.7	&+24 57 41	&...	&...	&17.59(0.01)	&1.21(0.05)	&oly	&19.1(2.3)	&7.35(0.29)	&...	&...\\
321203	&22 13 03.0	&+28 04 20	&18.31(0.04)	&18.19(0.02)	&...	&...	&...	&16.4(2.3)	&7.83(0.29)	&...	&...\\
321434	&22 47 44.0	&+30 45 12	&...	&...	&15.80(0.00)	&1.43(0.02)	&oly	&14.1(2.3)	&7.62(0.34)	&...	&...\\
320926	&22 55 58.5	&+26 10 07	&18.41(0.03)	&18.02(0.02)	&16.90(0.01)	&1.08(0.05)	&domi	&39.2(2.3)	&8.92(0.12)	&7.68(0.15)	&-0.98( 0.21)\\
333351	&23 00 32.5	&+30 42 21	&...	&...	&...	&...	&...	&13.3(2.3)	&7.55(0.36)	&...	&...\\
333363	&23 07 41.5	&+30 07 16	&...	&...	&...	&...	&...	&12.5(2.3)	&7.51(0.38)	&...	&...\\
332912	&23 12 01.0	&+27 17 41	&19.67(0.06)	&19.28(0.03)	&...	&...	&...	&12.7(2.3)	&7.31(0.37)	&...	&...\\
12613	&23 28 36.2	&+14 44 34	&15.71(0.01)	&14.71(0.01)	&14.42(0.01)	&1.79(0.06)	&pbphot	&0.9(0.0)	&6.62(0.07)	&...	&...\\
333210	&23 48 41.8	&+25 54 40	&20.25(0.11)	&19.38(0.05)	&...	&...	&...	&23.3(2.3)	&7.96(0.21)	&...	&...\\
12791	&23 48 49.3	&+26 13 14	&16.73(0.02)	&16.42(0.01)	&...	&...	&...	&12.1(2.3)	&8.33(0.38)	&...	&...\\
333214	&23 51 37.9	&+27 28 10	&19.84(0.07)	&19.68(0.04)	&...	&...	&...	&40.3(2.3)	&8.30(0.13)	&...	&...\\
\hline
\enddata
\end{deluxetable}

\end{document}